**Advanced Methods for Analyzing In-Situ Observations of Magnetic Reconnection**

H. Hasegawa[1,*], M.R. Argall[2], N. Aunai[3], R. Bandyopadhyay[4], N. Bessho[5,6], I.J. Cohen[7], R.E. Denton[8], J.C. Dorelli[6], J. Egedal[9], S.A. Fuselier[10,11], P. Garnier[12], V. Génot[12], D.B. Graham[13,*], K.J. Hwang[10], Y.V. Khotyaintsev[13], D.B. Korovinskiy[14], B. Lavraud[12,23], Q. Lenouvel[12], T.C. Li[8], Y.-H. Liu[8], B. Michotte de Welle[3], T.K.M. Nakamura[14,24], D.S. Payne[15], S.M. Petrinec[16], Y. Qi[17], A.C. Rager[6], P.H. Reiff[18], J.M. Schroeder[9], J.R. Shuster[2], M.I. Sitnov[7], G.K. Stephens[7], M. Swisdak[15], A.M. Tian[19], R.B. Torbert[20,21], K.J. Trattner[17], S. Zenitani[14,22]

First and corresponding author E-mail: hase@stp.isas.jaxa.jp
Second corresponding author E-mail: dgraham@irfu.se

[1] Institute of Space and Astronautical Science, Japan Aerospace Exploration Agency, Sagamihara, Kanagawa 252-5210, Japan
[2] Space Science Center, Institute for the Study of Earth, Oceans, and Space, University of New Hampshire, Durham, NH 03824, USA
[3] CNRS, Ecole polytechnique, Sorbonne Université, Université Paris Sud, Observatoire de Paris, Institut Polytechnique de Paris, Université Paris-Saclay, PSL Research Univsersity, Laboratoire de Physique des Plasmas, Palaiseau, France
[4] Department of Astrophysical Sciences, Princeton University, Princeton, NJ 08544, USA
[5] Department of Astronomy, University of Maryland, College Park, Maryland 20742, USA
[6] Heliophysics Science Division, NASA Goddard Space Flight Center, Greenbelt, Maryland 20771, USA
[7] Applied Physics Laboratory, The Johns Hopkins University, Laurel, MD, USA
[8] Department of Physics and Astronomy, Dartmouth College, Hanover, NH, USA
[9] Department of Physics, University of Wisconsin-Madison, Madison, Wisconsin 53706, USA
[10] Southwest Research Institute, San Antonio, TX, USA
[11] University of Texas at San Antonio, San Antonio, TX, USA
[12] Institut de Recherche en Astrophysique et Planétologie, CNRS, Université Paul Sabatier, CNES, Toulouse, France
[13] Swedish Institute of Space Physics, Uppsala, Sweden
[14] Space Research Institute, Austrian Academy of Sciences, Graz, Austria
[15] Institute for Research in Electronics and Applied Physics, University of Maryland,

College Park, MD, USA
[16] Lockheed Martin ATC, Palo Alto, CA, USA
[17] Laboratory for Atmospheric and Space Physics, University of Colorado, Boulder, CO, USA
[18] Rice Space Institute, Rice University, Houston, Texas, USA
[19] Shandong Key Laboratory of Optical Astronomy and Solar-Terrestrial Environment, School of Space Science and Physics, Institute of Space Sciences, Shandong University, Weihai, Shandong 264209, People's Republic of China
[20] Southwest Research Institute, Durham, NH, USA
[21] Physics Department, University of New Hampshire, Durham, NH, USA
[22] Research Center for Urban Safety and Security, Kobe University, Kobe 657-8501, Japan
[23] Laboratoire d'Astrophysique de Bordeaux, Université Bordeaux, CNRS, Pessac, France
[24] Krimgen LLC, Hiroshima 732-0828, Japan

**Abstract**

There is ample evidence for magnetic reconnection in the solar system, but it is a nontrivial task to visualize, to determine the proper approaches and frames to study, and in turn to elucidate the physical processes at work in reconnection regions from in-situ measurements of plasma particles and electromagnetic fields. Here an overview is given of a variety of single- and multi-spacecraft data analysis techniques that are key to revealing the context of in-situ observations of magnetic reconnection in space and for detecting and analyzing the diffusion regions where ions and/or electrons are demagnetized. We focus on recent advances in the era of the Magnetospheric Multiscale mission, which has made electron-scale, multi-point measurements of magnetic reconnection in and around Earth's magnetosphere.

**Keywords**: magnetic reconnection, data analysis techniques, in-situ measurements, magnetosphere, electron diffusion region

**Contents**

## 1 Introduction

Magnetic reconnection occurring in geospace is in the collisionless regime, so that the reconnection and surrounding regions have multi-scale structures: magnetohydrodynamic (MHD) regions where both the ion and electron fluids satisfy the frozen-in condition, ion diffusion regions (IDRs) where ions are demagnetized but electrons remain magnetized, and electron diffusion regions (EDRs) where both ions and electrons are demagnetized and magnetic topology changes (e.g., Daughton et al. 2006; Paschmann et al. 2013). Since its launch in 2015, the Magnetospheric Multiscale (MMS) mission (Burch et al. 2016a) has been making electron- or sub-ion-scale (unprecedented high spatial- and temporal-resolution) measurements of these regions in and around Earth's magnetosphere, especially in the magnetotail and at the magnetopause (Figure 1), to elucidate the microphysics of magnetic reconnection. A number of novel techniques for analyzing electromagnetic field and plasma data taken in and around the reconnection regions have been developed in preparation for and during the MMS mission.

The present review provides an overview of updated data analysis methods for in-situ observations of magnetic reconnection in space. This includes a range of prior applications of the methods, so that it can be used by the community and early career researchers to decide whether some of the methods is appropriate for their research. Thorough reviews of various single- and multi-spacecraft methods for analyzing MHD- and ion-scale aspects of reconnection and other space plasma processes were given by Paschmann and Daly (1998, 2008) in the era of the Cluster mission (e.g., Escoubet et al. 1997; Paschmann et al. 2005). Magnetic reconnection also involves inherently multi-dimensional structures and often occurs in highly nonuniform environments, as in the case at the magnetopause with substantial jumps across the current sheet in the plasma density, temperature, and magnetic field intensity (Figure 1). Thus, analysis methods previously reviewed and those assuming a uniform or weakly nonuniform background, such as wave analysis techniques (e.g., Narita 2017), are not covered in this review, except for some essential ones.

The analysis of a magnetic reconnection event may proceed as follows: (1) identification of electric current sheets or localized plasma bulk flows where reconnection may occur, (2) revealing the large-scale and local context of the reconnection event, based on upstream solar wind and geomagnetic field conditions, and the geometry (Figure 1) and structures of the reconnecting current sheet, and (3) detection and analysis of microscopic regions key to the reconnection process, such as the diffusion and energy conversion regions. For single or a few event analysis, step (1) can be done by identifying rapid magnetic field rotations, current density enhancements, and/or Alfvénic plasma velocity changes, which are intermittently seen in time series data. On the other hand, steps (2) and (3) require in-depth data analysis, empirical modeling, and/or numerical simulation performed specifically for the event of interest; they are the main topic of this review.

The rest of the paper is organized as follows. Section 2 presents a variety of methods for both large-scale and local contexts, including estimation of the coordinate system and frame velocity of the current sheet, and reconstruction of two- or three-dimensional plasma and electromagnetic field structures around the diffusion regions. Section 3 focuses on methods to identify and analyze the diffusion regions, including estimation of the reconnection rate and electric field (Figure 1). Section 4 gives a brief summary and outlook. In Appendix A, an overview is given of methods for the purpose of mission operations and automated identification of plasma regions and current sheets in and around the magnetosphere. Appendix B briefly explains how higher time resolution plasma moments are computed from the MMS data. In Appendix C we provide, as a quick user guide, tables (Tables 1-7) that summarize for each of the methods (1) required input data, (2) output, (3) fundamental theory, concept, or technique(s) that underlies the method, (4) model or underlying assumption(s), (5) relevant references, etc.

For a general overview of magnetic reconnection as a plasma physical process and primary scientific results from the MMS mission, many of which were obtained by use of either of the methods discussed in this review, see Chapter 1 of this issue (e.g., Norgren et al. (2024)), Fuselier et al. (2024), Hwang et al. (2023), and Oka et al. (2023). We also note that many of the methods summarized in this review can be used for the analysis of in-situ measurements in other regions of the solar system, and some may be applicable to reconnection observed in laboratory and solar (i.e., remotely sensed) plasmas. In particular, the method for estimating the reconnection rate, discussed in section 3.3.3, can be applied to imaging observations as well, and the concept underlying the one reviewed in section 3.3.2 is common to a method for analyzing reconnection observed during solar flares (Qiu et al. 2007).

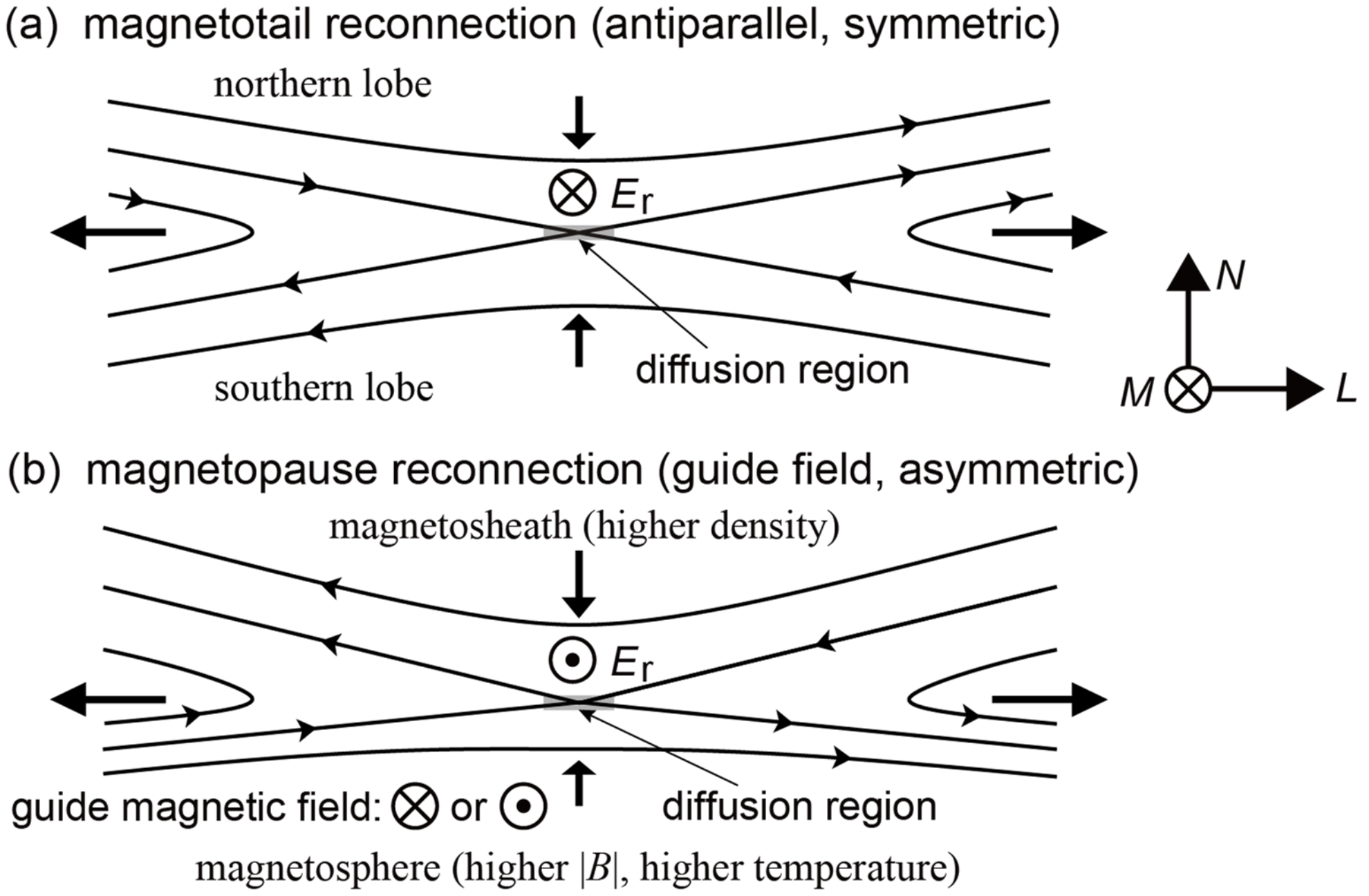

**Fig. 1** Typical geometry (magnetic field and plasma inflow and outflow pattern) of (a) magnetotail (approximately antiparallel and symmetric) reconnection and (b) magnetopause (normally guide-field and asymmetric) reconnection. Note that the reconnection electric field $E_r$ in LMN coordinates typically has a different polarity for the two cases.

## 2 Methods for Context

Reconnection regions on kinetic scales (of order 1-500 km) are much smaller than the size of the geospace or the magnetosphere (of order $10^5$ km) and are localized in space and often in time. It is thus important to understand a large-scale context and boundary conditions of spacecraft observations of reconnection-related phenomena and the field geometry and structures around the observing spacecraft. This section gives an overview of methods for revealing such contexts.

### 2.1 Large-Scale Context

#### 2.1.1 Maximum Magnetic Shear Model

The maximum magnetic shear model can predict the location on an empirical model magnetopause where the magnetic shear across the magnetopause current sheet is large or maximized, which is a plausible location of magnetopause reconnection, for interplanetary magnetic field (IMF) and geomagnetic dipole tilt conditions given as input (Trattner et al. 2021 and references therein). It was originally developed as a product of a polar cusp study using data from the NASA Polar satellite. The study determined the dayside magnetopause reconnection location (Trattner et al. 2007) for southward IMF conditions by using time-of-flight characteristics of cusp ions and the low-velocity cutoff method originally developed by Onsager et al. (1990, 1991) for the magnetotail reconnection location.

Figure 2 shows the general geometry of the low-velocity cutoff method that is used to estimate the dayside magnetopause reconnection location from cusp observations. Shown are the geomagnetic field lines (green), the reconnection location at the magnetopause (X), the satellite position in the cusp (ϴ), the ionospheric magnetic mirror point on the cusp field line (M), and the orbit path of a satellite passing through the cusp (red curve). A cusp-traversing satellite simultaneously observes slower magnetosheath ions that arrive from the site of magnetopause reconnection (incident ion beam) and faster magnetosheath ions that reached the ionospheric mirror point and returned to the high-altitude cusp-traversing satellite (mirrored ion beam).

The color inlay of Figure 2, centered along the cusp field line, shows an $H^+$ velocity distribution acquired by the TIMAS (Toroidal Imaging Mass Angle Spectrometer) instrument (Shelley et al. 1995) on board Polar in the cusp on 20 October 1997 from 14:05:59 to 14:06:11 UT. The $H^+$ distribution is presented in magnetic field-aligned coordinates after removing the effect of the $H^+$ bulk flow transverse to the magnetic field, and shows the incident magnetosheath ions injected at the location of magnetopause reconnection in addition to the mirrored ions that returned from the ionospheric mirror points.

The distance $X_\mathrm{r}$ along the field line between a satellite in the cusp and the magnetopause reconnection site can be computed by

$$X_\mathrm{r}/X_\mathrm{m} = 2\,V_\mathrm{e}/(V_\mathrm{m} - V_\mathrm{e}), \qquad (1)$$

derived from equating the flight times of the incident and mirrored ion beams. Here $V_\mathrm{e}$ and $V_\mathrm{m}$ are the cutoff velocities of the incident and mirrored beams, respectively, and $X_\mathrm{m}$ is the distance between the satellite and the mirror point (Figure 2). To determine the cutoff velocities, the peaks of the ion beams are fit with Gaussian distributions. The cutoff velocities are defined at the low-speed side of the peaks where the ion flux is 1/e of the peak flux (e.g., Fuselier et al. 2000; Trattner et al. 2007, 2005). The low velocity cutoffs are marked with black dashed lines in the color inlay of Figure 2.

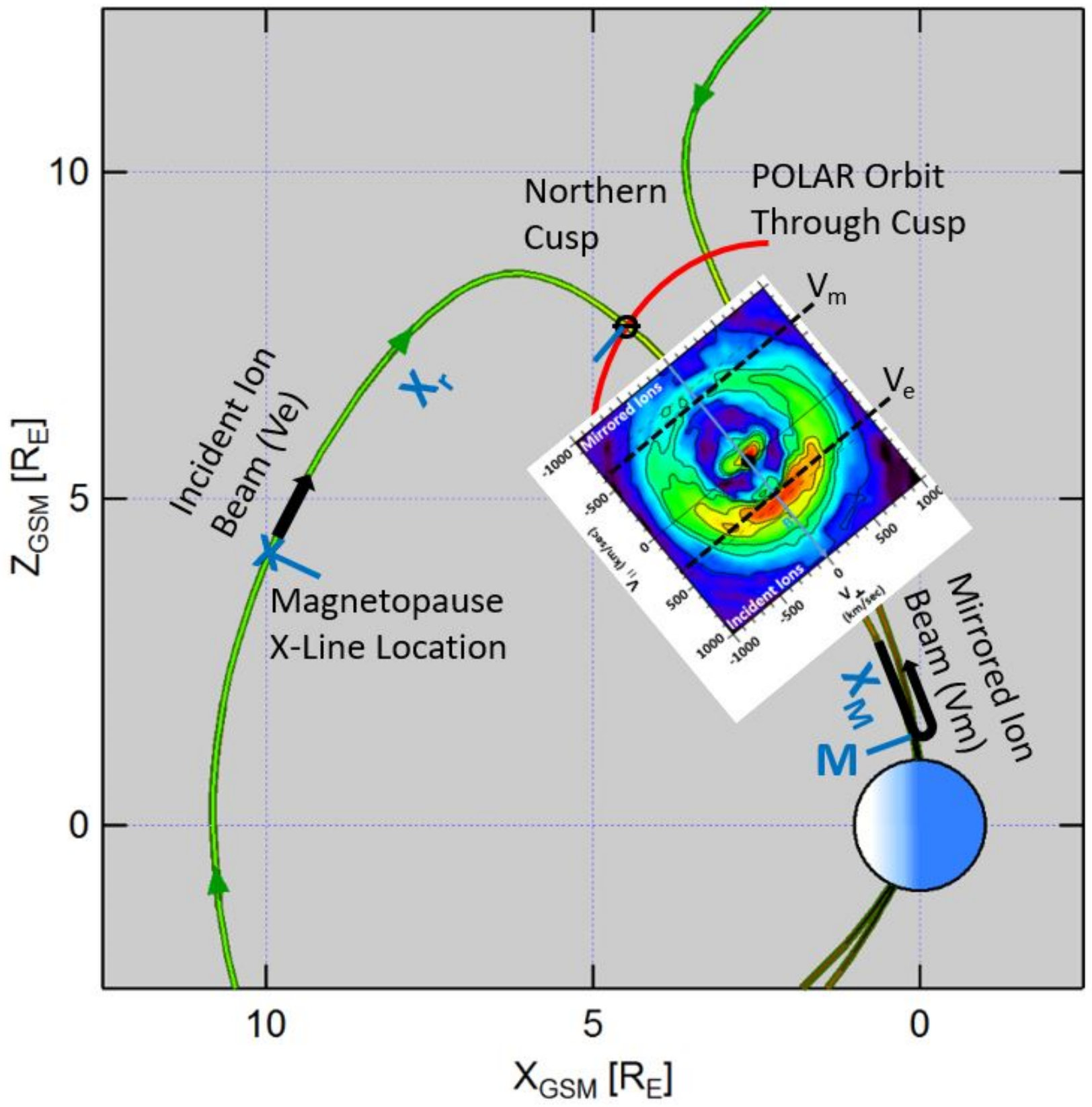

**Fig. 2** Schematic of the northern cusp region with the Polar satellite simultaneously observing incident ions on newly opened magnetic field lines which originate at the magnetopause reconnection location and mirrored ions returning from the ionosphere. The color distribution function shows the cutoff velocities $V_e$ and $V_m$ of the incident and mirrored ion beams, respectively.

To determine $X_m$, the geomagnetic field line at the satellite position in the cusp is traced down to the ionospheric mirror point by using the T96 model (Tsyganenko 1995). The model field line is also used to trace the calculated distance $X_r$ back to the reconnection location on the magnetopause. These end points of the field line traces mark the location of dayside magnetopause reconnection where the magnetosheath plasma enters the magnetosphere (e.g., Fuselier et al. 2000; Trattner et al. 2007, 2012, 2021).

Figure 3 (top panel) shows the distance to the reconnection site derived from Eq. (1) versus the Polar/TIMAS observation time during the cusp crossing on 11 April 1996. The distance to the reconnection site ranges from about 6 to 12 RE, which is most likely caused by changes in the satellite local time position. The uncertainties in the distance calculation are determined by those in measuring the low-velocity cutoff velocities. It is defined as 1/2 the difference between the velocity at the peak and the low-velocity cutoff (Fuselier et al. 2000; Trattner et al. 2007).

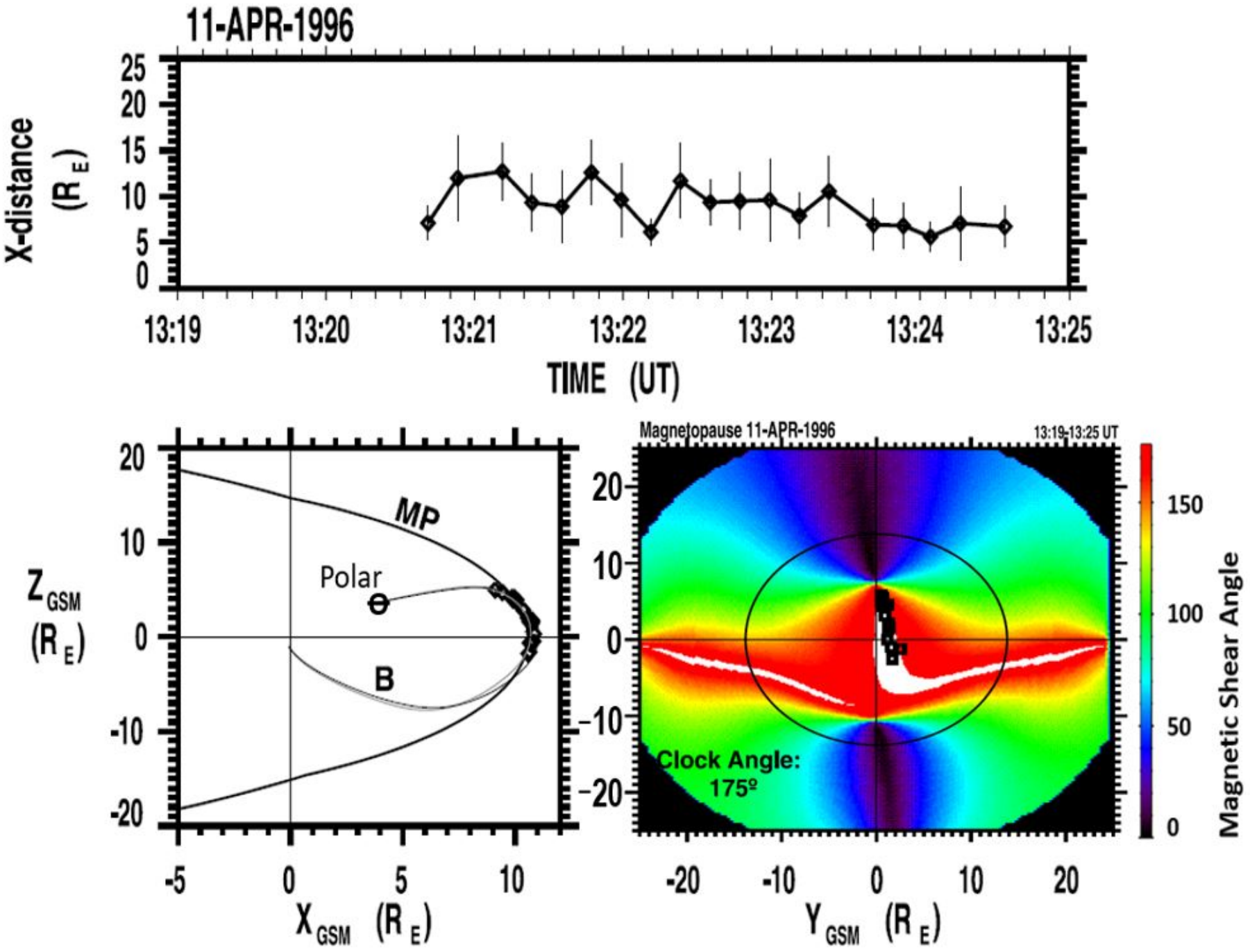

**Fig. 3** The distance to the magnetopause reconnection site from the cusp position of the Polar satellite on 11 April 1996 (top panel). The location of the magnetopause reconnection site as seen from dawn (bottom left panel). The magnetopause magnetic shear angle with the reconnection site location (black squares) as seen from the Sun for the 11 April 1996 Polar cusp crossing (bottom right panel).

The lower left panel of Figure 3 shows the magnetopause shape as viewed from dawn. Starting at the cusp location of the Polar satellite (ϴ), the distance to the reconnection site is traced along the T96 model geomagnetic field lines to the magnetopause where their end points are marked with black diamonds.

The lower right panel of Figure 3 shows the plot of magnetic shear angle on the magnetopause for the Polar cusp crossings. The shear angles are estimated by using the T96 model (internal field) together with the Kobel and Flückiger (1994) magnetosheath magnetic field draping model (external field), merged at the dayside ellipsoidal magnetopause shape of the Sibeck et al. (1991) model (e.g., Trattner et al. 2007, 2021). In the magnetic shear angle plot, the red areas represent the magnetopause antiparallel reconnection region with magnetic shear angles >160°. The white areas in the shear angle plots represent regions where the model magnetic fields are within 3° of being exactly antiparallel. The black circle represents the terminator plane at the magnetopause with the black squares showing the plasma entry points at the magnetopause, the end points of the cusp field line traces.

Because of the southward IMF conditions (IMF clock angle of 175°), the antiparallel reconnection region (red) covers most of the dayside magnetopause with the white regions for the highest magnetic shear shifted to the southern hemisphere due to the tilt of Earth's magnetic dipole. An exception is the dusk region close to local noon where the field-line trace points are also located.

The maximum magnetic shear model has been tested and validated for IMF conditions with $|B_x|/B < 0.7$ (e.g., Trattner et al. 2017). The model predicts long continuous X-lines that extend over the dayside magnetopause (e.g., Fuselier et al. 2002; Phan et al. 2006; Trattner et al. 2007; Dunlop et al. 2011; Trattner et al. 2021). For dominant IMF $B_y$ conditions, the model merges a component reconnection tilted X-line near the subsolar magnetopause with the two branches of the antiparallel reconnection regions, starting at the cusps and continuing towards the magnetotail along the flanks. The model was expanded to northward IMF conditions using observations by Trenchi et al. (2008, 2009) and confirming the existence of a dayside X-line down to an IMF clock angle of 50° (see also Gosling et al. 1990; Trattner et al. 2017). It highlighted the importance of antiparallel reconnection in constraining the location of the component reconnection line (Trattner et al. 2018).

The maximum magnetic shear model shows anomalies for dominant IMF $B_x$ conditions ($|B_x|/B > 0.7$) which are the result of the limitations of the IMF draping models used to determine the magnetopause magnetic shear. As shown by Michotte de Welle (2022), using a global three-dimensional and exclusively data driven model for the magnetopause magnetic shear, the local magnetic shear can differ significantly from the magnetic shear determined from the currently used numerical models (section 2.1.4), causing the anomalies in predicting the location of the dayside X-line. In addition, large magnetopause surveys (Trattner et al., 2007, 2017, 2021), comparing observed X-line locations with the predicted locations from the maximum magnetic shear model, also showed anomalies for events at the spring and fall equinoxes, specifically for events with IMF clock angles around 120 and 240 degrees, respectively. The fact that the equinox anomalies occur for specific narrow parameter ranges points to a currently unknown effect influencing the location of the magnetopause X-line under these conditions.

### 2.1.2 Event-Specific Global MHD Modeling

Global MHD models for simulating the solar wind-magnetosphere interaction can be used to provide an important large-scale context and connectivity information to assist in interpreting electron-scale observations of MMS. Runs on demand are readily requested through the Community Coordinated Modeling Center (CCMC) (Table 1 in Appendix C). Reiff and coauthors have now run the "SWMF" (Space Weather Modeling Framework) model (BATS-R-US with "Rice Convection Model" (RCM)) (Toth et al. 2005; see also Chapter 5.2 of this issue) for twelve instances where MMS observed crescent-shaped electron velocity distributions (sections 3.1.6 and 3.3.1), both in the dayside magnetopause region and in the tail (Reiff et al. 2018; Marshall et al. 2020, 2022). In each case, the SWMF model placed an X-line (or its neighboring separatrix sheet) within 1 $R_E$ and 2 minutes of the time of the MMS encounter.

MHD models that include RCM (e.g., SWMF) appear to do a better job in predicting the location of the reconnection sites in the tail. For example, the model predicted that MMS should be at the lobe-plasma sheet boundary layer interface near the X-line for an event on 23 June 2015 (Figure 5A,B in Reiff et al. (2016)). It also predicted that the X-line for that event would be patchy across the tail (Figure 5B in Reiff et al. (2016)). For a 11 July 2017 substorm event, the model not only accurately predicted the near-earth neutral line location but also predicted a huge plasmoid which was observed by MMS (Torbert et al. 2018; Reiff et al. 2018). See Figure 7 of Fuselier et al. (2024) for comparison images for that event of CCMC versus other global MHD models such as GGCM (Geospace General Circulation Model) (Raeder et al. 2017) and LFM (Lyon-Fedder-Mobarry model) (Lyon et al. 2004).

CCMC models have also been helpful in determining the time for field line reconfiguration, stretching and distortion for dayside events (Reiff et al. 2018). In a recent study, the location of the dayside X-line on 24 December 2016 moved dramatically as a result of a change in the Y-component of the IMF, with the model predicting not only reconnection at MMS but also connection of the Geotail spacecraft in the magnetosheath to the open field line on which MMS was situated (Figure 4). The Geotail data showed $O^+$ fluxes just a few minutes after the model predicted a connection to the northern polar cap and a close conjunction with MMS.

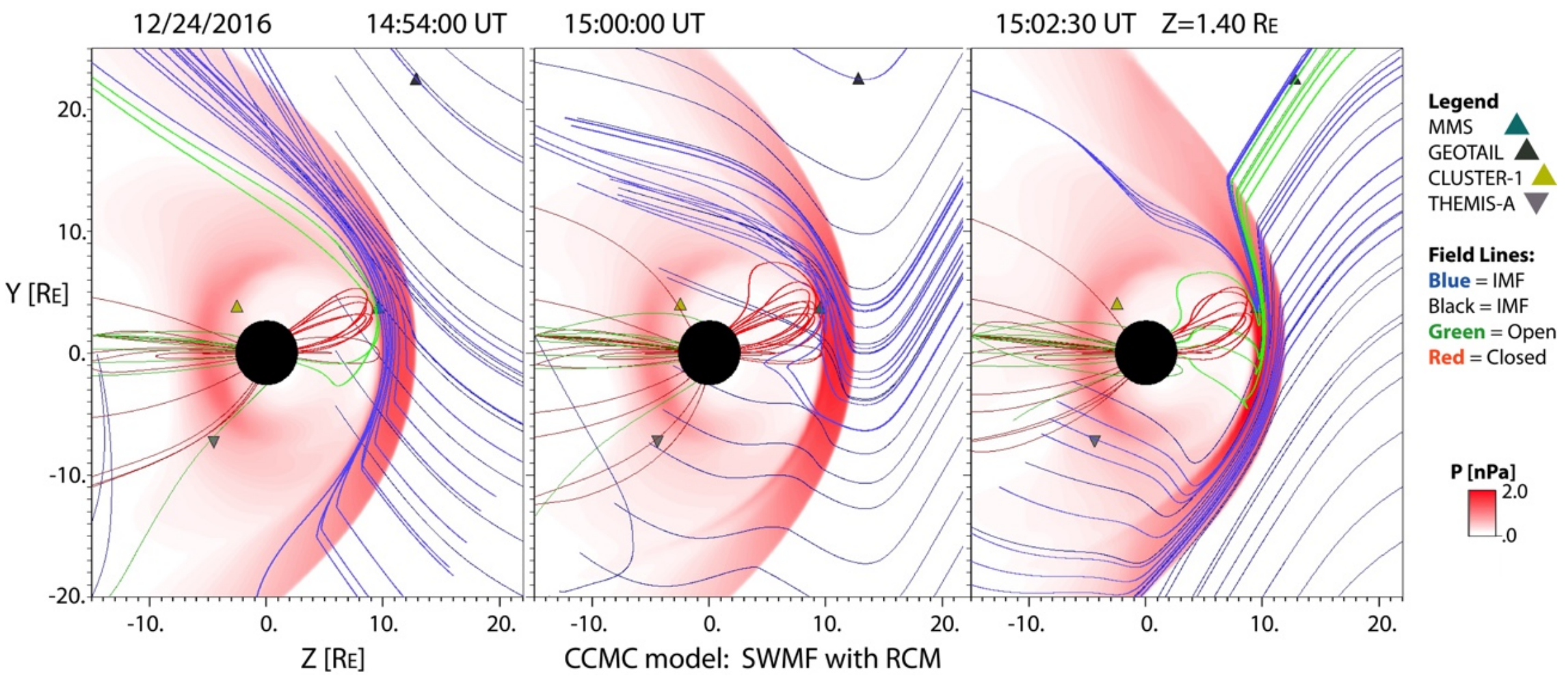

**Fig. 4** Evolution of the magnetic field line topology and spacecraft locations for an MMS dayside magnetopause reconnection event on 24 December 2016. Field lines are traced from MMS and from Geotail, and from other start locations. At time 14:54 UT, MMS was predicted by the SWMF model to be near the X-line, and on open field lines (green) connected to the southern cusp. Then the IMF $B_y$ changed sign, and at 15:20:30 UT, both MMS and Geotail were predicted to be on open field lines (green) connected to the northern polar cap, and their mapped field lines passed less than a half RE (RE: Earth radius) apart at the magnetopause. About a minute after the predicted connection, Geotail started observing O+ presumably from the magnetosphere, evidence of that connection.

In another study, an X-line that appears locally quite two-dimensional shows a dramatic difference in connection to the northern and southern ionospheres by field lines quite close on either side of the X-line, and electron fluxes correspondingly show a dramatic change in pitch angle (Marshall et al. 2022).

MHD models have also been run with embedded Particle-In-Cell (PIC) simulations, to overcome the inherent limitations of the MHD in reproducing kinetic-scale physics, e.g., MHD-EPIC (Chen et al. 2020). A recent CCMC workshop had dozens of presentations on linking CCMC models to solar, interplanetary, PIC and ionosphere/atmosphere models, many using open source modules [https://ccmc.gsfc.nasa.gov/ccmc-workshops/ccmc-2022-workshop/].

### 2.1.3 Data-Mining Approach to Reconstruction of the Global Reconnection Structure

The major problem in the global empirical reconstruction of the magnetosphere is data paucity: At any moment the huge volume of the magnetosphere ($\gtrsim 10^5\ \mathrm{R_E^3}$) is usually probed by less than a dozen spacecraft (e.g., Sitnov et al. 2020). In the past 15 years, it has been understood that this sparse data problem can be resolved or at least substantially mitigated due to the recurrent nature of the main space weather actors, storms and substorms. The storm occurrence depends on their intensity, as well as the strength and phase of the solar cycle. For medium intensity storms the recurrence period is about two weeks (Reyes et al. 2021). The recurrence time of periodic substorms is 2–4 h, while other substorm types have longer recurrence times depending on solar wind conditions (Borovsky and Yakymenko 2017). As a result, the historical records of spaceborne magnetometer observations can be organized using a multi-dimensional state-space, formed from the global storm and substorm activity indices and the solar wind input parameter. This allows the magnetic field for the event of interest to be reconstructed from its nearest neighbors in this state-space and not only from observations during the event. A specific data-mining (DM) technique leveraging this repeatability, the k-Nearest Neighbor (kNN) classifier (Wettschereck et al. 1997; Sitnov et al. 2008), combined with flexible and extensible magnetic field architectures (Tsyganenko and Sitnov 2007, Stephens et al., 2019), helped organize multi-decade archives of spaceborne magnetometer data to reconstruct storms (Tsyganenko and Sitnov 2007; Sitnov et al. 2008) and substorms (Stephens et al. 2019; Sitnov et al. 2019; hereafter referred to as SST19 model). The DM approach outlined in Figure 5 can be summarized as follows:

(a) First, a big database of historical magnetometer measurements (8.6 million points in the work by Stephens et al. (2023)) is mined in a global parameter state-space consisting of averaged values of the solar wind induced electric field $u_{sw}B_s^{IMF}$ ($u_{sw}$ is the solar wind velocity and $B_s^{IMF}$ is the southward interplanetary magnetic field: $B_s^{IMF} = -B_z^{IMF}$ when $B_z^{IMF} < 0$ and $B_s^{IMF} = 0$ otherwise, where $B_z^{IMF}$ is the north-south component of the IMF in geocentric solar magnetospheric (GSM) coordinates), the averaged *Sym-H* and *AL* (geomagnetic activity) indices, and the *Sym-H* and *AL* time derivatives. The mining procedure selects a small subset of moments at present but mostly in the past (red circles in Figure 5a), for which these global parameters are close to the event of interest in the state space (blue circle in Figure 5a). Events in this subset are called the nearest neighbors.

(b) The resulting subset of the magnetic field database (gray dots in Figure 5b), which is much larger than the handful of actual satellites available at that moment, is used to fit the free parameters of a very flexible magnetic field architecture ($\sim 10^3$ free parameters) and to reveal details of the magnetosphere such as the formation of new X-lines (at the earthward part of the $B_z = 0$ isocontour in Figure 5b).

(c) The obtained empirical model allows one to reconstruct a detailed 3D magnetic field structure, as is shown in Figure 5c for the 11 July 2017 MMS EDR event (Torbert et al. 2018).

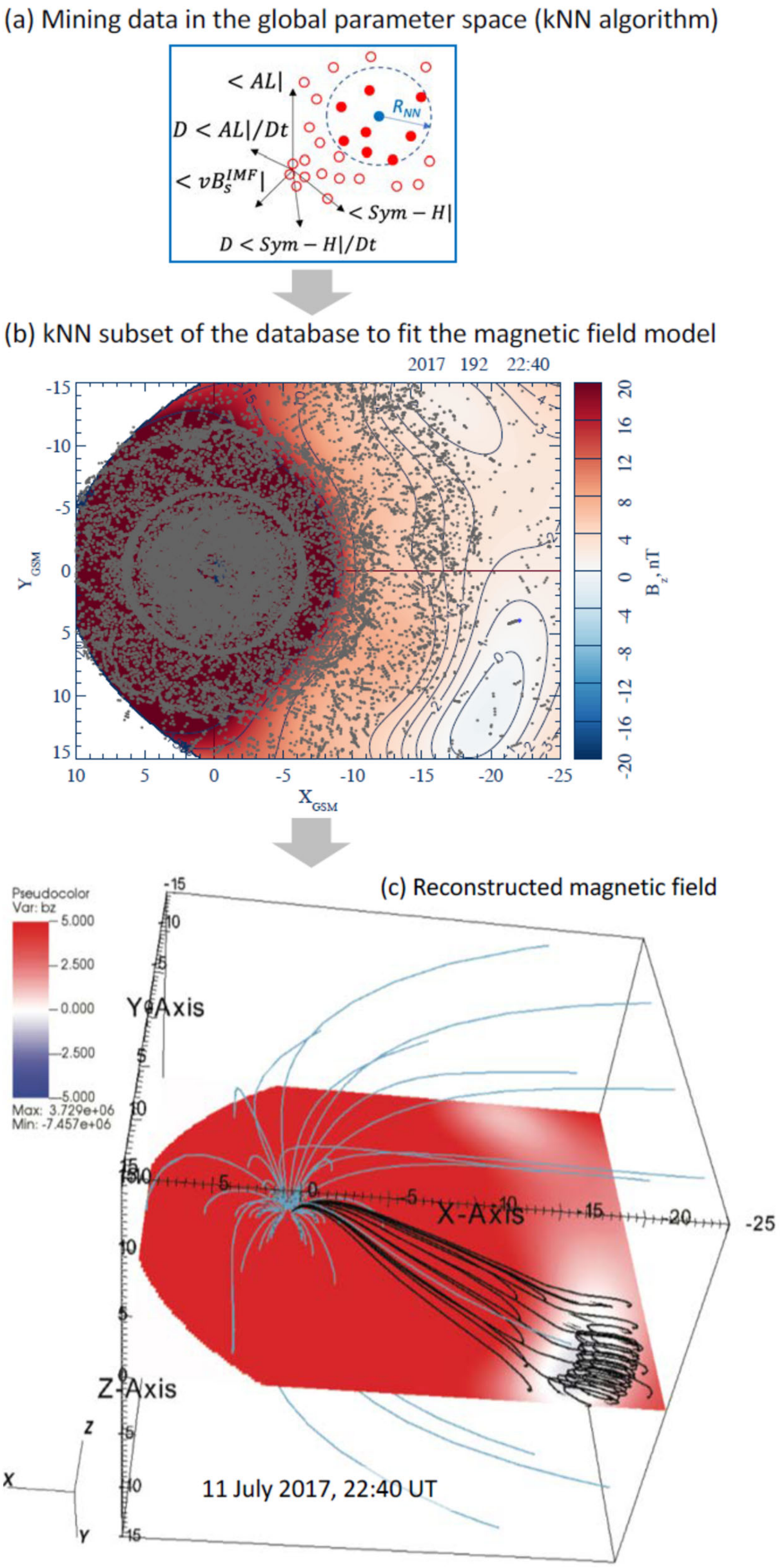

**Fig. 5** The kNN DM method outline (Sitnov et al. 2021): (**a**) selecting nearest neighbors for the event of interest (blue circle) in the 5D global parameter state-space; (**b**) finding

the corresponding subset in the magnetic field database (gray dots overplotted on the color-coded equatorial $B_z$ distribution) and using it to fit the magnetic field model and to yield 2D magnetic field distributions (here the equatorial slice) as well as (**c**) 3D magnetic field distributions. The example shown here is for the 11 July 2017 MMS EDR event (Torbert et al. 2018)) with the color-coded equatorial $B_z$ (in nT, saturated at 5 nT for better visualization) and a few sample field lines. Panels (**a**) and (**b**) are adapted from Sitnov et al. (2019).

The magnetospheric state shown in Figure 5a is characterized using geomagnetic indices and solar wind conditions. It can be described by a 5-D state-space vector, $\mathbf{G}(t) = (G_1,...,G_5)$, formed from the geomagnetic storm index (*Sym-H*), substorm index (*AL*), their time derivatives, and the solar wind electric field parameter ($u_{sw}B_s^{IMF}$). Most recently (Stephens et al., 2023), the *Sym-H* and *AL* indices have been replaced by the *SMR* and *SML* indices provided by the SuperMag project (Gjerloev 2012). The global binning parameters $G_{1\text{–}5}(t)$ are normalized by their standard deviations, smoothed over storm or substorm scales, and sampled at a 5-min cadence, as is detailed in Stephens and Sitnov (2021). Including the time derivatives of these activity indices allows the DM procedure to differentiate between storm and substorm phases as well to capture memory effects of the magnetosphere as a dynamic system (Sitnov et al. 2001). The space magnetometer archive contains data from 22 satellites (including four MMS probes) spanning the years 1995–2020 resulting in 8,649,672 magnetic field measurements after being averaged over 5 or 15 min time windows (Stephens et al. 2023).

Every query moment in time $t = t_q$ corresponds to a particular point in the 5-D state-space, $G_{(q)} = G(t_q)$. Its $k_{NN}$ nearest neighbors (NNs) will be other points, $G_{(i)}$, in close proximity to it: $R_i = |G_{(i)} - G_{(q)}| < R_{NN}$ (in the Euclidean metric). The specific choice of $k_{NN}$ (and hence $R_{NN}$), is determined by a balance between over- and under-fitting. Stephens and Sitnov (2021) found the optimal number to be $k_{NN} = 32{,}000$, corresponding to ~1% of the total database (~$10^7$ sampling cases). The resulting set is composed of a very small number (~1–10) of real (available at the moment of interest) and a much larger number (~$10^5$) of virtual (from other events in the database) satellites.

The large number of NNs provided by such synthetic satellite observations enables the use of new magnetic field architectures (Tsyganenko and Sitnov 2007; Stephens et al. 2019), which differ from classical empirical models with custom-tailored modules (e.g., Tsyganenko and Sitnov 2005) by utilizing regular basis function expansions for the major magnetospheric current systems. In particular, the equatorial current system, which was previously described by ring and tail current modules, is now described by two expansions representing arbitrary current distributions of thick and thin current sheets with different thicknesses. This architecture accounts for the multiscale structure of the tail current sheet with an ion-scale thin current sheet (TCS), with a thickness $D_{TCS}$, forming inside a much thicker current sheet, with a thickness $D \gg D_{TCS}$, during the substorm growth phase and then decaying during the expansion phase (e.g., Sergeev et al. 2011). The independence of the current sheet expansions is provided by the constraint $D_{TCS} < D_0 < D$, where $D_0$ is the ad hoc parameter ~$1R_E$. The proper reconstruction of substorms also requires a flexible description of the field-aligned currents, which is provided in the SST19 model using a set of distorted conical modules (Tsyganenko 1991) distributed in latitude and local time, as is discussed in more detail in Sitnov et al. (2017).

To improve the reconstructions, while fitting the magnetic field model with the NN subset, the spacecraft data were additionally weighted: in the real space, to mitigate the inhomogeneity of their radial distribution (Tsyganenko and Sitnov 2007), and in the state-space, to reduce the uncertainty and bias toward weaker activity regions (Sitnov et al. 2020; Stephens et al. 2020).

The SST19 model successfully describes the TCS buildup during the substorm growth phase and its decay during the expansion phase accompanied by the formation of the substorm current wedge (McPherron et al. 1973). It also identifies X-lines in the tail (Sitnov et al. 2019), which match in-situ MMS observations (Stephens et al. 2023), as is described in more detail in Chapter 3.1 of this issue (Fuselier et al. 2024). The model has been extensively validated using both in-situ observations (Sitnov et al. 2019; Stephens et al. 2019, 2020, 2023) and uncertainty quantification using DM binning statistics (Sitnov et al. 2019; Stephens et al. 2023).

#### 2.1.4 Global 3D Structure of the Magnetosheath Using In Situ Measurements: Application to Magnetic Field Draping

The dynamics of the Earth's magnetosphere and its coupling to the solar wind importantly depends on how the solar wind interacts at the bow shock and, in particular, on how the plasma is decelerated, heated and deflected there and on how the interplanetary magnetic field drapes around the magnetospheric obstacle in the magnetosheath. The specific structure of the draping, in particular, plays a major role for the reconnection of magnetic field lines at the magnetopause. Magnetic field draping was thus the focus of a study by Michotte de Welle et al. (2022) that permits, based on large-scale statistics of in situ spacecraft measurements, to reconstruct the global 3D structure of the magnetosheath.

Magnetic field draping is a fairly well understood concept, resulting from the frozen-in condition ruling the evolution of magnetized plasmas on large scales. However, our knowledge of the 3D global draping structure in the Earth's magnetosheath is very limited and is mostly described by analytical and numerical models. Michotte de Welle et al. have recently succeeded in reconstructing the 3D structure of the magnetic draping over the whole dayside of the magnetosphere, using only in situ observations, and as a function of the IMF orientation. Two decades of data from Cluster, Double Star, THEMIS and MMS missions have been used for that purpose. The measurements made in the magnetosheath were extracted automatically using a Gradient Boosting Classifier trained to classify magnetosphere, magnetosheath and solar wind data points (Nguyen et al. 2022a). About 50 million measurements were extracted and then associated with a causal solar wind and IMF conditions from OMNI data using a solar wind propagation method (Safrankova et al. 2002). The position of each data point relative to the bow shock and the magnetopause at the time of the measurement is then estimated using a Gradient Boosting Regression model of the boundaries, parameterized with solar wind and IMF conditions. All points are then repositioned between a standard bow shock and magnetopause boundary, determined for average solar wind conditions, and rotated into the solar wind interplanetary (SWI) magnetic field coordinate system (Zhang et al. 2019) in which the upstream IMF direction is parallel to the XY plane. This last step is crucial to ensure each point falls in the right sector of the magnetosheath (quasi-parallel or quasi-perpendicular bow shock sides) with respect to its causal IMF. Magnetic field lines are then integrated with a standard ordinary differential equation integrator, using at each step the weighted average of the k-Nearest Neighbor magnetic field measurements close to the current iteration step position (with k=45000).

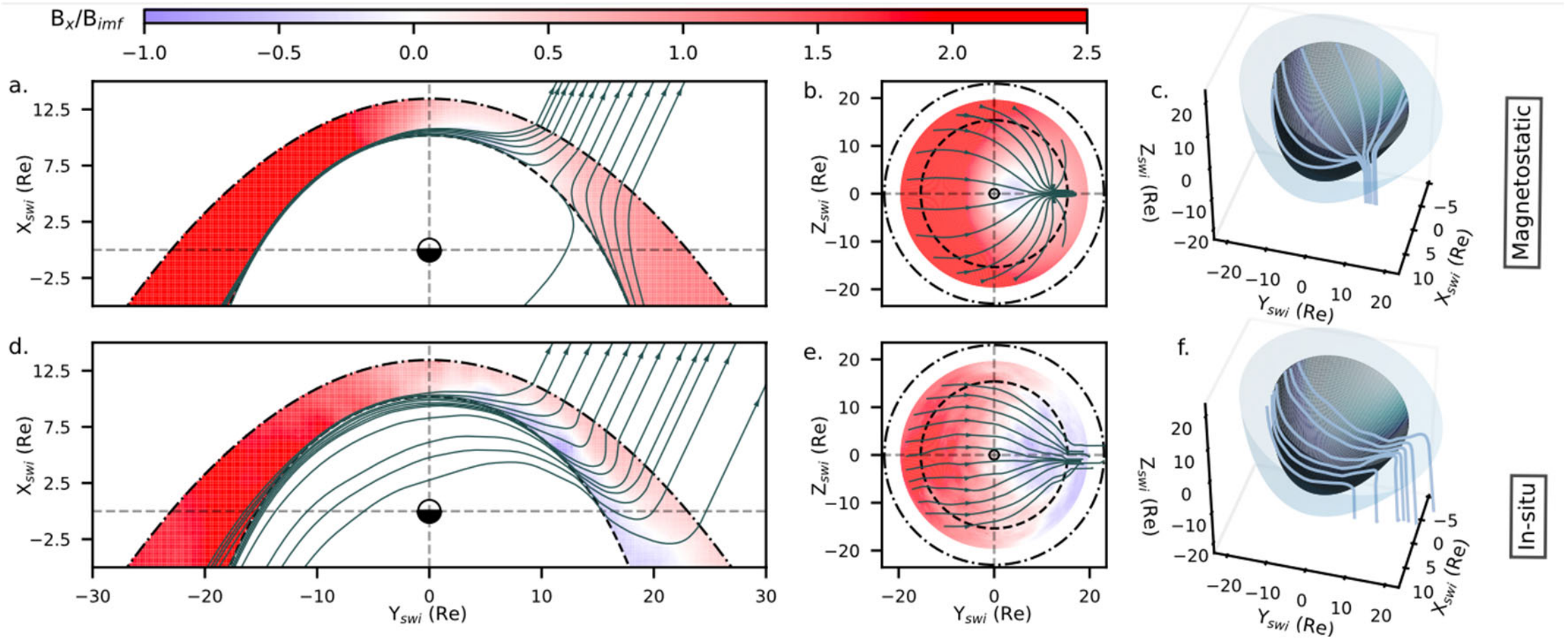

**Fig. 6** From left to right: representation of the magnetic field lines in the XY (left), YZ (middle) planes and in 3D (right) as predicted by the KF94 magnetostatic model (top panels) or reconstructed from in situ data (bottom panels). On the four leftmost panels, the color codes the value of the $B_x$ component of the magnetic field. Coordinates are from the SWI system.

The bottom three panels of Figure 6 show the obtained draping in the XY and YZ planes of the SWI coordinate system and in 3D on the rightmost panel. As a comparison, the top three panels show, for the same points of view, the draping obtained with the magnetostatic model of Kobel and Fluckiger (1994) (referred to as the KF94 model). In this configuration, the represented data is the subset of all measurements for which the associated IMF cone angle falls between 20° and 30° from the Sun-Earth axis. The figure reveals that the observed draping is fundamentally different from the modeled one. In the modeled draping, the magnetic field appears to diverge as it approaches the magnetopause in the region downstream of the quasi-parallel bow shock. This is the result of the only two constraints imposed by the model to the magnetic field. Indeed, on the one hand, the magnetic field in the quasi-parallel (positive Y) region is mostly conserved as it crosses the bow shock. On the other hand, the field must be tangential to the magnetopause. While these two constraints also apply in reality (if one neglects magnetopause reconnection as a first approximation), magnetic flux is also bound to the plasma as it flows upstream of the bow shock and circumvents the magnetopause in the magnetosheath. In other words, fluid elements connected to a magnetic flux tube entering the quasi-parallel region must remain connected to those that entered earlier on the quasi-perpendicular bow shock (negative Y) side. The considerable slowing down of the flow in the subsolar region forces all field lines entering in the quasi-parallel magnetosheath to head to the dayside where the flux piles up, rather than diverge partly to the nightside as the magnetostatic model predicts. This large-scale kink in magnetic field lines is thus associated with a macroscopic current sheet at mid-depth of the quasi-parallel side of the magnetosheath, an effect that is not seen with the vacuum magnetostatic model. The main consequence is that for this range of IMF cone angles, a large part of the magnetopause on the quasi-parallel region sees a magnetic shear that is vastly different from that predicted using the KF94 model, potentially adding difficulties to the maximum shear angle reconnection model in that cone angle regime.

Interestingly, when the IMF becomes quasi-radial ($B_x$ dominant), in practice when the cone angle is less than about 12°, magnetically connected solar wind elements are so far apart along the Sun-Earth line that by the time fluid elements arrive on the quasi-parallel side of the bow shock, connected fluid elements entered in the subsolar region have long ago re-accelerated and joined the nightside of the system. As a result, field lines are not kinked anymore and rather diverge on the magnetopause, which coincidentally qualitatively agrees with the vacuum magnetostatic model prediction (Michotte de Welle et al. 2022). For IMF cone angles larger than 45°, i.e., field lines arriving rather perpendicular to the Sun-Earth axis, the draping is also found to qualitatively match that predicted by the KF94 model.

This study was limited to the reconstruction of the field line draping. The method is precise enough to reconstruct the overall dependency of the magnetic topological properties on the IMF orientation. Local and detailed quantitative properties of the field such as its divergence-free character cannot be ensured, although on average $|\nabla \cdot \mathbf{B}|$ is on the order of $0.01 B_{\mathrm{IMF}}/\mathrm{R_E}$, where $B_{\mathrm{IMF}}$ is the upstream field intensity and Earth's radius $\mathrm{R_E}$ is comparable to the scale of field variations. Similar analysis can be made to reconstruct the global distribution of any physical quantities, providing the capability to reconstruct the global 3D structure of the solar wind–dayside magnetosphere interaction globally. The amount of data now available and modern statistical learning methods will prove useful to understand how physical parameters distribute on and around critical regions such as the magnetopause for various upstream conditions, which is a topic of on-going work.

## 2.2 Coordinate System, Frame Velocity, and Spacecraft Trajectory Estimation

Current sheets where reconnection may occur are never strictly stationary, their local normal direction can be highly variable in space and time, and the X-line, possibly embedded in those current sheets, may be moving, depending on the external conditions and instabilities excited in the current sheets. It is thus indispensable, for each current sheet crossing or reconnection event, to be able to obtain a proper coordinate system and frame velocity of the current sheet structure or reconnection regions. In this section, we briefly review various methods to estimate the characteristic orientations and motion of the structures from in-situ measurements.

### 2.2.1 Dimensionality and Coordinate Systems

Here we review methods for estimating the dimensionality and coordinate systems of magnetic or plasma structures in space from in-situ data. Since an overview was given by Sonnerup et al. (2006a) and Shi et al. (2019) on various single- and multi-spacecraft analysis methods for estimating the orientation and motion of plasma discontinuities (one-dimensional (1D) structures, such as planar current sheets), we focus only on recent developments.

**Minimum Directional Derivative** In Minimum Directional Derivative (MDD) analysis, a multi-spacecraft method applicable to four-spacecraft measurements at any instant of the magnetic (or any vector) field, one takes the gradient of the magnetic field vector, multiplies the resulting matrix by its transpose, and then solves for the eigenvectors of the resulting matrix, finding time-dependent maximum, intermediate, and minimum gradient eigenvalues, $\lambda_{max}$, $\lambda_{int}$, and $\lambda_{min}$, which represent the squared gradient in the respective time-dependent directions, $\hat{\mathbf{e}}_n$, $\hat{\mathbf{e}}_l$, and $\hat{\mathbf{e}}_m$, respectively (Shi et al. 2005, 2019) (alternatively, the eigenvalues can be defined as the square root of these quantities, proportional to the gradient in the respective directions). If $\lambda_{max} \gg \lambda_{int}, \lambda_{min}$, the system is roughly one dimensional with variation mainly in the maximum gradient ($\hat{\mathbf{e}}_n$) direction. If $\lambda_{max}, \lambda_{int} \gg \lambda_{min}$, the system is roughly two-dimensional (2D) with variation mainly in the maximum and intermediate gradient directions ($\hat{\mathbf{e}}_n$ and $\hat{\mathbf{e}}_l$, respectively). If all eigenvalues are comparable, the system is 3D with variation in all three directions, $\hat{\mathbf{e}}_n$, $\hat{\mathbf{e}}_l$, and $\hat{\mathbf{e}}_m$ (Shi et al. 2019). Rezeau et al. (2018) introduced dimensionality parameters that are useful for determining the dimensionality of the system, $D_{1D} = (\lambda_{max} - \lambda_{int})/ \lambda_{max}$, $D_{2D} = (\lambda_{int} - \lambda_{min})/ \lambda_{max}$, and $D_{3D} = \lambda_{min}/ \lambda_{max}$; $D_{1D}$, $D_{2D}$, and $D_{3D}$ quantify the degree to which the system is 1D, 2D, or 3D, respectively.

In practice, when studying magnetic reconnection events, $\lambda_{max}$ is often significantly greater than the other two eigenvalues in the vicinity of the current sheet ($D_{1D}$ close to unity). However, the system can be somewhat two-dimensional if $\lambda_{int} \gg \lambda_{min}$ so that the variation in the minimum gradient direction can be neglected relative to that in the other two directions, yielding a system that can be analyzed as quasi-2D. (Unfortunately, the minimum MDD eigenvalue direction is not always the *M* direction as defined below (Denton et al. 2016, 2018).) The coordinate system usually used to describe magnetic reconnection (the so-called LMN coordinate system) has the *L* direction in the direction of the reconnecting magnetic field and the *N* direction normal to the current sheet; the *M* direction completes the triad. Because $\lambda_{max}$ is often very large, the MDD maximum gradient direction, $\hat{\mathbf{e}}_{N'}$, found from the time dependent $\hat{\mathbf{e}}_n$ direction, is often the most accurately determined direction in the system, and can usually be used to define the normal direction across the current sheet. Then in order to define the reconnection coordinate system, it remains to find one more direction.

**Hybrid Methods** Denton et al. (2016, 2018), studying the 16 October 2015 magnetopause reconnection event of Burch et al. (2016b), determined the *L* direction as the maximum variance direction of Minimum Variance Analysis (MVA) (Sonnerup and Cahill 1967; Sonnerup and Scheible 1998) of the magnetic field. This is reasonable seeing as the reconnection magnetic field reverses across the current sheet, leading to large variance. The *M* direction can be taken to be the direction of the cross product between $\hat{\mathbf{e}}_{N'}$ defined by MDD and $\hat{\mathbf{e}}_{L'}$ defined by MVA, but if $\hat{\mathbf{e}}_{N'}$ and $\hat{\mathbf{e}}_{L'}$ are not exactly orthogonal, a choice must be made to determine the *N* and *L* directions. For instance, one could take $\hat{\mathbf{e}}_N = \hat{\mathbf{e}}_{N'}$, and find *L* from $\hat{\mathbf{e}}_L = \hat{\mathbf{e}}_M \times \hat{\mathbf{e}}_N$, which is what Denton et al. (2016) did. Denton et al. (2018) proposed a hybrid method weighting the influence of $\hat{\mathbf{e}}_{N'}$ and $\hat{\mathbf{e}}_{L'}$ based on the ratio of the maximum MDD eigenvalue to the maximum MVA eigenvalue.

Genestreti et al. (2018) found that Denton et al.'s (2018) method did not work well for the 11 July 2017 magnetotail reconnection event studied by Torbert et al. (2018). Instead, Genestreti et al. used the maximum variance direction of MVAVe (Minimum Variance Analysis of the electron bulk velocity $\mathbf{u}_\mathrm{e}$) to determine the *L* direction. Large variance in the velocity moments along the *L* direction are expected since the reconnection outflow will be along that direction. (Another possibility is to use MVAE, using the variance of the electric field.) Heuer et al. (2022) recently proposed a hybrid system similar to that of Denton et al. (2018), except that $\hat{\mathbf{e}}_{L'}$ is determined from MVAB (MVA using the magnetic field) only when the spacecraft have a significant velocity component across the current sheet in the frame of the magnetic structure. If the velocity of the spacecraft relative to the magnetic structure is mostly in the *L* direction, they recommend using MVAVe to determine $\hat{\mathbf{e}}_{L'}$.

**Magnetic Configuration Analysis** Among the analysis methods that are enabled by four-spacecraft measurements are those that allow the determination of the geometrical properties of the magnetic field. Following the main ideas of the magnetic MDD (Shi et al. 2005) and magnetic rotational analysis procedure (Shen et al. 2007), Fadanelli et al. (2019) derived a new method named the "magnetic configuration analysis" (MCA). The method in effect determines the main axes of the magnetic field rotation rate in space, in a normalized fashion, and permits the categorization of magnetic field geometries in terms of planarity and elongation properties, for instance. MCA is thus designed to estimate the spatial scales on which the magnetic field varies locally and to determine the actual magnetic field shape and dimensionality from multi-spacecraft data. Case studies using MMS data showed that the method is capable of determining, for example, the planar and cigar shapes of structures such as current sheets and small flux ropes, respectively. An interesting property of such a method is that the determination is made very locally, at the scale of the inter-spacecraft separation, which is much smaller than that of the current sheet or flux rope itself.

Fadanelli et al. (2019) also statistically applied the MCA method to magnetic field observations in different near-Earth regions (magnetosphere, magnetosheath, and solar wind). The findings show that the magnetic field structure is typically elongated at small scales (cigar and blade shapes), is less frequently planar (pancake shapes generally associated with current sheets), but rarely shows an isotropic variance in the magnetic field rotation rate. The occurrence frequency of the type of magnetic geometries observed and, most importantly, their scale lengths, strongly depend on the region sampled and plasma β. Interestingly, the most invariant direction is statistically aligned with the electric current, suggesting that electromagnetic forces are fundamental in determining the magnetic field configuration at small scales.

### 2.2.2 Velocity of the Magnetic Structure

In addition to determining the coordinate system, it is beneficial to determine the velocity of the magnetic structure in order to determine a reference frame in which the magnetic structure is approximately time stationary (what Shi et al. (2019) call the "proper reference frame"). In homogeneous regions, such a velocity can be the $E \times B$ velocity or the ion velocity perpendicular to the background magnetic field for MHD-scale structures, and in IDRs the perpendicular components of the electron velocity can be used. A related approach, applicable to inhomogeneous regions, is deHoffmann-Teller (HT) analysis, which finds a frame with minimum electric field, and hence the frame in which ion or electron flows are roughly aligned with the spatially varying magnetic field (deHoffmann and Teller 1950; Khrabrov and Sonnerup 1998). However, these approaches are unreliable in the EDR.

Four spacecraft timing analysis (Dunlop and Woodward 1998), which assumes that the spatial structure varies in only one direction, can yield the velocity component along that direction, namely, the velocity normal to the plane along which spatial gradient is negligible. However, results may vary depending on the input quantity used. Minimum Faraday residue analysis (MFR) is another approach to get the normal velocity of MHD discontinuities from single-spacecraft data (Terasawa et al. 1996; Khrabrov and Sonnerup 1998).

Shi et al. (2006) introduced the Spatio-Temporal Difference (STD) method, which solves for the structure velocity from the convection equation for steady magnetic structures ($\partial \mathbf{B}/\partial t = 0$)

$$\frac{d\mathbf{B}}{dt} = (\mathbf{V}_{\mathrm{sc}} \cdot \nabla)\mathbf{B} = -(\mathbf{V}_{\mathrm{str}} \cdot \nabla)\mathbf{B}, \qquad (2)$$

using instantaneous values of the magnetic gradient at one time, and a centered time step around that time for the total time derivative $d\mathbf{B}/dt$ observed by the spacecraft. Here $\mathbf{V}_{\mathrm{sc}}$ is the spacecraft velocity relative to the magnetic structure, and $\mathbf{V}_{\mathrm{str}}$ is the structure velocity relative to the spacecraft. Using Eq. (3) assumes that the velocity is constant on the spatial scale of the four spacecraft and on the time scale for motion across that spatial scale. However, in most cases, only one or at most two velocity components can be determined, because when the gradient is very small, the velocity component in that direction is unreliable (Shi et al. 2019; Denton et al. 2021). Here we examine the 16 October 2015 magnetopause reconnection event of Burch et al. (2016b) and introduce a modification of STD to get as much information as possible from STD.

Figure 7a shows the MDD eigenvalues normalized to $(0.1\ \mathrm{nT}/d_{\mathrm{sc}})^2$, where 0.1 nT is the maximum calibration error of the MMS magnetometers and $d_{\mathrm{sc}}$ is the average spacecraft spacing; $(0.1\ \mathrm{nT}/d_{\mathrm{sc}})^2$ represents a reasonable minimum value required for accuracy of the squared gradient. In Figure 7a, the maximum and intermediate eigenvalues are always well above this value, suggesting that they can be satisfactorily determined. However, where the minimum eigenvalue becomes significantly smaller than this value, the STD velocity component in that direction has unrealistically large values. Where all three normalized eigenvalues are significantly above unity, however, as sometimes happens for this event (like after $t = 2.5$ s in Figure 7a), we may be able to determine a three-dimensional velocity. For the calculations leading to Figure 7, we required that an eigenvalue be at least $20(0.1\ \mathrm{nT}/d_{\mathrm{sc}})^2$ in order to include the STD velocity component associated with that eigenvalue. (A large value was required to yield consistent velocities.) Otherwise, that component was set equal to zero. Figures 7b-d show the velocity components calculated in the time-dependent MDD intermediate, minimum, and maximum gradient directions, $\hat{\mathbf{e}}_L$, $\hat{\mathbf{e}}_M$, and $\hat{\mathbf{e}}_N$, respectively. In Figure 7c, the dotted green curve is the instantaneous STD velocity component for the MDD local minimum gradient ($\hat{\mathbf{e}}_m$) direction. Black circles mark the data points where all three velocity components are determined, and at data points without black circles, the dotted green curves drop to zero.

Figures 7f–h show the resulting velocity in the fixed LMN coordinate system that was found using the hybrid method of Denton et al. (2018). The $N$ direction agrees well with the instantaneous MDD $\hat{\mathbf{e}}_n$ direction, but the $L$ direction is found from the direction of maximum variance of the magnetic field. Note that whereas the $l$ and $m$ components of the velocity reverse at $t = 2.3$ s in Figures 7b and 7c, the velocity components in the fixed $L$ and $M$ directions are well behaved in Figures 7f and 7g. Thus in Figure 7, we have calculated a three-dimensional structure velocity at the times indicated by the black circles. The velocity is less reliable at the other times because of the omission of the minimum gradient component.

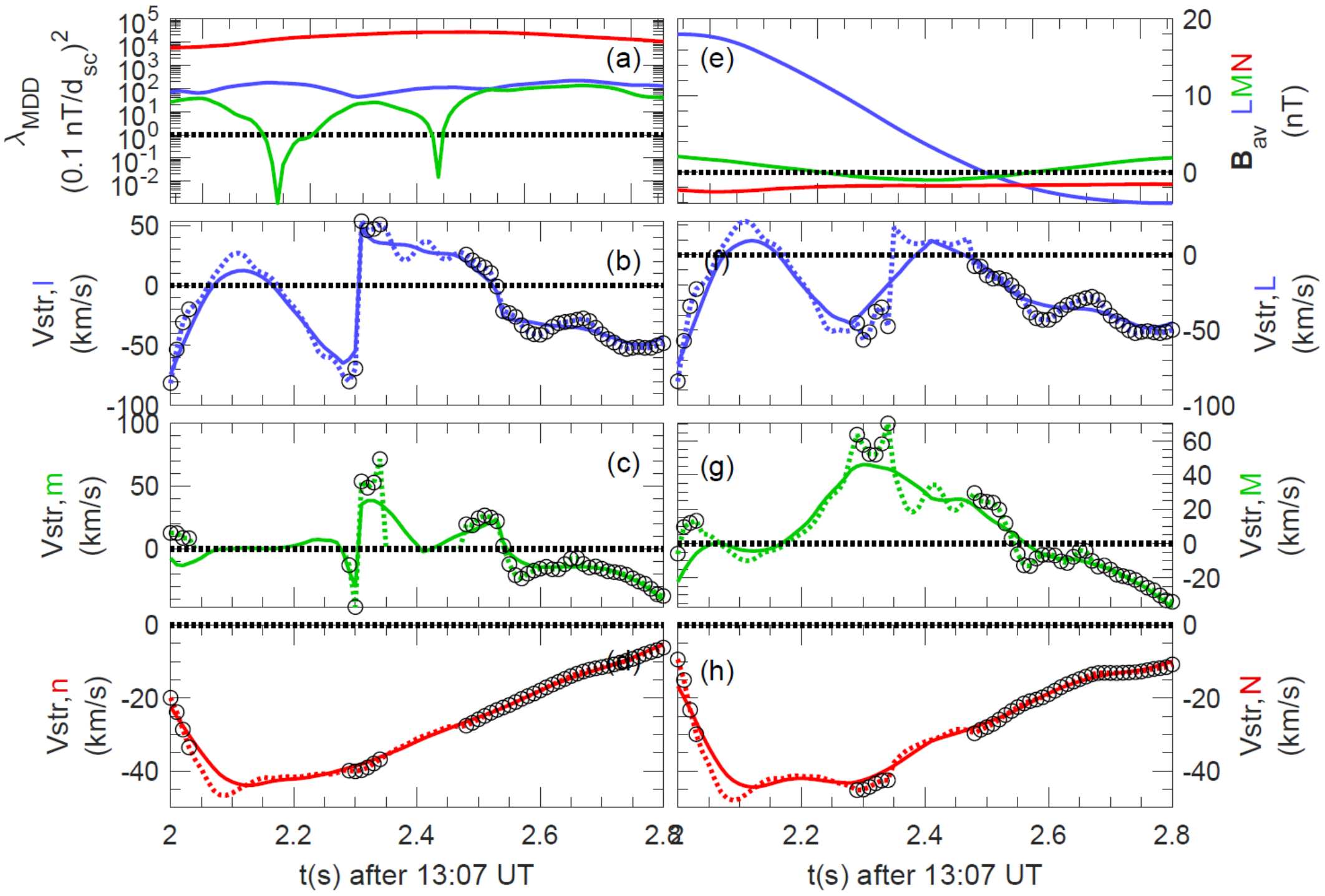

**Fig. 7** STD analysis for the 16 Oct 2015 magnetopause reconnection event. (**a**) MDD eigenvalues, (**b–d**) STD structure velocity components in the local MDD gradient directions, $\hat{\mathbf{e}}_l$, $\hat{\mathbf{e}}_m$, and $\hat{\mathbf{e}}_n$ respectively, (**e**) magnetic field averaged over the four MMS spacecraft, (**f–h**) STD structure velocity in the fixed LMN coordinates. The dotted curves are the instantaneous velocities, and the solid curves are smoothed over a time scale of 0.5 s.

Denton et al. (2021) used polynomial reconstruction (section 2.3.1) to track the motion of the X-line, and the resulting velocity was in rough agreement with the STD velocity in the *LN* plane for times for which the X line was less than two $d_{\mathrm{sc}}$ from the centroid of the MMS spacecraft. Another approach is to match MMS observations to simulation data in order to find a velocity though the simulation fields (Shuster et al. 2017; Nakamura et al. 2018b; Egedal et al. 2019; Schroeder et al. 2022; section 2.2.3).

### 2.2.3 Spacecraft Trajectory Estimation

MMS observations of a magnetotail EDR can be directly compared to 2D kinetic PIC simulation, for determination of spacecraft trajectories in the event specific *LN*-plane. As will be described in detail in Section 2.3.3, the trajectory of the MMS constellation further allows for reconstruction of the MMS data in a 2D format, which can in turn be compared against the simulation data. Note that these methods rely on spacecraft observations with sufficient features to well constrain the trajectory. The MMS event considered here (10 August 2017) features strong electron pressure anisotropy followed by large electric field gradients that indicate a path that closely follows a separatrix layer. For events that do not exhibit such features, one may find difficulty in accurate determination of the spacecraft trajectory.

**Spacecraft trajectory optimization** The spacecraft paths are found through a $\chi^2$-optimization procedure, in which a penalty function made up of a sum of squared deviations of spacecraft measurements from corresponding simulation quantities is minimized, similar to that laid out in Egedal et al. (2019). However, some adjustments are made to suit the MMS event at hand. Similarly to the previous method, PIC simulation units are converted to physical MMS units using two parameters, the ratio of PIC to MMS densities and the ratio of PIC to MMS temperatures. Once simulation units are converted, a direct numerical optimization scheme to fit the spacecraft path is tractable.

To ensure a well-fit magnetic field profile the path is optimized in such a way that it is constrained to be on PIC simulation contours that match the $B_L$ values measured by MMS1. This amounts to optimizing the MMS1 position along a unique simulation $B_L$ contour at each time point, reducing a 2D problem to a 1D problem and largely simplifying the numerical method. The choice of MMS1 in optimizing the path is arbitrary; one may choose any other spacecraft or use mean magnetic field value at the centroid and can achieve near identical results. The event-specific LMN axes (section 2.2.1) combined with the converted simulation units allows for determination of the spacecraft positions relative to MMS1 in the simulation *LN*-plane.

At each time point, a penalty function $h(r)$ is evaluated, where $h$ is a sum of weighted $\chi^2$-differences between spacecraft measurements and corresponding simulation data parameterized by $r$, the distance along the given $B_L$ contour that matches MMS1 data. The signals included in the penalty function are all components of the electromagnetic fields (except for $B_{L,MMS1}$), electron and ion flow velocities, parallel and perpendicular electron pressures, and the ratio of parallel to perpendicular electron pressures. An additional contribution to the penalty function, $g(r)$, penalizes solutions whose positions $r$ are too far away from that at the previous time step $r_{\text{previous}}$ and ensures a continuous trajectory. This contribution takes the exact form $g(r) = (r - r_{\text{previous}})^2/\sigma_g^2$, where $\sigma_g$ is an adjustable weight to enforce a smooth trajectory.

The optimization problem is solved by stepping through time and taking the MMS1 position to be the $r$-value corresponding to the minimum value of the penalty function. For further details of this method, see Supporting Information of Schroeder et al. (2022).

### 2.3 Methods for Reconstructing 2D/3D Structures

#### 2.3.1 Field Reconstruction Using Quadratic Expansion

To understand the context of reconnection events, it is desirable to have a reconstruction of the magnetic field in the vicinity of the spacecraft. Without an explicit reconstruction, researchers map the location of the spacecraft by comparing the time series of **B** to the nominal diffusion region picture seen in many 2D simulations (e.g., see Torbert et al. 2018), often using the LMN coordinate system as determined in different ways and described in section 2.2. MMS provides new measurements that allow reconstructions that depend only on the data and the vanishing divergence of the magnetic field. This is made possible because of: 1) the very high fidelity of the current density measurements using only particle data (Pollock et al. 2016; Phan et al. 2016); and 2) the very high accuracy of the magnetometers (Russell et al. 2016), assisted with independent measurements of the field magnitude by the Electron Drift Instrument (EDI) (Torbert et al. 2016b). Using a “modified” curlometer (see Dunlop et al. (1988) for the original method), which employs both temporal and spatial variations of **B** to estimate the current density, Torbert et al. (2017) showed that the particle data matched the magnetic variations at the highest cadence available on MMS within an EDR, where the current density is far from uniform.

If the current density **j** from the particle measurements at four spacecraft locations is assumed correct, then one can extend the linear curlometer approximation (Dunlop et al. 1988) to the second order and reconstruct the magnetic topology in the vicinity of the MMS tetrahedron to sense the locations of X-lines and the four spacecraft within the diffusion region. Torbert et al. (2020) implemented such a reconstruction, using a 24-parameter Taylor expansion around the barycenter of the tetrahedron. Given that there are 24 knowns (3 components times 4 spacecraft measurements of **B** and **j**), this gives a solution for the field that exactly matches the data, with the divergence of **B** zero everywhere. However, such an exact solution requires the addition of at least one cubic term in the expansion because of the constraint that the divergence of the current density, which in the expansion is computed from the curl of **B**, must be zero (neglecting the displacement current for a nonrelativistic system). The measured current values almost never have this property, for the primary reason that they are taken at separated spatial locations where the current may be highly varying. Torbert et al. (2020) assumed that the second derivative in the MDD minimum gradient direction was zero, and arrived at an exact fit for spacecraft measurements of **B** and **j** by using a superposition of cubic terms weighted by the inverse of the coefficient required for each term.

Given that MMS can measure the electron distribution and compute the current density every 30 ms (and sometimes every 7.5 ms (Rager et al. 2018; see also Appendix A)), a reconstruction can be computed for every such time step. As an example, such a reconstruction is given in Figure 8, for a time when the MMS constellation approached an EDR at the magnetopause on 16 October 2015, as reported by Burch et al. (2016b). The field is computed in a 3D cubic lattice, and the field lines are traced in this lattice. The field lines are then projected into the shown *LN* plane. The field topology is insensitive to the actual weighting of the 18 solutions using different cubic terms. Using such a reconstruction with synthetic data from simulation as input, Torbert et al. (2020) showed that these reconstructions are very representative of the simulation data within a volume whose linear extent is about twice that of the spacecraft tetrahedron.

Denton et al. (2020) implemented a modification of this technique, using only quadratic terms, based on scaling arguments for the various terms and the concern that the exact solutions may lead to over-fitting of the data and show spurious X-lines when far from the tetrahedron. The number of terms in the expansion is reduced using estimates of their relative scaling, and the coefficients are then determined by a least-squares fitting procedure with an assumed weighting between the **B** and **j** values, depending on their accuracies. Although the data cannot exactly match the model for reasons described above, these reconstructions appear to give better results without false X-lines when sensing the presence of X-lines out further from the tetrahedron.

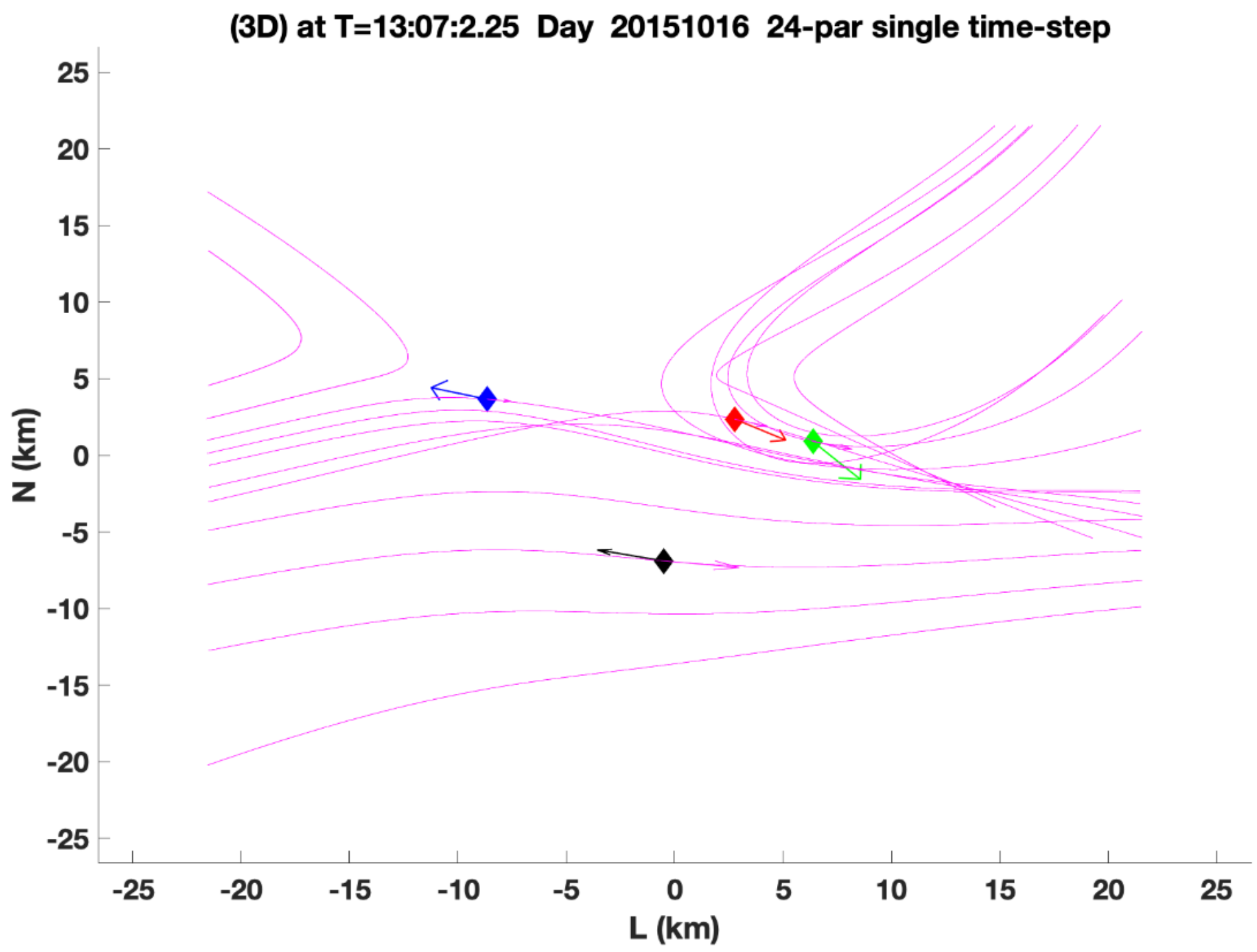

**Fig. 8** A reconstruction of the magnetic field lines as MMS approached an electron diffusion region at 13:07:02.25 UT on 16 October 2015. The four MMS spacecraft locations are colored diamonds: (black, red, green, blue) are the standard colors for (MMS1, 2, 3, 4) respectively. The colored arrows show the projection of the electron flow velocity into this *LN* plane. The purple arrows at each spacecraft show the direction of the field and the lengths the relative magnitude of the magnetic field at that location.

Figure 9 shows the reconstructed magnetic field close to the MMS spacecraft using the reduced quadratic model of Denton et al. (2020) at the same time as that plotted in Figure 8. Figure 9 shows some interesting features, the sheared field with an X-line close to MMS4, the field line of MMS2 approaching the X-line even closer, and the tilt of the magnetic island structure toward more positive *L* at more positive *M*, suggesting that the invariant direction has an *L* component. Some features are possibly unrealistic. For instance, the flux rope in the island might well be larger than Figure 9 suggests. Also, some features of the reconstruction are sensitive to details of the reconstruction procedure, such as the amount of smoothing and adjustment of the electron density (scaling of the electron density from the particle instruments to better agree on average with the current density from the curl of the magnetic field, as described by Denton et al. (2020)), so the exact field line structure is not known. However, the reconstruction well shows the positions of the MMS spacecraft relative to the X line.

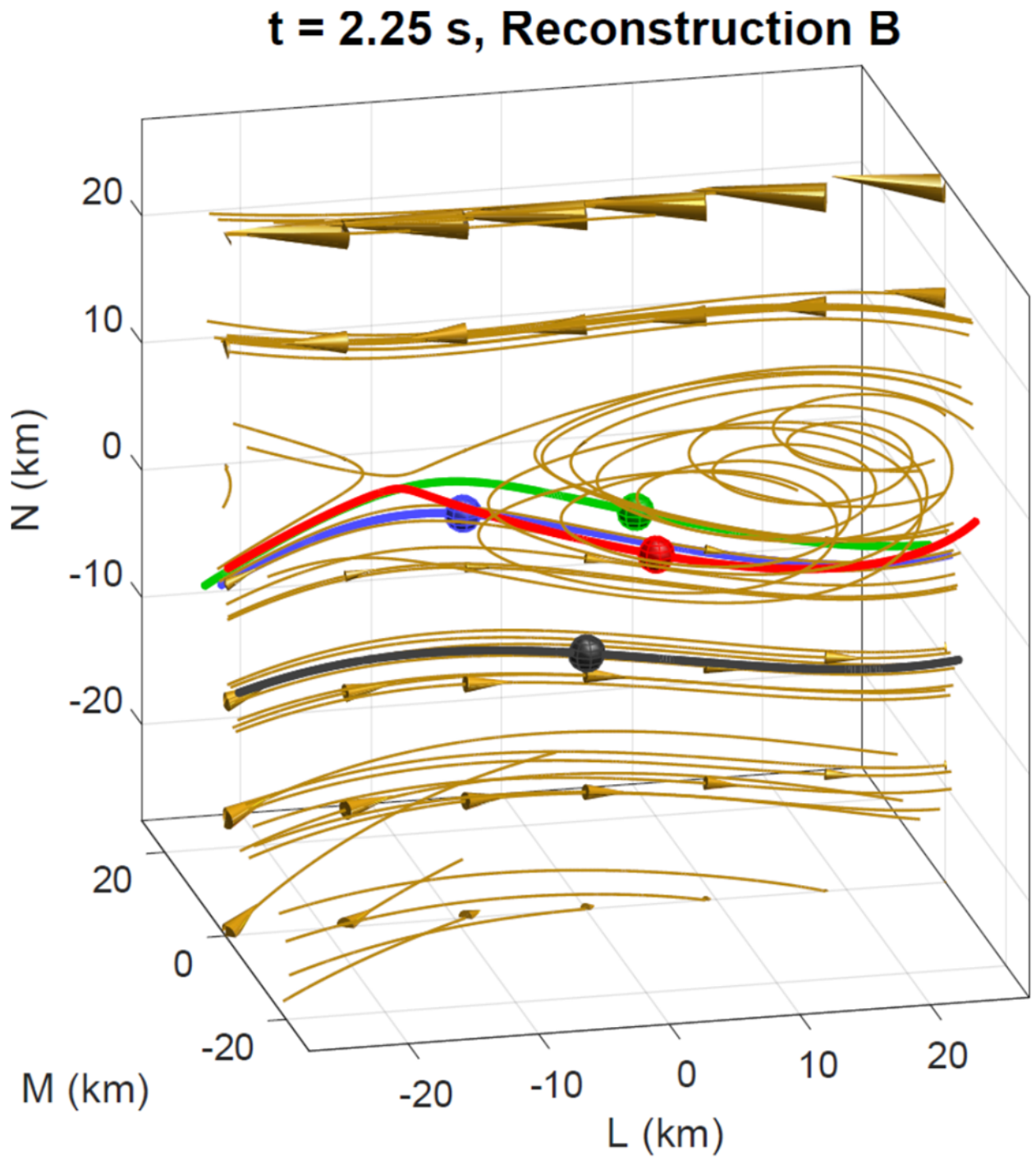

**Fig. 9** Reduced quadratic reconstruction for 16 October 2015, 13:07:02.25 UT. The black, red, green, and blue spheres and curves show the positions and magnetic field lines passing through MMS1, 2, 3, and 4, respectively. The gold curves are other magnetic field lines with the cones indicating the direction and magnitude.

Another approach to the over-fitting problem is to use the data at multiple times and assume that the magnetic topology has not changed over this time interval. The assumption is that the spacecraft are moving through a semi-stationary structure. After all, this is the essence of what researchers have done in the past to draw cartoons of the reconnection regions from time series data over much longer intervals. In producing a data-based reconstruction, the velocity of the spacecraft relative to the structure is required. This can be estimated from time-of-flight analysis or STD, or found from the best fit to the data. The best fit method requires an iterative procedure. For example, for the encounter of the EDR seen in Figure 8, Burch et al. (2016b) estimated that over an interval of about 0.2s, the spacecraft were moving through the structure with $V_{sc,N} = 45$ km/s. For a multiple time-step reconstruction, a least-squares fit is required for the 72 data elements (24 at 3 times around 13:07:02.25 UT) of the 24-parameter expansion. The iterative solution produced a different velocity ($V_{sc,N} = 21$ km/s, while hardly moving in the $L$ direction with $V_{sc,L} = 1 \pm 5$ km/s). The reconstruction at an earlier time of ~13:07:02.05 UT showed $V_{sc,N} =$ ~55 km/s, closer to that determined by Burch et al.

(2016b). Research is ongoing into the accuracy of the velocity determined in this way, but the multiple time-step solution appears to be more stable than the single one.

A reduced quadratic reconstruction using the method of Denton et al. (2022) with multiple input times (as described above) yields a result similar to that in Figure 9, except that the field line passing through MMS2 wraps around the magnetic island (not shown). Differences in the path of this field line are not surprising considering that the field line passing through MMS2 comes very close to the X-line in Figure 9. A similar result is found with a complete quadratic reconstruction using the Denton et al. (2022) method (not shown).

### 2.3.2 3D Empirical Reconstruction Using Stochastic Optimization Method

Zhu et al. (2022) developed a new model for empirical reconstruction of the 3D magnetic field and current density field using a stochastic optimization method called simultaneous perturbation stochastic approximation (SPSA) (e.g., Spall 1998; Zhu and Spall 2002; Spall 2003). The model employs an empirical approach by fitting the prescribed analytic functions for the magnetic field to the point-wise measurements from a constellation of spacecraft using physical constraints derived from a set of Maxwell equations. The fitness of the reconstruction is defined by a general loss function ($G$), which consists of both the differences between the model and in-situ measurements and the model deviations from linear or nonlinear physical constraints. While most applications of SPSA utilize loss functions that include only the differences between the modeled and measured quantities (e.g., Chin 1999; Spall 2003), the new model characterizes the physical robustness of the reconstructed fields. The SPSA approach also has an additional feature that the algorithm includes the effects of random measurement errors. Zhu et al. (2022) demonstrated this new model using MMS measurements of the magnetic field and current density ($\hat{\mathbf{B}}, \hat{\mathbf{j}}$) for the 11 July 2017 magnetotail EDR event (Torbert et al. 2018), which was previously explored by a least-squares method (e.g., Denton et al. 2020; Torbert et al. 2020) introduced in section 2.3.1.

The generalized loss function ($G$) used in this new empirical reconstruction model has the form of

$$G = G_O + w_A \varepsilon_A G_A + w_B \varepsilon_B G_B + w_C \varepsilon_C G_C, \tag{3}$$

where the components of the loss function ($G_O, G_A, G_B, G_C$) are defined as

$$G_O = \frac{1}{12} \sum_{a=1}^{4} \sum_{i=1}^{3} \left[B_i(\mathbf{r}_a) - \hat{B}_{a,i}\right]^2, \tag{4}$$

$$G_A = \frac{1}{12} \sum_{a=1}^{4} \sum_{i=1}^{3} \left[j_i(\mathbf{r}_a) - \hat{j}_{a,i}\right]^2, \tag{5}$$

$$G_B = \frac{1}{9}[\delta^2(\mathbf{r}_0) + \sum_{a=1}^{4} \delta^2(\mathbf{r}_a) + \sum_{a=1}^{4} \delta^2(\mathbf{r}_{Fa})] \text{ or } G_B^* = \frac{1}{5}[\delta^2(\mathbf{r}_0) + \sum_{a=1}^{4} \delta^2(\mathbf{r}_a)],$$

(6) and

$$G_C = \frac{1}{4}\sum_{a=1}^{4}\left[\mu_0 \mathbf{j}(\mathbf{r}_{Fa}) \cdot \left(\Delta\mathbf{r}_{\beta\gamma} \times \Delta\mathbf{r}_{\beta\delta}\right) - (\bar{B}_{\beta\gamma} \cdot \Delta\mathbf{r}_{\beta\gamma} + \bar{B}_{\gamma\delta} \cdot \Delta\mathbf{r}_{\gamma\delta} + \bar{B}_{\delta\beta} \cdot \Delta\mathbf{r}_{\delta\beta})\right]^2.$$

(7)

Here, $\Delta\mathbf{r}_{\beta\gamma} = (\mathbf{r}_\gamma - \mathbf{r}_\beta)$ is the edge vector connecting the vertices $\mathbf{r}_\beta$ and $\mathbf{r}_\gamma$ and $\bar{\mathbf{B}}_{\beta\gamma} = (\hat{\mathbf{B}}_\beta + \hat{\mathbf{B}}_\gamma)$ is the mean magnetic field on the edge $\Delta\mathbf{r}_{\beta\gamma}$ calculated using the measured $\hat{\mathbf{B}}$ field by applying a linear approximation between the two spacecraft observations along that edge. $G_O$ and $G_A$ each comprises twelve terms and quantifies the model-measurement difference at each vertex of the tetrahedron ($\mathbf{r}_a$). $G_B$ comprises nine physical constraints and requires minimization of $\delta^2(\mathbf{r}) = (\nabla \cdot \mathbf{B})^2$ at nine spatial points across the tetrahedron (i.e., the barycenter $\mathbf{r}_0$, each of the four vertices $\mathbf{r}_a$, and the center of each of the four faces $\mathbf{r}_{Fa}$). The face centers can be disregarded by replacing $G_B$ with $G_B^*$. $G_C$ comprises four approximate physical constraints derived from applying Stokes' theorem to Ampere's law $\left(\mu_0 \iint_S \tilde{\mathbf{j}} \cdot d\mathbf{S} = \oint_C \hat{\mathbf{B}} \cdot dl\right)$ on each of the four faces of the tetrahedron; the current density components normal to the tetrahedron faces ($\tilde{\mathbf{j}}$) are derived from the curlometer method with the measured $\hat{\mathbf{B}}$. $G_C$ results from minimizing the difference between $\tilde{\mathbf{j}}$ and $\mathbf{j}$ projecting onto the normal of each of the four tetrahedron faces. Specification of the weighting factors $(w_A, w_B, w_C)$ in $G$ determines which loss function components are included in the reconstruction. The scaling parameters $(\varepsilon_A, \varepsilon_B, \varepsilon_C)$ are dependent on the spatial separations of the four spacecraft and are defined so that the different components of the loss function are of the same order of magnitude. The applied SPSA approach also determines the model parameters that minimize a dimensionless loss function that includes a random perturbation that captures the effects of measurement errors.

Zhu et al. (2022) validated the empirical reconstruction by introducing indices $(\gamma_B, \gamma_j)$ defined as the normalized magnitude of the differences between the measured $(\hat{\mathbf{B}}, \hat{\mathbf{j}})$ and modeled $(\mathbf{B}, \mathbf{j})$ fields. These sets of indices $(\gamma_B, \gamma_j)$, shown in Figure 10, provide a qualitative measure of the accuracy to the reconstructed fields. Additionally, a model quality indicator, $Q_{\text{model}}$, is introduced – based on the quality indicator $Q_{\text{curl}}$, which is a measure of the ratio $|\nabla \cdot \hat{\mathbf{B}}|/|\nabla \times \hat{\mathbf{B}}|$, introduced by Dunlop et al. (1988) – to provide a quantitative assessment of the robustness of the modeled field in terms of the physical property of $\nabla \cdot \mathbf{B} = 0$. These indices respectively represent the two sets of constraints applied to the model-measurement differences and the deviations of the model considered when designing the applied generalized loss function. Zhu et al. (2022) examined the error sources in the reconstructed fields previously noted by studies applying the curlometer method and found that these curlometer-calculated errors in the current density primarily arose from the application of the linear approximation to what is in reality a nonlinear configuration of the 3D magnetic fields.

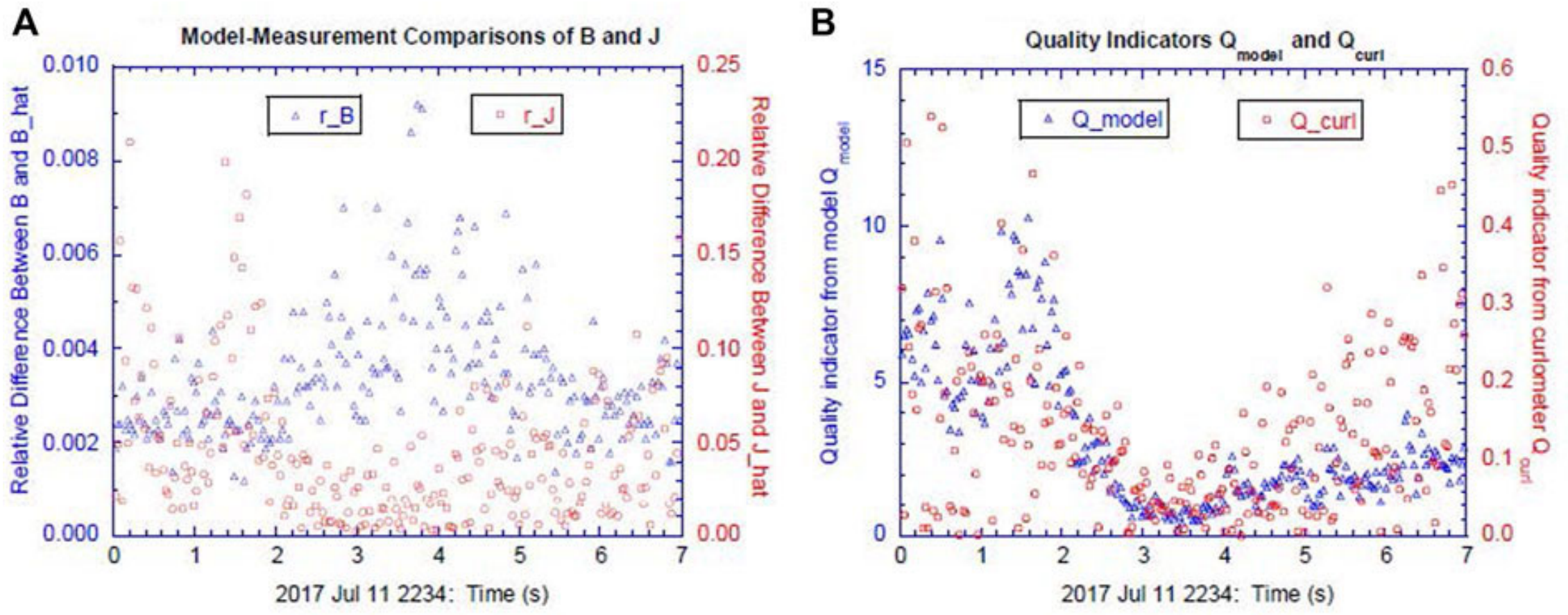

**Fig. 10** (**a**) Relative differences ($\gamma_B, \gamma_j$) and (**b**) quality indicators ($Q_{\mathrm{model}}, Q_{\mathrm{curl}}$) for a sensitivity run with weighting factors $w_B$ and $w_C$ both set to 0. The very small relative difference values ($\ll 1$) highlights that the empirical model results in a very good fit between the modeled and measured fields at the prescribed spatial points.

### 2.3.3 2D Reconstruction of Reconnection Events Assisted by Simulation

For some spacecraft events a 2D reconstruction can shed light on physics of interest or validate models. For an MMS magnetotail reconnection event (on 10 August 2017) shown in this section, reconstruction allowed for revealing whether the time series of data is consistent with a laminar 2D reconnection geometry or if 3D dynamics are required to explain the observations. Here we present an interpolative method that assumes steady-state reconnection. For the given event, it allows for reconstruction of a physical area extending about $40d_e \times 10d_e$ (where $d_e = c/\omega_{pe}$ is the electron inertial length) around the x-line, an area much larger than that allowed in methods that rely on Taylor expansion (section 2.3.1) or electron magnetohydrodynamics equations (section 2.3.4).

For a given event, once one optimizes spacecraft trajectories parameterized in 2D space and time, using the method as introduced in section 2.2.3, the MMS signals can be used to construct 2D field maps, as shown in Figure 11. In the event considered here, the trajectories closely follow the magnetic separatrix of the reconnection geometry (shown as the nearly horizontal thick black curve in Figure 11). There it can be assumed that electron-scale gradients are mostly perpendicular to the magnetic separatrix because electrons thermally stream along the field lines. Therefore, a grid is defined to have cells elongated approximately parallel with the separatrix in order to capture variation in signals across the topological boundary. To achieve this, the spacecraft trajectories are rotated clockwise by an angle $\theta = 17°$ to a coordinate system in which the magnetic separatrix followed by the spacecraft becomes nearly horizontal. These rotated coordinates are defined as ($L'$, $N'$); the $L'$-coordinate approximately represents the distance along the separatrix, and the $N'$-coordinate approximately represents the distance perpendicular to the separatrix.

Raw MMS data are then distributed spatially according to the optimized spacecraft trajectories. Data are placed into spatial bins with lengths $\Delta L' \simeq 2d_e$ and $\Delta N' \simeq 0.08d_e$, such that the aspect ratio of each cell is approximately 25. All bins through which neither of the trajectories passes are left as empty cells (NaN values). Thus, each spatial grid cell contains data values from all times when either of the MMS spacecraft paths falls within its area. The final value for each cell is calculated by a simple average of all data values contained in that cell.

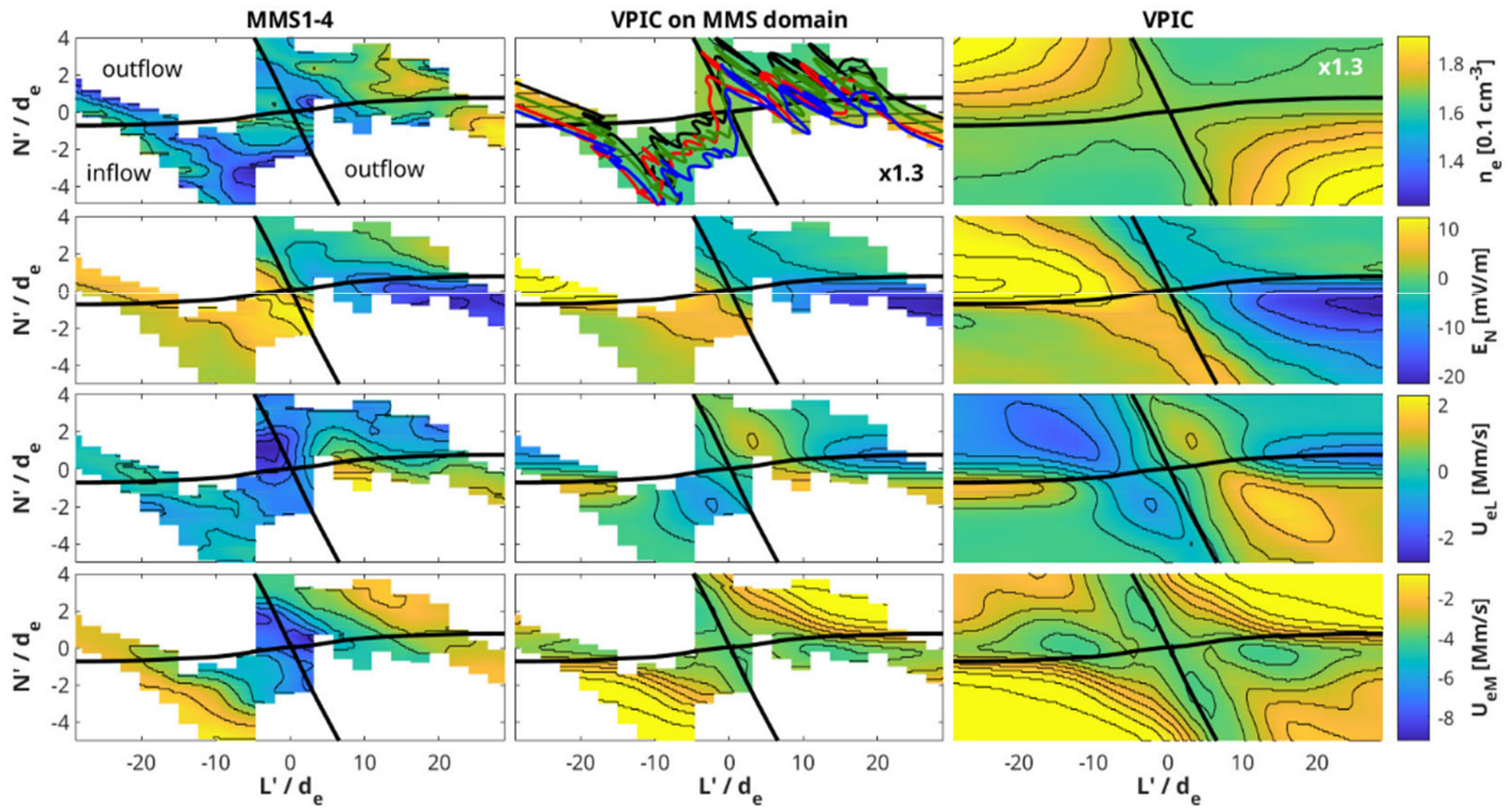

**Fig. 11** The left column shows fields measured by MMS constructed into a 2D map based on the spacecraft trajectory. The middle column shows PIC simulation data in the same region of the reconnection geometry for comparison, while the right column shows the simulation data plotted over the entire domain. Adapted from Schroeder et al. (2022).

The particle-in-cell data inherently fills out the entire 2D simulation domain (right panels of Figure 11). However, to allow for direct comparison between the measurement maps and simulation, the simulation data are binned and averaged on the same grid as the spacecraft data (middle column of Figure 11). We note that $\nabla \cdot \mathbf{B} = 0$ is not guaranteed by this method, because the spacecraft paths are found by the method discussed in section 2.2.3, so that the measured and thus reconstructed field values do not strictly agree with the simulation values.

### 2.3.4 Grad-Shafranov and Electron Magnetohydrodynamics Reconstruction

Fundamental equations, such as a set of MHD equations, can be used to reconstruct steady, 2D structures around the path of an observing spacecraft from in-situ measurements of the electromagnetic field and plasma. In standard numerical simulations, a set of equations governing the system is solved as an initial value problem for studying temporal evolution of the system or physical quantities. This can be done by setting the initial and boundary conditions at every part of the simulation domain. On the other hand, the reconstruction as explained below solves a time-independent form of the governing equation(s) as a spatial initial value problem to get 2D field maps of physical quantities. This is possible by setting, based on the measurements, the initial conditions at points along the spacecraft path and solving the equation(s) for spatial development of the corresponding quantities. The first such method, Grad-Shafranov (GS) reconstruction technique, was introduced by Sonnerup and Guo (1996), and was further extended to include MHD (Sonnerup and Teh 2008) and Hall-MHD effects (Sonnerup and Teh 2009). An overview and reviews of these earlier types of reconstruction were given by Sonnerup et al. (2006b, 2008) and Hasegawa (2012). Here we describe more recent developments of the reconstruction techniques along the same line.

For GS reconstruction schemes, in which 2D and steady structures are assumed, one needs to find a proper or comoving frame of reference in which the structure looks approximately time-independent, and a reconstruction plane that is perpendicular to the invariant axis ($\hat{\mathbf{z}}$) along which the structure has negligible spatial gradients. The 2D maps of plasma and magnetic fields are recovered on that plane. From single spacecraft observations, the velocity of the proper frame ($\mathbf{V}_{\mathrm{str}}$) can be obtained by the HT analysis (Khrabrov & Sonnerup 1998; section 2.2.2). If observations from four spacecraft are available, the STD method (Shi et al. 2006; section 2.2.2) can also be used to determine the velocity of the structure.

The invariant axis ($\hat{\mathbf{z}}$) can be determined by rotating one of the eigenvectors from Minimum Variance Analysis (MVA) (Sonnerup and Scheible 1998), taken as a trial invariant axis, by some angle until measured data points in the parameter plane of a field line invariant (such as the axial component of the magnetic field $B_z$ and the transverse pressure $P_t = p + B_z^2/(2\mu_0)$ where $p$ is the plasma pressure) versus partial vector potential $A$ (out-of-plane component of the vector potential) are approximately expressed by a single curve, namely, an exponential or polynomial function: $B_z = B_z(A)$ and $P_t =$ $P_t(A)$ (e.g., Hu and Sonnerup 2002). By ingesting multi-spacecraft data, the optimal axis could also be found in such a way that the correlation coefficient between the reconstructed magnetic fields based on one spacecraft data and the measured magnetic fields from other spacecraft reaches the maximum value (Hasegawa et al. 2004). Using magnetic field data from four-point measurements, the minimum gradient direction from the MDD analysis can be taken as the invariant axis (Shi et al. 2005; section 2.2.1). In some cases, the results of MDD and MVA can be combined to provide a reconstruction coordinate system, in which not only the invariant axis (parallel to $\hat{\mathbf{e}}_M$) but also the *L* and *N* axes are properly defined (Denton et al. 2016, 2018; Hasegawa et al. 2017; Tian et al. 2020; section 2.2.1). The *x* axis is defined as being antiparallel to the projection of the structure velocity $\mathbf{V}_{\rm str}$ onto the plane perpendicular to the invariant axis ($\hat{\mathbf{z}}$), thus representing the spacecraft path in the reconstruction (*xy*) plane. A right-handed orthogonal system is formed by $\hat{\mathbf{y}} = \hat{\mathbf{z}} \times \hat{\mathbf{x}}$.

**Grad-Shafranov Reconstruction with pressure anisotropy effects** In collisionless plasma there can be pressure anisotropy, that is $p_\perp \neq p_\parallel$, where $p_\perp$ and $p_\parallel$ are the thermal pressures perpendicular and parallel to the magnetic field, respectively. Taking the effects of pressure anisotropy and field aligned flow into account, Sonnerup et al. (2006b) derived a new GS equation by considering the double-polytropic energy laws (Hau and Sonnerup 1993) $\mathrm{d}\{p_\perp/(\rho B^{\gamma_\perp - 1})\}/dt = 0$ and $\mathrm{d}(p_\parallel B^{\gamma_\parallel - 1}/\rho^{\gamma_\parallel})/dt = 0$, where $\rho$ is the mass density, *B* is the magnetic field strength, $\gamma_\perp$ and $\gamma_\parallel$ are polytropic exponents, which can be inferred from observations. Different values of $\gamma_\perp$ and $\gamma_\parallel$ represent different thermodynamic conditions (e.g., Hau et al. 2020). Chen and Hau (2018) developed a GS code for anisotropic and field-aligned flow for the first time and benchmarked it with an analytical model. The application of this code to a magnetopause crossing event showed that the recovered magnetic islands inside the magnetopause had larger widths than that from the GS reconstruction for isotropic plasma.

There are also some space plasma structures with anisotropic pressure in quasi-static equilibrium, such as mirror-mode structures and magnetospheric ultra-low frequency compressional waves (drift mirror-mode wave). Tian et al. (2020) reconstructed the magnetic field structure of the ultra-low frequency compressional wave by the GS method including the pressure anisotropy effect. They call it the reduced GS-like method, because the corresponding GS-like equation,

$$\nabla \cdot [(1-\alpha)\nabla A] = \mu_0 \rho \left[ T_\perp \frac{dS_\perp}{dA} + T_\parallel \frac{dS_\parallel}{dA} - \frac{dH}{dA} \right] - B_z \frac{dC_z}{dA}, \qquad (8)$$

can be derived by removing terms containing the bulk velocity in the equations given by Sonnerup et al. (2006b), which contain both the anisotropy and field-aligned flow effects.

Here, $S_\perp = c_{v\perp} \cdot \ln(p_\perp/\rho B^{\gamma_\perp - 1})$ and $S_\parallel = c_{v\parallel} \cdot \ln(p_\parallel B^{\gamma_\parallel - 1}/\rho^{\gamma_\parallel})$ are the perpendicular and parallel pseudo entropies, respectively, $H = [p_\perp/\{(\gamma_\perp - 1)\rho\}] + [\gamma_\parallel p_\parallel/\{(\gamma_\parallel - 1)\rho\}]$ is the total enthalpy, $C_z = (1-\alpha)B_z$, $\alpha = (p_\parallel - p_\perp)\mu_0/B^2$ is the pressure anisotropy factor, $\mu_0$ is the vacuum permeability, $B_z$ is the magnetic field component along the invariant axis, $c_{v\perp} = R/(\gamma_\perp - 1)$ and $c_{v\parallel} = R/(\gamma_\parallel - 1)$ with the ordinary gas constant $R = c_{p\parallel} - c_{v\parallel}$. All of $S_\perp$, $S_\parallel$, $H$ and $C_z$ are field line invariants and are functions of $A$ only. Reduced auxiliary equations,

$$\mathbf{M}\mathbf{X}^T = \mathbf{Y}^T, \tag{9}$$

are used to spatially advance quantities $\alpha$, $\rho$, $p_\perp$, $p_\parallel$, $B_z$, $B^2$ and $\partial^2 A/\partial^2 y$ in $y$, along with spatial integration of $A$ and $B_x$. Here, the superscript $T$ denotes the matrix transpose, and **M** is a 7×7 matrix expressed as follows:

$$\mathbf{M} = \begin{bmatrix} \frac{c_{v\perp}}{p_\perp} & 0 & -\frac{c_{v\perp}}{\rho} & 0 & -\frac{R}{2B^2} & 0 & 0 \\ 0 & \frac{c_{v\parallel}}{p_\parallel} & -\frac{c_{p\parallel}}{\rho} & 0 & \frac{R}{2B^2} & 0 & 0 \\ \frac{c_{v\perp}}{R\rho} & \frac{c_{p\parallel}}{R\rho} & -\frac{(c_{v\perp}p_\perp + c_{p\parallel}p_\parallel)}{R\rho^2} & 0 & 0 & 0 & 0 \\ 0 & 0 & 0 & 1-\alpha & 0 & -B_z & 0 \\ -1 & 1 & 0 & 0 & -\alpha & -B^2 & 0 \\ 0 & 0 & 0 & B_z & -0.5 & 0 & B_x \\ 0 & 0 & 0 & 0 & 0 & \frac{B_x}{\alpha - 1} & 1 \end{bmatrix}, \tag{10}$$

$$\mathbf{X} = [\frac{\partial p_\perp}{\partial y}, \frac{\partial p_\parallel}{\partial y}, \frac{\partial \rho}{\partial y}, \frac{\partial B_z}{\partial y}, \frac{\partial B^2}{\partial y}, \frac{\partial \alpha}{\partial y}, \frac{\partial^2 A}{\partial y^2}], \tag{11}$$

$$\mathbf{Y} = [B_x \frac{dS_\perp}{dA}, B_x \frac{dS_\parallel}{dA}, B_x \frac{dH}{dA}, B_x \frac{dC_z}{dA}, B_y \frac{\partial B_x}{\partial x}, \frac{Q}{1-\alpha}], \tag{12}$$

where $Q$ in **Y** is RHS$-\partial[(1-\alpha)\partial A/\partial x]/\partial x$, and RHS is the right-hand side quantity in Eq (8). One difficulty in this method exists in determining the proper polytropic exponents $\gamma_\perp$ and $\gamma_\parallel$. Hau et al. (2020) inferred these parameters by using the measured magnetosheath data, and recovered the 2D topology of a mirror-mode structure observed in the magnetosheath.

Aiming at reconstruction of anisotropic plasma structures, Teh (2019) developed another simple extended GS equation. He did not use the double polytropic energy laws, but assumed that parameters $\alpha$, $p_\perp$, and $p_\parallel$ are functions of the magnetic field strength $B$ only to derive a relatively simple GS-like equation. This assumption might not be valid. Nevertheless, basic features of magnetic mirror-mode and flux rope with pressure anisotropy were revealed by this extended GS solver (Teh 2019; Teh and Zenitani 2020).

**Electron Magnetohydrodynamics Reconstruction** The electron

magnetohydrodynamics (EMHD) reconstruction is a single-spacecraft method for the reconstruction of steady, 2D electromagnetic fields and electron streamlines in regions where ions are fully demagnetized and thus electron dynamics dominates. It was developed to recover the field geometry in and around the EDR of magnetic reconnection, where electrons are demagnetized. The original version (Sonnerup et al., 2016) is based on an inertia-less and time-independent form of the electron MHD equation (Kingsep et al. 1990 and references therein) and assumes uniform electron density (electron incompressibility) and temperature. A recent version incorporates electron inertia effects in the streamline reconstruction, and the effects of nonuniform density and temperature and a guide magnetic field component $B_z$ in the reconnection region (Hasegawa et al. 2021). The density and temperature are, however, assumed to be preserved along the magnetic field lines, because such assumptions are roughly satisfied around the EDR of symmetric, antiparallel reconnection (Korovinskiy et al. 2020). These conditions are not well satisfied for guide-field or asymmetric reconnection; therefore, further model developments are needed. Nonetheless, under the above assumptions, the magnetic field can be reconstructed by use of a Grad-Shafranov-type equation

$$\nabla^2 A = -\mu_0 j_z(A) = \mu_0 e n_\mathrm{e}(A) u_{\mathrm{e}z}(A). \tag{13}$$

Reconstruction of the EDR requires some kind of dissipation term and, for antiparallel reconnection, makes use of a term corresponding to the component of the divergence of the electron pressure tensor $\mathbf{P}$ in the direction $\hat{\mathbf{z}}$ of reconnection electric field (or X-line) (see review by Hesse et al. (2011))

$$(\nabla \cdot \mathbf{P}) \cdot \hat{\mathbf{z}} = n_\mathrm{e}\sqrt{2 m_\mathrm{e} k_\mathrm{B} T_\mathrm{e}}\frac{\partial u_{\mathrm{e}L}}{\partial L}. \tag{14}$$

In the case of guide-field reconnection, see Hasegawa et al. (2021) for some recipes. The reconstruction is performed in the rest frame of magnetic field structures, as introduced in the second paragraph of section 2.3.4.

Figure 12 shows 2D maps of the magnetic field and electron streamlines from the EMHD reconstruction with electron inertia effects for the magnetotail EDR event on 11 July 2017, first reported by Torbert et al. (2018). Magnetic field, electric field, and electron moment data taken by the MMS3 spacecraft, which made the closest approach to the X-line, are used to set the initial conditions on the *x* axis. The final frame velocity was determined by a multi-spacecraft method (Hasegawa et al. 2017), in which the correlation coefficient is maximized between the components of the magnetic fields and electron velocities measured by the three spacecraft (MMS1, MMS2, and MMS4) not used as input and those predicted from the maps along the spacecraft paths. The final *z* axis was optimized by a method based on the *y* component of a time-independent and 2D form of Ampère's law, $-\partial B_z/\partial x = \mu_0 j_y$; when the $z$ axis is properly chosen, this relation approximately holds for the particle current density and magnetic field data taken along the spacecraft path ($x$ axis) (Hasegawa et al. 2019).

Figure 12a shows a clear X-type magnetic field geometry, as seen in simulations. We note that the information on the separatrix opening angle gained from the reconstructed field map, as seen in Figure 12a, can be used to estimate the reconnection electric field by a method explained in section 3.3.3. Figure 12b essentially shows expected patterns of electron inflow and outflow and corresponding Hall magnetic fields $B_{\rm z}$. Interestingly, the electron stagnation point is displaced in the earthward (outflow) direction by ~$3d_{\rm e}$ ($d_{\rm e}$~27 km/s) from the X-point, a new feature revealed by the reconstruction. The method has also been successfully applied to an EDR of magnetopause reconnection (Hasegawa et al. 2017) and an ion-scale magnetic flux rope in the magnetopause current sheet (Hasegawa et al. 2023).

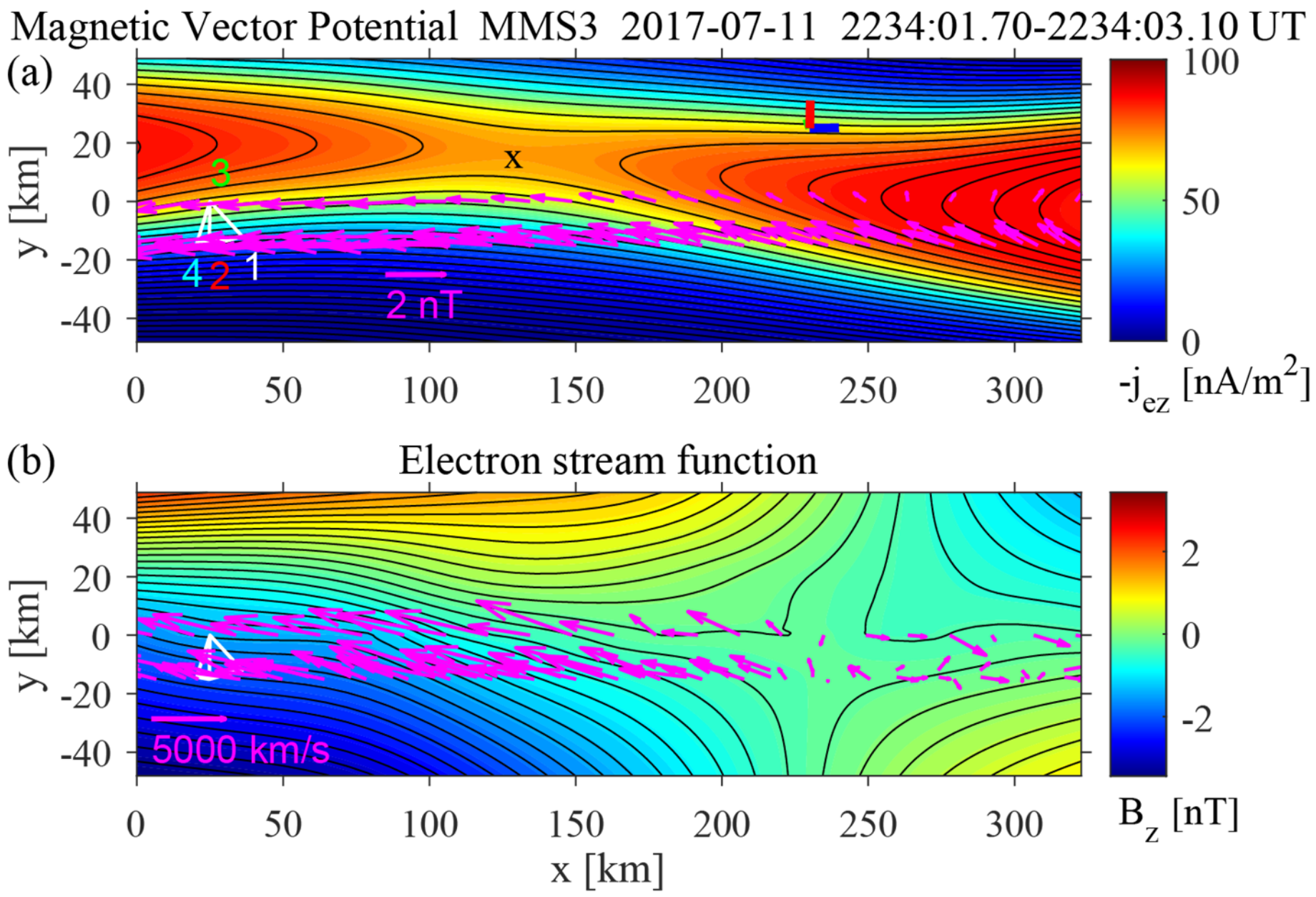

**Fig. 12** 2D maps of the magnetic field (**a**) and electron streamlines (**b**) from the EMHD reconstruction applied to data from the magnetotail EDR encounter by MMS3 on 11 July 2017. The arrows show the projection onto the reconstruction plane of the measured magnetic fields (**a**) and electron velocities in the structure-rest frame (**b**). The axial current density (**a**) and axial magnetic field component (**b**) are shown in color. The blue and red bars in Figure 12 are the projection of the GSM $x$ and $z$ axes, respectively. Adapted from Hasegawa et al. (2021).

**MMS-Tailored Electron Magnetohydrodynamics Reconstruction** An alternative

approach to the EMHD GS reconstruction of EDR is represented by the MMS-tailored model, introduced by Korovinskiy et al. (2021). Adopting an assumption of steady-state two-dimensional magnetoplasma configuration and assuming additionally the uniform number density $n$, the problem is reduced to calculation of two quantities: the magnetic potential $A$ of the in-plane magnetic field and the out-of-plane magnetic field component $B_z$. The out-of-plane component of Ampère's law makes the magnetic potential to obey the equation,

$$\nabla^2 A = \mu_0 e n u_{\mathrm{e}z}, \qquad (15)$$

where the right-hand side is represented, in general, by a function of two variables (contrary to Eq. (13)). Then, with neglected ion current and $\partial/\partial z = 0$, the in-plane components of Ampère's law reveal that the quantity $-B_z/(\mu_0 e)$ serves as a stream function for the in-plane electron flow $n\mathbf{u}_{\mathrm{e}\perp} = (nu_{\mathrm{e}x}, nu_{\mathrm{e}y})$: $\mu_0 e n u_{\mathrm{e}x} = -\partial B_z/\partial y$ and $\mu_0 e n u_{\mathrm{e}y} = +\partial B_z/\partial x$. With uniform number density, this yields the equation for $B_z$,

$$\nabla^2 B_z = \mu_0 e n\left(\partial u_{\mathrm{e}y}/\partial x - \partial u_{\mathrm{e}x}/\partial y\right) = Q. \qquad (16)$$

Here, the model function $Q$ represents the contribution of the electron inertia and anisotropy (see Eq. 14–16 of Korovinskiy et al. 2021).

Since the Jacobian of a variable transform $(x, y) \to (A, B_z)$ may turn to zero or infinity at a manifold of Lebesgue measure zero only (see Eq. 9 of Korovinskiy et al. (2021)), the right-hand sides of Eqs. (15,16) can be considered as the functions of $(A, B_z)$. Assuming that Cartesian coordinates correspond to the local co-moving LMN coordinate system (e.g., Denton et al. 2018), where the $x$ axis coincides with $\hat{\mathbf{e}}_L$ and the $y$ axis coincides with $\hat{\mathbf{e}}_N$, one can note that the stretched configuration of EDR, dictated by a low reconnection rate $\varepsilon \sim 0.1$ (e.g., Liu et al. 2017), brings the ratio $\partial/\partial x \ll \partial/\partial y$. The analogous scaling ratio $\partial/\partial B_z \ll \partial/\partial A$ (see, e.g., Eqs. 12 and 13 of Korovinskiy et al. (2021)) is valid in the variable space $(A, B_z)$. Omitting the minor dependence on $B_z$, one can consider the zeroth-order reconstruction model, where the model functions depend on $A$ only. This way Eq. (15) turns into the Grad-Shafranov equation (13) and in Eq. (16) we have $Q = Q(A)$. Notably, under this equality the mathematical self-consistency of the solution demands an extra term $\sim B_z$ on the right-hand side of Eq. (15); however, the contribution of this term is found to be negligible (see Eq. 21 and Fig. 7 of Korovinskiy et al. 2021), so it can be omitted.

The model functions $u_{\mathrm{e}z}(A)$ and $Q(A)$ are evaluated from the boundary conditions. Particularly, the latter is calculated by using the data of the four MMS probes. Thus, the reconstruction model development is completed and the only problem left is the solution of the ill-posed problem, stated by Eq. (15,16), with the boundary conditions specified along the probe trajectory. In an approach discussed by Korovinskiy et al. (2020), the problem regularization was performed by utilizing the so-called boundary layer approximation (BLA) (Schlichting 1979). Namely, the second-order small terms $\partial^2/\partial x^2 \sim \varepsilon^2\, \partial^2/\partial y^2$ are omitted, reducing the problem to the system of ordinary differential equations of the second order. To benefit from the simplicity of this method, the local LMN coordinate system must be accurately determined. Besides that, this method is less universal, as it is not applicable to EDR crossings in the direction normal to the current sheet.

The described model, named 'Model 2' in Korovinskiy et al. (2021), was tested by reconstruction of the MMS event on 11 July 2017 (Torbert et al. 2018). The advantages of this model are the following. First, it does not depend on the out-of-plane electric field $E_z$, which is assumed to be constant, while in reality it may be considerably oscillating (see Fig. 2a in Korovinskiy et al. (2021)). The nonuniformity of $E_z$ brings appreciable uncertainty to the reconstruction results obtained by utilizing Eq. (14), representing the approximation derived by Hesse et al. (1999) for the electron pressure anisotropy (see 'Model 3' of Korovinskiy et al. (2021)). Since this approximation is not used, reconstruction errors, which can appear due to its possible inaccuracy (see Fig. 2 in Korovinskiy et al. (2020), and the corresponding discussion), are also eliminated. Second, with BLA the model allows evaluation of the small terms $B_y$ and $u_{\mathrm{e}y}$, but these quantities do not affect other computations, particularly, the computation of $B_z$ and $u_{\mathrm{e}x}$. This property brings advantage when the translational symmetry of the configuration is corrupted, since in this case $\partial B_x/\partial z \neq 0$, and hence the formula $\mu_0 enu_{\mathrm{e}y} = \partial B_z/\partial x$, used in the two-dimensional models, fails.

Thus, the major advantages of the discussed MMS-tailored technique consist in its comparative simplicity and increased accuracy due to the reduced sensitivity to violations of the ideal theoretical conditions. This is achieved at the expense of the lost universality, since the model requires current sheet crossing in the direction with a nonzero angle with respect to $\hat{\mathbf{e}}_N$ and multi-spacecraft data for evaluating the model function $Q$. The major disadvantage − the assumption of uniform number density, which limits the model applicability to the internal EDR − can be relaxed by substituting $n = n(A)$ in Eqs. (12−16) of Korovinskiy et al. (2021). Another limitation is related to the adopted assumption of a not-too-small guide field value. For a case of small or zero guide field, the corresponding (rather straightforward) modifications are required; in particular, the symmetry considerations demand $Q = Q(A, B_z)$, while the zeroth-order model equation $j_{\mathrm{e}z} = j_{\mathrm{e}z}(A)$ stays unchanged. The extended compressible model and discussion of the ways to further improve the EMHD GS reconstruction technique can be found in

Korovinskiy et al. (2023).

### 2.3.5 3D Field Reconstruction Using Modified Radial Basis Functions

Another method for the reconstruction of 3D local magnetic field structures is based on the use of toroidal and poloidal magnetic potentials that are expressed as linear combinations of radial basis functions (e.g., Buhmann, 2003). In this approach, the magnetic field is represented by

$$\mathbf{B} = \nabla \times \left(\frac{\psi_1}{r}\mathbf{r}\right) + \nabla \times \nabla \times (\psi_2 \mathbf{r}), \qquad (17)$$

where $\psi_1$ and $\psi_2$ are the toroidal and poloidal potentials, respectively. Instead of using orthogonal basis functions, the potentials are individually expanded into a series $\psi_j = \sum_i \alpha_{ji}\chi_i$ of a modified form of radial basis functions (Andreeva and Tsyganenko 2016; Tsyganenko and Andreeva 2016)

$$\chi_i(\mathbf{r}; D, \mathbf{L}) = \left\{\left(\frac{x - R_{i,x}}{L_x}\right)^2 + \left(\frac{y - R_{i,y}}{L_y}\right)^2 + \left(\frac{z - R_{i,z}}{L_z}\right)^2 + D^2\right\}^{1/2}. \qquad (18)$$

Here, the vectors $\mathbf{R}_i$ are coordinates of meshwork grid nodes, $\mathbf{L} = \{L_x, L_y, L_z\}$ a characteristic length set equal to the node separation along each axis, and $D$ an adjustable regularization parameter. An advantage of the magnetic field thus defined is its divergence-free nature (Stern 1976). The expansion coefficients $\{\alpha_i\}$ can be determined by optimally fitting to magnetic field data taken by multiple spacecraft, such as Cluster and MMS, for a time interval.

Chen et al. (2019) tested this technique for 2D and 3D model magnetic fields, and applied it to magnetic field data from the Cluster mission. The structure velocity relative to the spacecraft was treated as a hyper parameter, and was tuned by minimizing the average magnitude of an error vector field $\Delta\mathbf{B} \equiv \mathbf{B}_{\text{data}} - \mathbf{B}_{\text{rec}}$, where $\mathbf{B}_{\text{data}}$ and $\mathbf{B}_{\text{rec}}$ are the measured and reconstructed magnetic field vectors, respectively. Their study suggests that the method can be used to identify and investigate the properties of characteristic structures in 3D magnetic reconnection, including magnetic nulls (section 3.1.7) and separator lines.

## 3 Detection and Analysis of In-Situ Observations of the Diffusion Regions

This section mostly focuses on the methods for detecting and analyzing in-situ observations of the electron diffusion region (EDR). See Phan et al. (2005) for a brief review of in-situ observations and analysis of ion diffusion regions (IDRs) with effects of the Hall term in the generalized Ohm's law (e.g., Liu et al. 2024).

### 3.1 Diffusion Region Identification

This section reviews several measures that can be used to identify the diffusion region or nearby regions. We stress that none of the following measures are uniquely non-zero or significant only in the diffusion region, and that none of them by themselves necessarily identify "dissipation" in the sense of an irreversible process, although some are conventionally called a dissipation measure or have been used to identify the region where a process leading to entropy increase may occur. Therefore, one should use these measures in the context of other measurements of the reconnection process to identify a candidate diffusion region; the candidate may or may not turn out to be an actual diffusion region after full quantitative analysis to make sure whether other reconnection and diffusion region signatures are observed.

**3.1.1 Plasma-Frame Dissipation Measure**

Since magnetic reconnection converts electromagnetic field energy to plasma kinetic energy, we expect the energy conversion rate $\mathbf{j} \cdot \mathbf{E}$ or its variants ($\mathbf{j}_\mathrm{i} \cdot \mathbf{E}$ and $\mathbf{j}_\mathrm{e} \cdot \mathbf{E}$) to be significant in magnetic reconnection. These quantities are important, but may not be useful to identify the EDR because they can be nonzero even in ideal regions (see Eq. (30) below).

One problem is that $\mathbf{j} \cdot \mathbf{E}$ is measured in the stationary or observer's (laboratory) frame; if the EDR is moving, it might be better to evaluate $\mathbf{j} \cdot \mathbf{E}$ in a particular frame. Starting from this, we consider the energy conversion in a moving frame that travels at a velocity $\mathbf{V}_\mathrm{r}$. In this reference frame, the electric current density and the electric field are given by

$$\mathbf{j}'_\mathrm{r} = \mathbf{j}'(\mathbf{V}_\mathrm{r}) = \mathbf{j} - \rho_\mathrm{c}\mathbf{V}_\mathrm{r}, \tag{19}$$

$$\mathbf{E}'_\mathrm{r} = \mathbf{E}'(\mathbf{V}_\mathrm{r}) = \mathbf{E} + \mathbf{V}_\mathrm{r} \times \mathbf{B}, \tag{20}$$

where $\rho_\mathrm{c}$ is the charge density and the primed quantities denotes those in the moving frame. The energy conversion rate in the $\mathbf{V}_\mathrm{r}$-moving frame yields

$$D_\mathrm{r} = D(\mathbf{V}_\mathrm{r}) \equiv \mathbf{j}' \cdot \mathbf{E}' = \mathbf{j} \cdot (\mathbf{E} + \mathbf{V}_\mathrm{r} \times \mathbf{B}) - \rho_\mathrm{c}(\mathbf{V}_\mathrm{r} \cdot \mathbf{E}). \tag{21}$$

Here, the reference velocity is arbitrary, so we call Eq. (21) the frame-independent dissipation measure. Employing the electron fluid velocity as the reference velocity, we define the electron-frame dissipation measure $D_\mathrm{e} = D(\mathbf{u}_\mathrm{e})$, and employing the ion fluid velocity, we obtain the ion-frame dissipation measure $D_\mathrm{i} = D(\mathbf{u}_\mathrm{i})$.

Importantly, these measures are Galilean invariants, in other words, frame-independent. By using the Lorentz factor $\gamma_\mathrm{r} = [1 - (V_r/c)^2]^{1/2}$, we obtain a Lorentz invariant form (Zenitani et al., 2011a)

$$D_\mathrm{r} = \gamma_\mathrm{r}[\mathbf{j} \cdot (\mathbf{E} + \mathbf{V}_\mathrm{r} \times \mathbf{B}) - \rho_\mathrm{c}(\mathbf{V}_\mathrm{r} \cdot \mathbf{E})]. \tag{22}$$

It is seen that as long as we choose a unique reference velocity $\mathbf{V}_\mathrm{r}$, Eq. (22) always gives the same result regardless of the observer's velocity or direction.

Let us discuss properties of the dissipation measures. We focus on the nonrelativistic limit of $\gamma_r \to 1$ for simplicity. In a plasma, the charge density and the electric current density are

$$\rho_c = \sum_s q_s n_s, \qquad \mathbf{j} = \sum_s q_s n_s \mathbf{u}_s, \tag{23}$$

where $s$ denotes the plasma species. We consider a charge-weighted sum of Eq. (20)

$$\sum_s q_s n_s \mathbf{E}'(\mathbf{u}_s) = \rho_c \mathbf{E} + \mathbf{j} \times \mathbf{B}. \tag{24}$$

Applying $\mathbf{j}\cdot$ to both sides, and using Eq. (23), we obtain

$$\mathbf{j} \cdot \sum_s q_s n_s \mathbf{E}'(\mathbf{u}_s) = \rho_c \mathbf{j} \cdot \mathbf{E} = \rho_c \left(\sum_s q_s n_s \mathbf{u}_s\right) \cdot \mathbf{E}, \tag{25}$$

$$\sum_s q_s n_s (\mathbf{j} \cdot \mathbf{E}'_s - \rho_c \mathbf{u}_s \cdot \mathbf{E}) = 0. \tag{26}$$

This provides a useful relation

$$\sum_s q_s n_s D_s = 0. \tag{27}$$

Next we consider the relevance to resistive MHD. We define the MHD quantities,

$$\rho_{\rm mhd} \equiv \sum_s m_s n_s, \qquad \mathbf{u}_{\rm mhd} \equiv \frac{\sum_s m_s n_s \mathbf{u}_s}{\sum_s m_s n_s} = \frac{\sum_s m_s n_s \mathbf{u}_s}{\rho_{\rm mhd}}. \tag{28}$$

Since Eq. (21) only uses linear operators, we obtain

$$D_{\rm mhd} = D(\mathbf{u}_{\rm mhd}) = D\left(\frac{\sum_s m_s n_s \mathbf{u}_s}{\rho_{\rm mhd}}\right) = \frac{\sum_s m_s n_s D(\mathbf{u}_s)}{\rho_{\rm mhd}} = \frac{\sum_s m_s n_s D_s}{\rho_{\rm mhd}}. \tag{29}$$

From the Ohm's law, we obtain the energy conversion rate

$$\mathbf{E} + \mathbf{u}_{\rm mhd} \times \mathbf{B} = \eta \mathbf{j}, \qquad \mathbf{j} \cdot \mathbf{E} = (\mathbf{j} \times \mathbf{B}) \cdot \mathbf{u}_{\rm mhd} + \eta j^2. \tag{30}$$

Rearranging the MHD-frame dissipation $D_{\rm mhd} = D(\mathbf{u}_{\rm mhd})$, we immediately obtain from Eq. (21)

$$\mathbf{j} \cdot \mathbf{E} = (\mathbf{j} \times \mathbf{B} + \rho_c \mathbf{E}) \cdot \mathbf{u}_{\rm mhd} + D_{\rm mhd}. \tag{31}$$

In a quasineutral ion-electron plasma, from Eqs. (27) and (29), we obtain

$$D_{\rm mhd} \approx D_{\rm i} \approx D_{\rm e}. \tag{32}$$

From Eqs. (30) and (31) and considering that $\rho_c$ is negligible in MHD, we see that the $D_{\rm mhd}$ term ($\approx D_{\rm e}$) plays the same role as the nonideal energy conversion rate $\eta j^2$.

PIC simulations have revealed that $D_{\rm e} > 0$ marks a compact physically-significant region surrounding the X-line (Zenitani et al. 2011a). Although the resolution was limited and it was only partially evaluated, Zenitani et al. (2012) reported $D_{\rm e} > 0$ during a magnetotail reconnection event observed by the Geotail spacecraft (Nagai et al. 2011). Recent observations by MMS unambiguously reported $D_{\rm e} > 0$ near the EDR (Burch et al. 2016b; Phan et al. 2018; Torbert et al. 2018). Based on these results, it is fair to say that $D_{\rm e} > 0$ is an important signature of the EDR.

We raise unsolved issues here. First, there is often a weakly negative region of $D_{\rm e} < 0$ in the downstream side of the EDR, where the electrons outrun the $\mathbf{E} \times \mathbf{B}$ velocity (Karimabadi et al. 2007; Nakamura et al. 2018b; Pritchett 2001; Shay et al. 2007). We often see $D_{\mathrm{e}} < 0$ at the jet termination region where the reconnected magnetic field is compressed (Payne et al. 2021), while $D_{\mathrm{e}} \lesssim 0$ is also seen inside the elongated electron jet (Zenitani et al. 2011b). By clarifying the underlying mechanisms, we may be able to predict the negative amplitude of $D_{\mathrm{e}}$. Second, one can split the electron-frame measure into its perpendicular and parallel contributions, $D_{\mathrm{e}} = \mathbf{j}' \cdot \mathbf{E}' \approx \mathbf{j}_{\perp} \cdot \mathbf{E}'_{\perp} + j_{\parallel} E_{\parallel}$. Inside the EDR, we expect that electron meandering motion provides $\mathbf{j}_{\perp} \cdot \mathbf{E}'_{\perp} > 0$ in antiparallel reconnection, and that the electron parallel motion leads to $j_{\parallel} E_{\parallel} > 0$ in guide-field reconnection. Wilder et al. (2018) evaluated the energy conversion rate $\mathbf{j} \cdot \mathbf{E}' = \mathbf{j}_{\perp} \cdot \mathbf{E}'_{\perp} + j_{\parallel} E_{\parallel}$ in the diffusion regions for multiple reconnection events seen by MMS. They organized the results as a function of the guide-field amplitude $B_{\mathrm{g}}/B_0$, and reported that the perpendicular contribution is dominant in the antiparallel cases ($B_{\mathrm{g}}/B_0 \leq 0.3$) while the parallel contribution is dominant in the guide-field cases ($B_{\mathrm{g}}/B_0 \geq 0.3$) (Figure 10 in Wilder et al. (2018)). However, the critical guide field between the two regimes has not been addressed before, theoretically or numerically. Further research is thus necessary to make a quantitative discussion. Third, strictly speaking, the term "dissipation" is ambiguously used in this section, because these measures do not always lead to an irreversible energy conversion in a collisionless plasma. Connection between the dissipation measures and a true irreversible dissipation process needs to be clarified. This might be better understood from the viewpoint of the entropy production in a kinetic plasma (Liang et al. 2019; Section 3.2.4).

### 3.1.2 Agyrotropy

Due to their small Larmor radii and short gyroperiods, electrons are closely tied to the magnetic field, much more so than heavier ions. As a result, they efficiently probe the field's structure so that locations of topological interest, such as reconnection X-lines and magnetic separatrices, should leave signatures in electron distribution functions. Vasyliunas (1975) identified one such signature by noting that for reconnection to occur, the electron pressure tensor must be non-gyrotropic at the X-point. Spacecraft measurements of non-gyrotropic electron distributions can hence be used as a measure of the proximity to the X-point.

Other authors (Scudder and Daughton 2008; Aunai et al. 2013) have proposed quantifications of agyrotropy, but the version discussed here follows the presentation in Swisdak (2016). A pressure tensor in a field-aligned coordinate system can always be put in the form

$$\mathbf{P} = \begin{pmatrix} P_{\parallel} & P_{12} & P_{13} \\ P_{12} & P_{\perp} & P_{23} \\ P_{13} & P_{23} & P_{\perp} \end{pmatrix}, \qquad (33)$$

where $P_{\parallel}$ and $P_{\perp}$ represent the pressure parallel and perpendicular to the field, respectively.

In the gyrotropic case the off-diagonal components vanish; measures of agyrotropy attempt to quantify the size of these components relative to the diagonal ones. Pressure tensors are positive semidefinite (i.e., have non-negative eigenvalues) and thus satisfy the inequalities

$$P_{12}^2 \leq P_{\parallel} P_{\perp}, \qquad P_{13}^2 \leq P_{\parallel} P_{\perp}, \qquad P_{23}^2 \leq P_{\perp}^2, \qquad (34)$$

which leads to a natural definition of a gyrotropy parameter

$$Q = \frac{P_{12}^2 + P_{13}^2 + P_{23}^2}{P_{\perp}^2 + 2P_{\perp} P_{\parallel}}. \qquad (35)$$

For gyrotropic distributions $Q = 0$, while maximally agyrotropy occurs when $Q = 1$. By using certain rotationally invariant quantities it is possible to calculate $Q$ while in any coordinate system (i.e., the pressure tensor needs not be in the form of Eq. (33)):

$$Q = 1 - \frac{4I_2}{(I_1 - P_{\parallel})(I_1 + 3P_{\parallel})}. \qquad (36)$$

The necessary factors are the trace, $I_1 = P_{xx} + P_{yy} + P_{zz}$, the sum of principal minors, $I_2 = P_{xx}P_{yy} + P_{xx}P_{zz} + P_{yy}P_{zz} - \left(P_{xy}^2 + P_{xz}^2 + P_{yz}^2\right)$, and the parallel pressure $P_{\parallel} = \hat{\mathbf{b}} \cdot \mathbf{P} \cdot \hat{\mathbf{b}} = b_x^2 P_{xx} + b_y^2 P_{yy} + b_z^2 P_{zz} + 2\left(b_x b_y P_{xy} + b_x b_z P_{xz} + b_y b_z P_{yz}\right)$. Here $\hat{\mathbf{b}}$ is the unit vector aligned with the magnetic field.

Figure 13 shows MMS data from a well known crossing of an EDR (Torbert et al. 2018) in the magnetotail with a small guide field ($B_y$ component). The top two panels, which show the components of the magnetic field and the electron velocity, respectively, exhibit the expected signatures of an EDR crossing: a null in $B_x$ associated with a minimum in $B$ and a divergence or reversal in $u_{ex}$. Within the EDR $\sqrt{Q_e}$ rises sharply from its background value of $\approx 0.05$ to a peak of $\approx 0.25$. The bifurcated structure in $\sqrt{Q}$ has been seen in PIC simulations of antiparallel reconnection (Swisdak et al. 2016). Calculations of $Q_i$ (not shown) exhibit a much broader peak, as expected, since ions decouple from the magnetic field, and hence can acquire non-gyrotropic distribution functions on the much larger scales of the IDR.

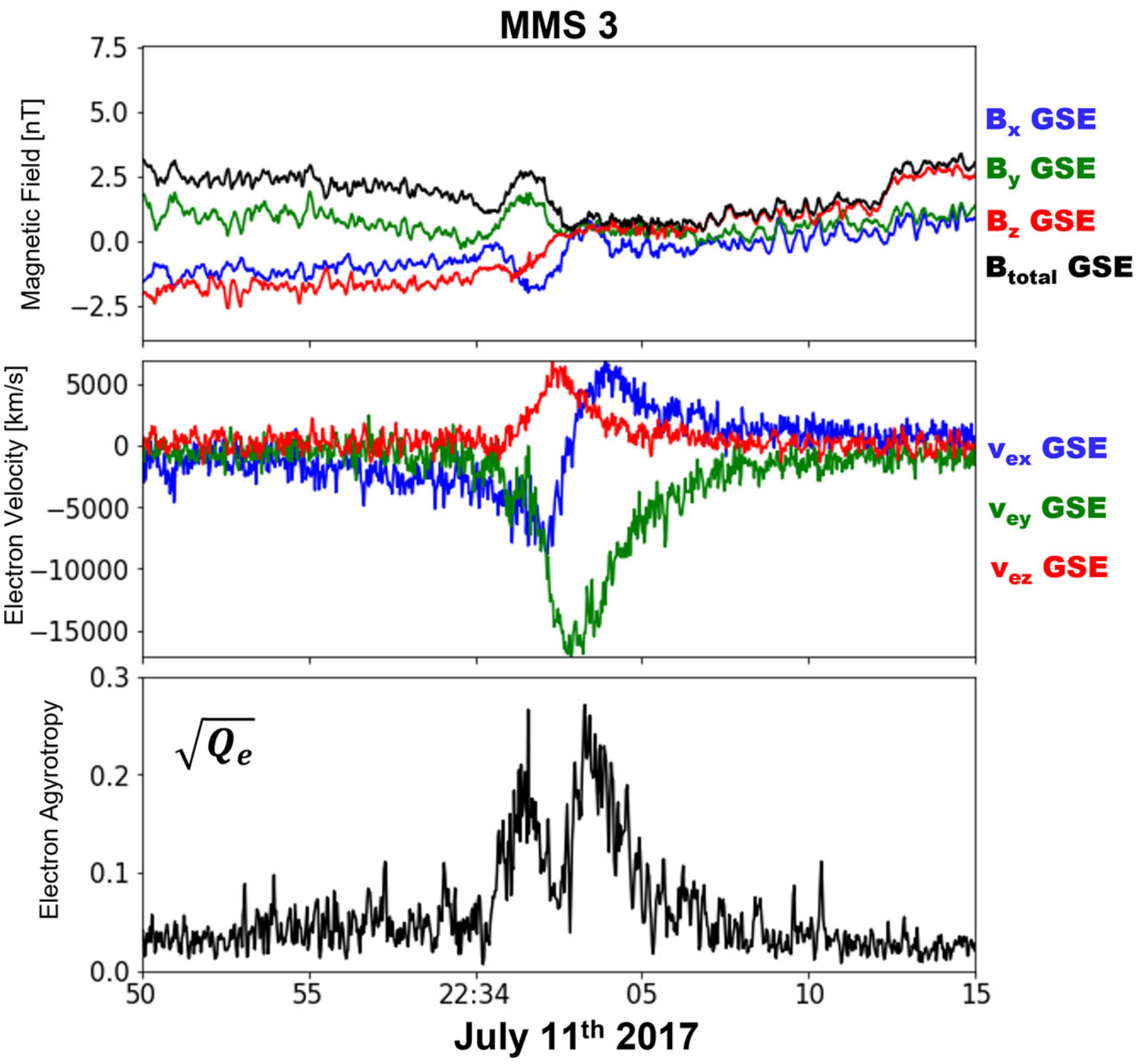

**Fig. 13** Magnetospheric Multiscale (MMS) spacecraft 3 observations of an electron diffusion region at 2233:50-2234:15 UT on 11 July 2017. The top panel shows the three components and the strength of the magnetic field in geocentric solar ecliptic (GSE) coordinates ($B_x$, blue; $B_y$, green; $B_z$, red; and $|\mathbf{B}|$, black). The second panel shows the components of the electron velocity in the same color scheme. The final panel shows $\sqrt{Q_e}$.

We note that non-gyrotropic electron velocity distributions are not a unique signature of magnetic reconnection or the EDR; they have been observed at a non-reconnecting magnetopause current sheet (Tang et al. 2019).

### 3.1.3 Pressure-Strain Interaction in Reconnection Diffusion Region

Energy conversion in magnetic reconnection is a topic of fundamental importance. Several energy conversion measures, such as the Zenitani measure (Zenitani et al. 2011; section 3.1.1), have been used to evaluate energy conversion at the reconnection site. Recent studies have revealed the role of pressure-strain interaction in the conversion of bulk kinetic energy to the random (or thermal, or internal) component in collisionless plasmas (Cerri 2016; Yang et al. 2017; Fadanelli et al. 2021). Here, we show the evaluation of pressure-strain interaction near reconnection X-lines and discuss related interpretations (Bandyopadhyay et al. 2021).

The equation of the thermal or random energy density $E_\alpha^{\mathrm{th}}$, computed from the moments of the Vlasov equations, is given by

$$\partial E_\alpha^{\mathrm{th}}/\partial t + \nabla \cdot \left(E_\alpha^{\mathrm{th}} \mathbf{u}_\alpha + \mathbf{h}_\alpha\right) = -(\mathbf{P}_\alpha \cdot \nabla) \cdot \mathbf{u}_\alpha, \tag{37}$$

where α indicates a specific charged species, $\mathbf{u}_\alpha$ is the fluid velocity, $\mathbf{P}_\alpha$ is the pressure tensor, $E_\alpha^{\mathrm{th}} = 3p^{(\alpha)}/2 = P_{ii}^{(\alpha)}/2$ with $P_{ii}^{(\alpha)}$ as the trace of the pressure tensor, and $\mathbf{h}_\alpha$ is the heat flux.

When integrated over a closed domain, the terms on the left-hand side within the divergence operator average to zero. Therefore, these may be interpreted as transport terms that do not contribute to net change in the form of energy but simply move energy from one location to another spatially. However, if we are concerned with quantifying conversion between different kinds of energy instead of transport, we see that the quantity on the right-hand side is responsible for the conversion of bulk kinetic energy to or from the thermal or random energy. We note that $-(\mathbf{P}_\alpha \cdot \nabla) \cdot \mathbf{u}_\alpha$ is not single-signed, and therefore it does not quantify irreversible conversion of energy. The net energy can convert in or out of the random component, depending on the sign of the pressure-strain.

The pressure-strain can be decomposed into compressive and incompressive components as

$$-(\mathbf{P}_\alpha \cdot \nabla) \cdot \mathbf{u}_\alpha = -p^{(\alpha)}\theta^{(\alpha)} - \Pi_{ij}^{(\alpha)} D_{ij}^{(\alpha)}, \tag{38}$$

where $-p^{(\alpha)}\theta^{(\alpha)}$ represents the energy conversion due to compressive motion, and $\theta^{(\alpha)} = \nabla \cdot \mathbf{u}_\alpha$ is the dilatation. Therefore, the remaining term $-\Pi_{ij}^{(\alpha)} D_{ij}^{(\alpha)}$ corresponds to incompressive energy conversion with the deviatoric pressure tensor $\Pi_{ij}^{(\alpha)} = P_{ij}^{(\alpha)} - p^{(\alpha)}\delta_{ij}$ and the traceless strain rate tensor $D_{ij}^{(\alpha)} = (1/2)\left(\nabla_i u_j^{(\alpha)} + \nabla_j u_i^{(\alpha)}\right) - (1/3)\theta^{(\alpha)}\delta_{ij}$ (Yang et al. 2017; Cassak and Barbhuiya 2022 and references therein).

In recent times, two major advances have facilitated the study of the pressure-strain interaction. First, PIC simulations have become sufficiently accurate to evaluate the pressure tensor, and supercomputers have become adequately powerful to perform plasma simulations with higher number of particles and larger systems. Secondly, an evaluation of $-(\mathbf{P}_\alpha \cdot \nabla) \cdot \mathbf{u}_\alpha$ requires accurate evaluation of the full pressure tensor as well as spatial derivatives of the fluid velocity down to kinetic scales. This was not possible observationally before the Magnetospheric Multiscale (MMS) Mission. The MMS mission, consisting of 4 spacecraft separated by a small distance, with high cadence instruments, provides the first and the only opportunity yet to study pressure-strain interactions using in-situ data.

Figure 14 shows an example of MMS observation of the reconnection diffusion region in the magnetopause current sheet, presented by Burch et al. (2016b). The top panel shows the magnetic field measurement in GSE coordinates. The next panel plots the electromagnetic energy conversion rate, as measured by the Zenitani measure (narrow black), and the energy conversion rate to internal energy, quantified by the total (ion+electron) pressure-strain rate (broad blue). We use the extension of the multi-spacecraft curlometer method (Dunlop et al. 1988; Paschmann & Daly 1998), along with the averaged pressure tensor from all 4 MMS spacecraft to measure $-(\mathbf{P}_\alpha \cdot \nabla) \cdot \mathbf{u}_\alpha$ for ions and electrons. Both conversion rates show an elevated signal in the diffusion region with similar magnitude, but the value of $-(\mathbf{P}_\alpha \cdot \nabla) \cdot \mathbf{u}_\alpha$ is smaller than $\mathbf{j} \cdot \mathbf{E}'$. This observation indicates that only part of the magnetic energy is being converted to random energy. The third panel shows that the electrons are responsible for the majority of the conversion. Finally, the bottom two panels show that for both electrons and ions, the compressive heating rate is stronger than the incompressive part.

These results show that the pressure-strain interaction can be used as an independent diagnostic of plasma energization in reconnection regions. Bandyopadhyay et al. (2021) show a few other examples of MMS reconnection events, including magnetosheath reconnection in thin current sheets and electron-only reconnection to analyze the role of the pressure-strain. Examples from turbulent PIC simulations are also shown by Bandyopadhyay et al. (2021) for comparison with MMS data. Broadly speaking, the simulations and different kinds of reconnection events sampled by MMS do not show any systematic difference in the pressure-strain interaction in the diffusion region. However, $-(\mathbf{P}_\alpha \cdot \nabla) \cdot \mathbf{u}_\alpha$ can be negative in some cases at the reconnecting X-lines, indicating that internal energy is locally being converted into kinetic energy. This contrasts with the electromagnetic energy conversion rate (as measured by $\mathbf{j}\cdot\mathbf{E}'$), which is positive for most of the cases. Understanding the ratio of ion to electron conversion between kinetic and internal energy also poses an intriguing challenge. Like the example shown here, most reconnection cases show that the electron energy conversion rate (as measured by $-(\mathbf{P}_\alpha \cdot \nabla) \cdot \mathbf{u}_\alpha$) is larger than the ion energy conversion rate. This is in contrast to the global energy conversion, which is dominated by the ions in the magnetosheath. A recent work by Barbhuiya and Cassak (2022) provided an explanation of this based on scaling analysis. Further statistical studies are required to fully understand the role of pressure-strain in heating due to magnetic reconnection.

Finally, it should be noted that the above works focused mainly on the $-(\mathbf{P}_\alpha \cdot \nabla) \cdot \mathbf{u}_\alpha$

term that relates the transfer of bulk kinetic energy to internal energy. However, as shown in Fadanelli et al. (2021), it is possible to look at all the terms that compose the electromagnetic, kinetic, and internal energy equations. Such a description allows to perform a point-by-point analysis of all energy conversion channels. While the study of how these various terms compare with and balance each other in the context of reconnection was done using a Hybrid-Vlasov simulation, this exercise remains to be done with spacecraft observations.

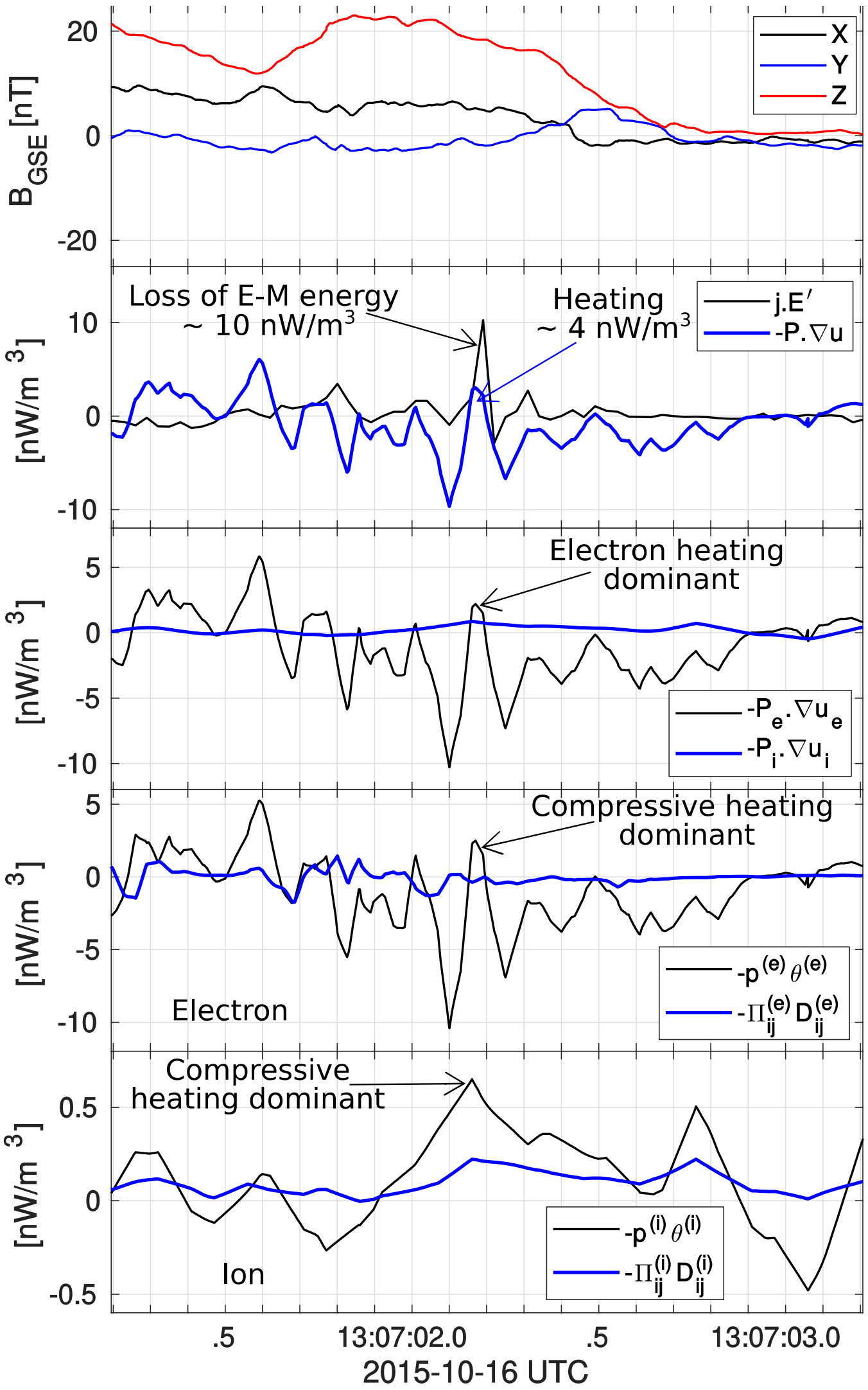

**Fig. 14** MMS data in the magnetopause reconnection event on 16 October 2015. The X-line crossing was around 13:07:02.4 UTC. From top the plotted quantities are the magnetic field in GSE coordinates ($\mathbf{B}_{\mathrm{GSE}}$), the Zenitani measure ($\mathbf{j}\cdot\mathbf{E}'$) and pressure-strain rate ($-(\mathbf{P}\cdot\nabla)\cdot\mathbf{u}$), electron ($-(\mathbf{P}_e\cdot\nabla)\cdot\mathbf{u}_e$) and ion ($-(\mathbf{P}_i\cdot\nabla)\cdot\mathbf{u}_i$) pressure-strain rates, compressive ($-p^{(e)}\theta^{(e)}$) and incompressive ($-\Pi_{ij}^{(e)}\, D_{ij}^{(e)}$) pressure-strain rates for electrons, and compressive ($-p^{(i)}\theta^{(i)}$) and incompressive ($-\Pi_{ij}^{(i)}\, D_{ij}^{(i)}$) pressure-strain rates for ions.

### 3.1.4 Electron Vorticity Indicative of the Electron Diffusion Region

The electron vorticity ($\mathbf{\Omega}_e = \nabla \times \mathbf{u}_e$) can be used as a proxy for delineating the EDR of magnetic reconnection. Figures 15a-d show the 11 July 2017 event (Torbert et al. 2018), during which MMS traversed a magnetotail current sheet along the trajectory shown in Figure 15e: (a) the four-spacecraft tetrahedral-averaged magnetic field components in boundary normal coordinates (LMN), (b) the current density calculated from the curlometer technique (Dunlop et al. 2002), (c and d) electron vorticity and its magnitude (black profile) compared to $\omega_{ce}$ (the electron cyclotron angular frequency; blue). The electron vorticity is enhanced around ‘b’ marked by the vertical dashed red line in Figures 15a-d and red arrow in Figure 15e.

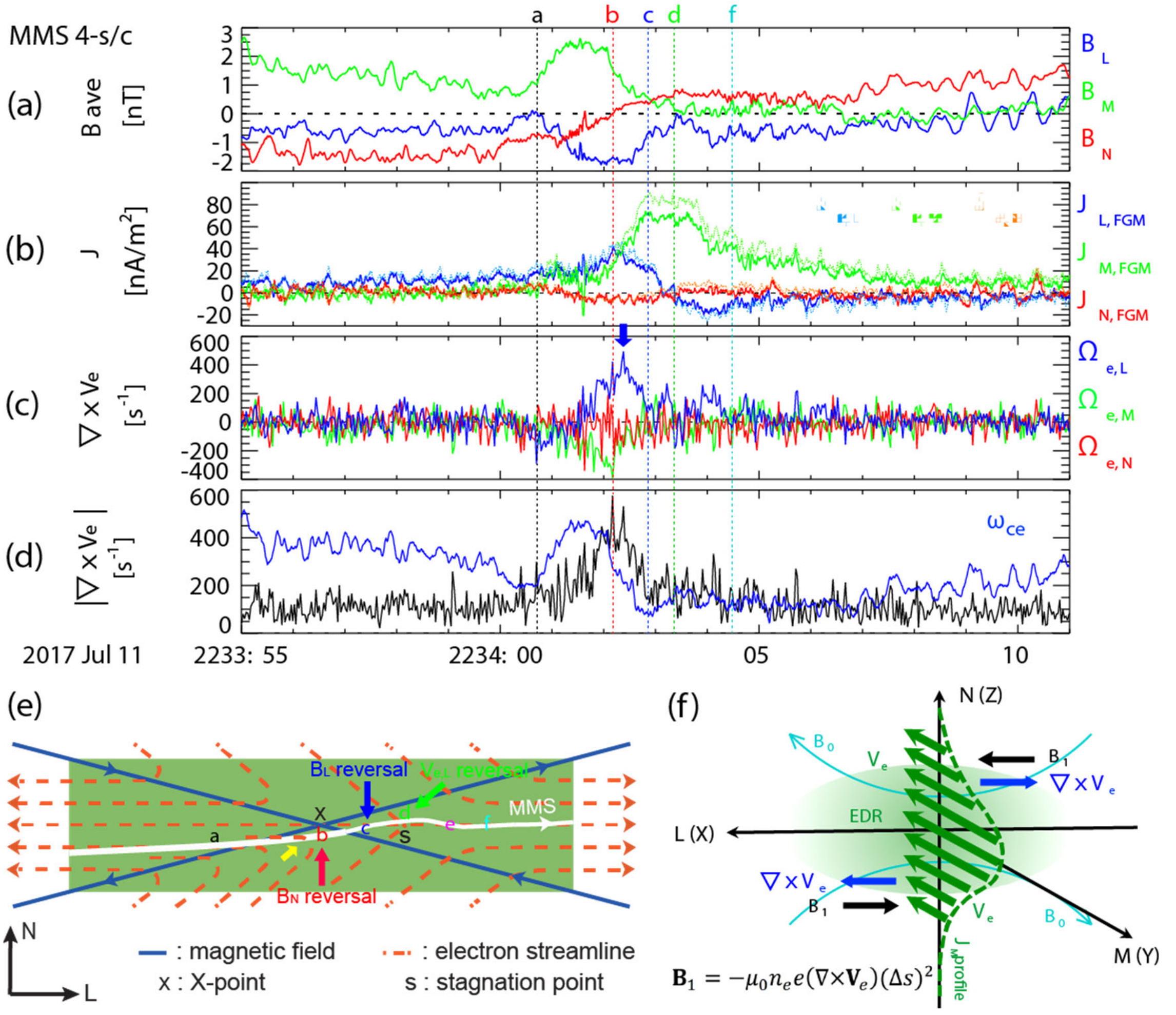

**Fig. 15** MMS observation of a magnetotail current sheet crossing along the trajectory shown in white in panel (**e**). (**a**) The tetrahedral-averaged magnetic field, (**b**) current density calculated from the curlometer technique, (**c**,**d**) electron vorticity and its magnitude compared to $\omega_{ce}$. Panel (**f**) illustrates the origin of the enhanced electron vorticity near the northern/southern edge of the EDR. Adapted from Hwang et al. (2019).

Electron velocity vectors and electron distribution functions measured at the four spacecraft (not shown) demonstrate that the enhanced electron vorticity is due to the intense shear of the velocity mostly along $-\hat{\mathbf{e}}_M$, which originates from the variation along $\pm\hat{\mathbf{e}}_N$ of the meandering electrons' velocity (Figure 15f). Since the meandering electrons carry the out-of-plane current ($j_M$ in Figure 15b), the electron vorticity enhancement should coincide with the strong gradient of the current density. Indeed, $\mathbf{\Omega}_e$ peaks (blue arrow in Figure 15c) are located at the edges of the current density ($j_M$) profile (Figure 15b).

In these observations, the largest component of electron vorticity, i.e., $\Omega_{e,L}$ (Figure 15c), can be approximately written as $\Omega_{e,L} \sim -\partial u_{e,M}/\partial N \sim (1/(en_e))\, \partial j_M/\partial N \sim (1/(en_e\mu_0))\, \partial^2 B_L/\partial N^2$. If $B_L$ changes from 0 at the neutral sheet to $B_{edge}$ at the southern/northern edge of an EDR with a thickness of $d_e$, $\Omega_{e,L} \sim (1/(en_e\mu_0))\, B_{edge}/{d_e}^2 \sim \omega_{ce}$. Thus, a peak of $\Omega_{e,L}$ that is comparable to or larger than $\omega_{ce}$ (Figure 15d) delineates the $N$-directional edge of the EDR of a reconnecting current sheets on the $d_e$ scale. This demonstrates why $|\mathbf{\Omega}_e|$ compared with $\omega_{ce}$ can be a physical measure for EDR identification (Hwang et al. 2019).

### 3.1.5 Magnetic flux transport method

The magnetic flux transport (MFT) method represents a novel way of detecting diffusion regions in situ. It is based on the definition of reconnection as the transport of magnetic flux across magnetic separatrices that intersect at an X-line (Vasyliunas et al. 1975). This method measures signatures of active reconnection in the in-plane velocity of magnetic flux, $\mathbf{U}_\psi$, and its divergence, $\nabla \cdot \mathbf{U}_\psi$. Previously derived in 2D (Liu et al. 2018a; Liu and Hesse 2016) from Faraday's law and the advection equation of magnetic flux, $\partial\psi/\partial t + \mathbf{U}_\psi \cdot \nabla_\perp \psi = 0$, $\mathbf{U}_\psi$ was simplified and adapted for application in 3D (Li et al. 2021, 2023) as

$$\mathbf{U}_\psi = (E_M/B_{LN})(\hat{\mathbf{e}}_M \times \hat{\mathbf{b}}_{LN}), \qquad (39)$$

where $E_M$ is the out-of-plane ($M$) component of the electric field in LMN coordinates, $B_{LN} = \sqrt{{B_L}^2 + {B_N}^2}$ is the magnetic field component in the 2D reconnection ($LN$) plane, $\hat{\mathbf{e}}_M$ is the unit vector in the $M$ direction, and $\hat{\mathbf{b}}_{LN} \equiv \mathbf{B}_{LN}/B_{LN}$ the unit vector of the in-plane magnetic field $\mathbf{B}_{LN}$. The underlying assumptions are that the advection equation does not have a source or loss term, i.e., no magnetic field generation or diffusion occurs, and that $k_M \ll k_\perp$, where $k_M$ and $k_\perp$ are the wavenumbers corresponding to the length scales of the magnetic field variation parallel and perpendicular to the *M* direction, respectively (Li et al. 2023, 2021). The latter essentially means that the scale of variation in the out-of-plane (*M*) direction is much larger than the current sheet thickness. Physically, it represents quasi-2D reconnection (Liu et al. 2018b, 2019; Li et al. 2020). The LMN coordinates can be determined by methods such as minimum variance analysis (Sonnerup and Scheible 1998), maximum directional derivative (Shi et al. 2019), or a combination of methods (Genestreti et al. 2018) (see section 2.2.1 for more details). In simulations, $\mathbf{U}_\psi$ can be calculated if the guide field (*M*) direction is known (e.g., Li et al 2023). Based on measured electromagnetic fields, MFT locates reconnection sites in diffusion regions without using information on plasma flows (Qi et al. 2022). This is ideal for identifying diffusion regions where ion and/or electron outflow jets are not well developed.

The MFT method has been demonstrated to accurately identify reconnection in 2D and 3D kinetic turbulence (Li et al. 2021, 2023) and 3D shock turbulence (Ng et al. 2022) simulations. Recent MMS observations further demonstrated the capability and accuracy of MFT statistically, by directly measuring MFT signatures for active reconnection throughout Earth's magnetosphere (Qi et al. 2022). Reconnection signatures in MFT are (i) co-existing Alfvénic inflow and outflow magnetic flux ($\mathbf{U}_\psi$) jets, and (ii) a significantly enhanced divergence of flux transport ($\nabla \cdot \mathbf{U}_\psi$) at an X-line exceeding the threshold of order $0.1\omega_{\mathrm{ce}}$. We note that the first signature should be observed in a proper frame in which the corresponding X-line is seen to be quasi-stationary.

Here we show an example of application of MFT to the Eriksson event (Eriksson et al. 2018) observed in the magnetosheath on 25 October 2015 by MMS in Figure 16. Panels (e,f) show the MFT quantities: $\mathbf{U}_\psi$ reveals bi-directional inflow MFT jets in the N direction (blue) and a super-Alfvénic outflow jet in the L direction (red), as a signature of reconnection; $\nabla \cdot \mathbf{U}_\psi$ is on the order of the electron gyro-frequency $f_{\mathrm{ce}} = \omega_{\mathrm{ce}}/(2\pi)$, exceeding the threshold of $0.1f_{\mathrm{ce}}$ for identification. The MFT signatures are clear despite the fact that no ion jets were detected along the spacecraft trajectory in this event, interpreted as reconnection in an extended current sheet. A total of 37 previously reported EDR or reconnection-line crossing events were analyzed, including well-known MMS events (e.g., Burch et al. 2016b; Torbert et al. 2018) and electron-only events (e.g., Phan et al. 2018). Almost all (≥95%) of the events can be identified through either of the MFT signatures (Qi et al. 2022). The range of the observed $\mathbf{U}_\psi$ is on the order of ion to electron

Alfvén speeds, and $\nabla \cdot \mathbf{U}_\psi$ is of order $0.1 f_{ce}$ or higher. This order of magnitude is consistent with simulations (Li et al. 2021, 2023). The MFT method can thus provide a clear identification of reconnection in diffusion regions in space.

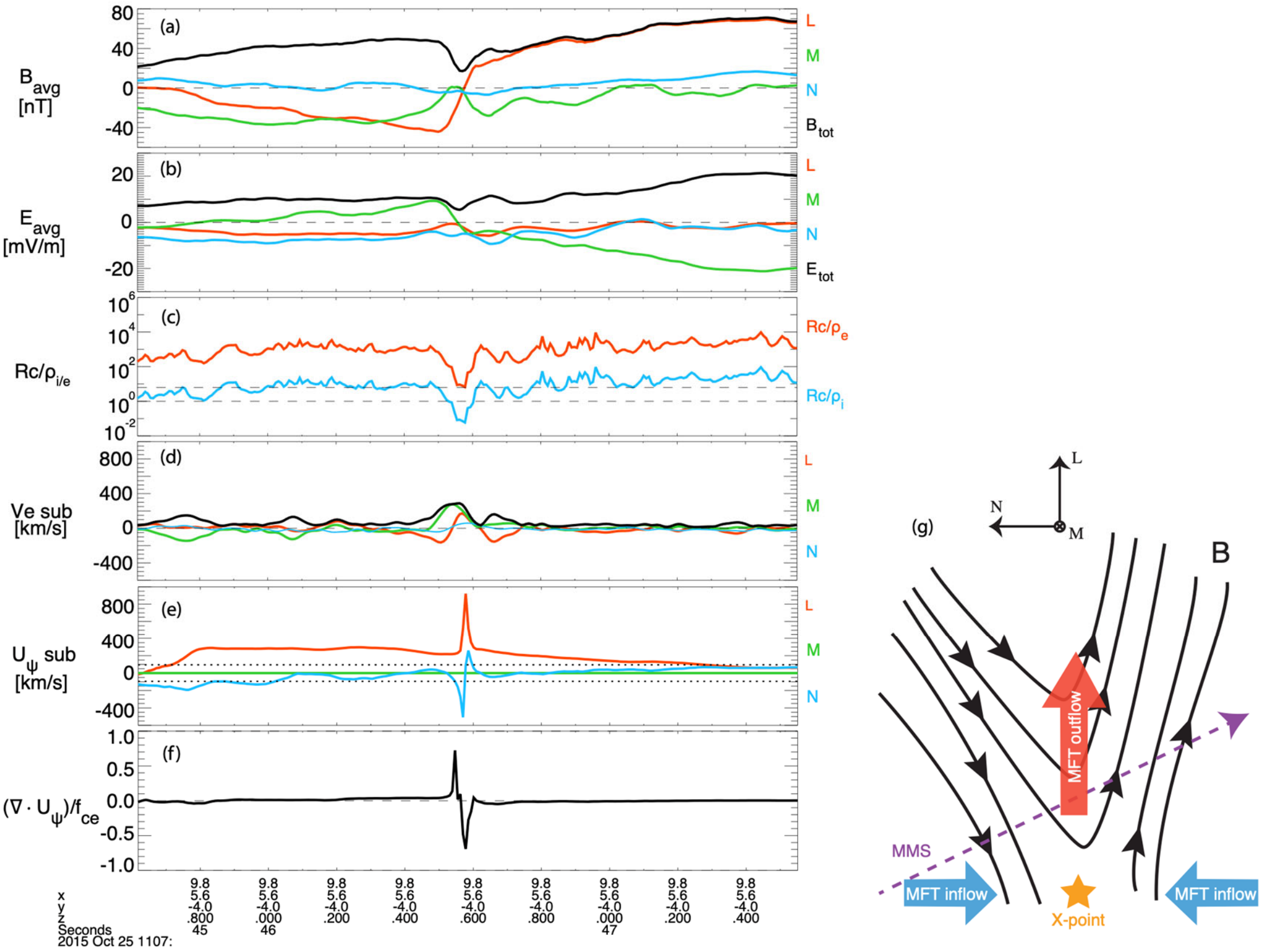

**Fig. 16** MMS observations of the Eriksson et al. event (adapted from Qi et al. (2022)). (**a**) Magnetic field and (**b**) electric field averaged over four spacecraft. (**c**) Radius of curvature $R_c$ normalized to the electron (red) and ion (blue) gyro-radius. (**d**) Electron bulk flow velocity and (**e**) MFT velocity $\mathbf{U}_\psi$, where the ion bulk flow velocity is subtracted. Dotted lines denote the upstream Alfvén speed. (**f**) $\nabla \cdot \mathbf{U}_\psi$ normalized to $f_{ce}$. (**g**) Sketch of the trajectory of MMS and expected MFT inflows and outflow, adapted from Eriksson et al. (2018).

### 3.1.6 Electron Diffusion Region Detection with Machine Learning methods

Machine learning methods recently became a useful tool for space physics data analysis and were employed for a variety of tasks, including classification, event detection, and prediction (sections A.3-A.5). In the following, we present some recent developments which apply this approach to the detection of EDRs. This type of event is "rare": Webster et al. (2018) reported 32 events, Lenouvel et al. (2021) identified 18 new events, while Fuselier et al. (2016) estimated 56 events for the first 2.5 years of the nominal MMS mission. Therefore, this, at first glance, does not argue for the use of machine learning for EDR detection. However, by using special features of the MMS measurements, namely the details of the electron distribution function, it is possible to extract valuable information which can be processed by machine learning algorithms.

The first algorithm is detailed in Lenouvel et al. (2021) and is only summarized here. It is a rather classical feed-forward MultiLayer Perceptron (e.g., Rumelhart et al. 1986), i.e., a neural network with multiple layers of neurons connected to each other, using MMS observations as features and 4 classes as outputs. Notably one of the key features is a scalar parameter specifically characterizing the asymmetry observed in the crescent-shaped electron distribution functions (hereafter referred to as ”electron crescents” or ”crescents” for notational simplicity) on the dayside magnetopause (Hesse et al. 2014; Bessho et al. 2016; see also section 3.3.1). The main drawback of this early model was the large number of false positives which needed to be visually analyzed to extract the best EDR candidates. It nevertheless enabled an increase in the number of EDR events and eventually led to a statistical analysis of the sign of the energy conversion rate $\mathbf{j} \cdot \mathbf{E}'$, and a discussion on the distinction between inner and outer EDRs (Lenouvel et al. 2021).

The second algorithm (Lenouvel 2022) has a different structure and is based on a Convolutional Neural Network (CNN) (Lecun et al. 1998), i.e., a deep learning algorithm specifically adapted to image recognition and classification. The idea of the architecture builds on the characteristic feature of electron crescents seen in the full distribution functions rather than reducing them to a scalar as in the first algorithm. The training set is based on all 50 events described in Webster et al. (2018) and a total of 214 crescents yielded by Lenouvel et al. (2021) using data from the four MMS spacecraft. The distribution functions are transformed into 32 by 32 pixel images where each pixel holds the value of the electron phase space density (PSD) for a given angle ($\theta = \arctan(v_{\perp 2}/v_{\perp 1})$) and energy ranges (see Figure 17 for details). The range of PSD on log scale (min-max for all images) is coded on 256 levels. Data augmentation is then used to increase the number of training samples, by first extracting 112 most clear crescents (exhibiting a clear left-right asymmetry in the $v_{\perp 1} - v_{\perp 2}$ plane) and then combining them with each other (by averaging two images after random small rotation and adding logarithmic noise), which produces a dataset of ${}_{112}C_2 = 112\,!/(110\,!\,2\,!) = 6126$ new synthetic crescents from all possible combinations. The CNN architecture is formed by a succession of dedicated layers which aim at extracting features or patterns in input data. Finally, applying this algorithm on the full MMS phase 1b of the prime mission resulted in the discovery of 17 new events (from the analysis of MMS 1 and 2 data only). For future studies, a list of events combining those obtained from both algorithms, with the addition of individual events analyzed in the literature during phases 1a and 1b, is available at Zenodo (https://zenodo.org/record/8319481).

To demonstrate how the detection works and performs, a model was trained after removing the distribution functions (736 distribution functions including real crescents, synthetic crescents and random distribution functions) from the EDR event reported by Burch et al. (2016b) that took place on 16 October 2015 at 13:07:02 UTC. The model could then be applied to the whole event (from 13:05:25 UTC to 13:07:44 UTC) without any bias that could be due to data leakage between the training dataset and the data from this example. In Figure 17, the red vertical lines correspond to times of the distribution functions labeled as "crescents" by the model; 26 are visible on the plot and, after visual inspection, 18 of them were considered as correctly identified by the algorithm (true positives).

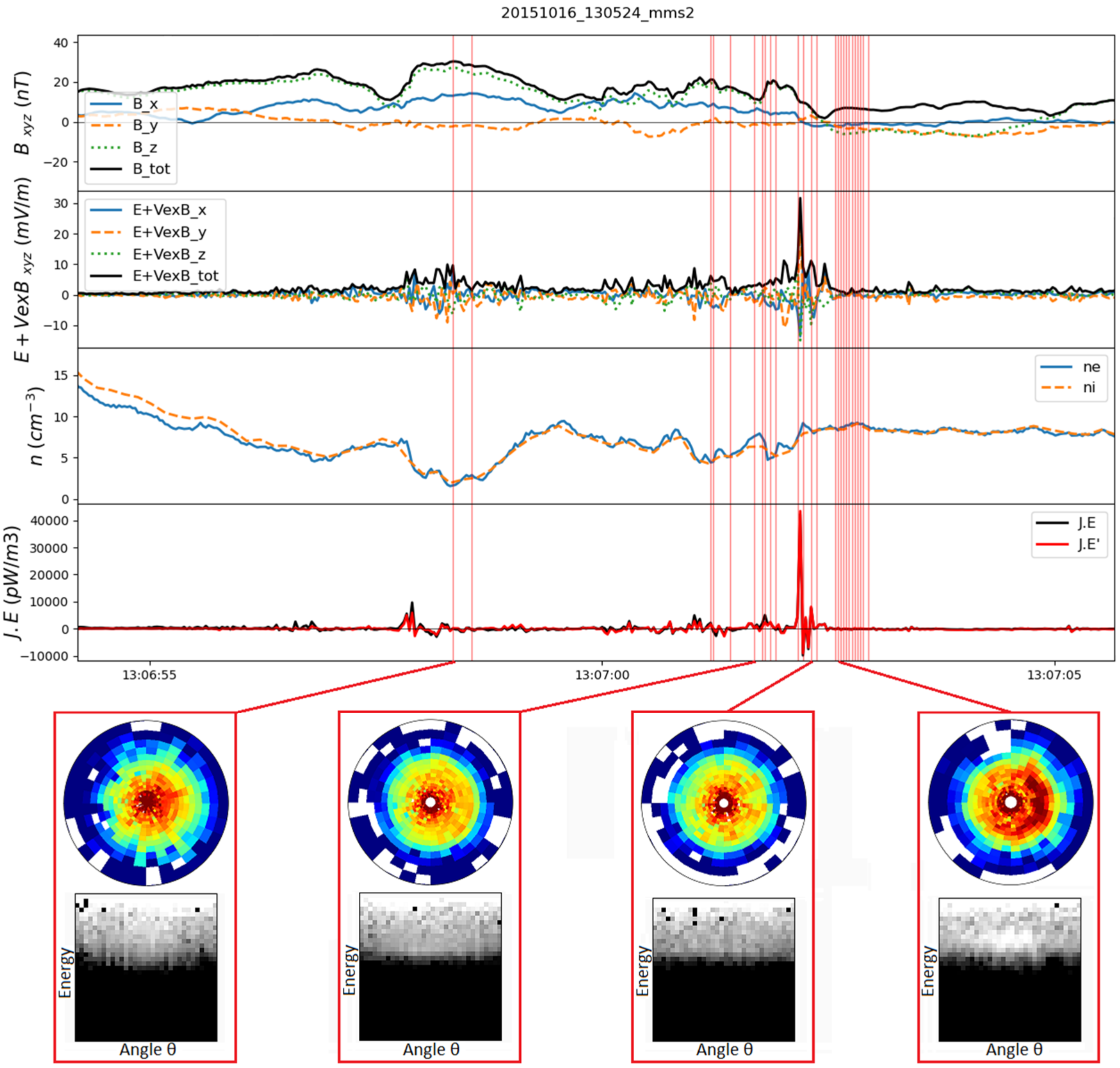

**Fig. 17** EDR event from Burch et al. (2016b) with detected crescent distribution functions. The top four panels show MMS2 observations of the magnetic field **B**, electron-frame electric field $\mathbf{E}'$ (showing the departure from the ideal conditions), electron and ion densities, and the energy conversion rate $\mathbf{j}\cdot\mathbf{E}'$. Four of the identified "crescents" are shown below the time series plots, both in classical phase space density units (top) and in transformed 32×32 images (bottom). The two-dimensional electron distribution slices are displayed in the $v_{\perp 1} - v_{\perp 2}$ plane where $v_{\perp 1}$ is directed along $(-\mathbf{u}_e \times \mathbf{B}) \times \mathbf{B}$ (approximately the $\mathbf{E}\times\mathbf{B}$ direction, where **E** is the electric field, and $\mathbf{u}_e$ is the electron bulk velocity), and $v_{\perp 2}$ is directed along $-\mathbf{u}_e \times \mathbf{B}$ (approximately the direction of **E**).

In this application, only the distribution function is used to make a prediction and no post-processing is applied, so false positives are to be expected. Removing these false positives would require the use of additional parameters to be associated with the distribution functions, such as the electron density (low densities are known to produce incomplete distribution functions that are interpreted as asymmetric by the model) or other key EDR parameters including the electron-frame electric field **E**' and $\mathbf{j} \cdot \mathbf{E}'$. One needs to bear in mind that peaks in the last two parameters may not occur at the same time as the presence of electron crescents (the separation time can go up to a few hundred milliseconds), so the automatic removal of false detections is not an easy matter and will be the topic of future work with a more advanced automatic EDR detection model. Nonetheless, the ability of these automatic EDR detection methods shows that they are valuable to identify and analyze rare and complex physical plasma processes.

### 3.1.7 Magnetic Nulls

Magnetic nulls are singularities or critical points where the magnetic field vanishes, and can be essential for characterizing 3D magnetic topology and understanding magnetic reconnection in 3D (Pontin and Priest 2022 and references therein). See recent works (e.g., Fu et al. 2015; Olshevsky et al. 2020; Guo et al. 2022; Ekawati and Cai 2023) for details about the methods to identify and analyze the magnetic nulls.

## 3.2 Analysis Methods for the Electron Diffusion Region

### 3.2.1 Estimation of Anomalous Resistivity, Viscosity, and Diffusion

In Graham et al. (2022) anomalous terms associated with lower hybrid waves were estimated from direct spacecraft observations in magnetopause reconnection events, and an assessment was performed to see whether the anomalous terms could contribute to the reconnection electric field $E_M$. The anomalous terms are based on expansions of the electron continuity and momentum equations:

$$\frac{\partial n}{\partial t} + \nabla \cdot (n\mathbf{u}) = 0, \qquad (40)$$

$$m\frac{\partial (n\mathbf{u})}{\partial t} + m\nabla \cdot (n\mathbf{u}\mathbf{u}) + \nabla \cdot \mathbf{P} + en(\mathbf{E} + \mathbf{u} \times \mathbf{B}) = 0, \qquad (41)$$

where $m$ is the electron mass, $e$ is the unit charge, $n$ is the number density, **u** is the bulk velocity, **P** is the pressure tensor, **E** is the electric field, and **B** is the magnetic field. To derive the anomalous terms the quantities in these equations are separated into fluctuating and quasi-stationary components: $Q = \delta Q + \langle Q \rangle$, where $\delta Q$ is the fluctuating component due to waves, and $\langle Q \rangle$ is an ensemble average over Q, and $\langle \delta Q \rangle = 0$. The anomalous terms are obtained by taking the ensemble average of the momentum equation. The ensemble average of the product of two quantities is $\langle QR \rangle = \langle Q \rangle \langle R \rangle + \langle \delta Q \delta R \rangle$. From the continuity equation (40) a cross-field diffusion coefficient can be defined as

$$D_{\perp} = -\frac{\langle \delta n \delta u_N \rangle}{\nabla \langle n \rangle_N}, \quad (42)$$

where **N** is the direction normal to the local boundary. From the momentum equation (41) we obtain:

$$\langle \mathbf{E} \rangle + \langle \mathbf{u} \rangle \times \langle \mathbf{B} \rangle = -\frac{\nabla \cdot \langle \mathbf{P} \rangle}{\langle n \rangle e} - \frac{m}{\langle n \rangle e} \nabla \cdot (\langle n \rangle \langle \mathbf{u} \rangle \langle \mathbf{u} \rangle) + \mathbf{D} + \mathbf{T} + \mathbf{I}.$$

(43)

The time derivative term in Eq. (41) is assumed to be small, so it is neglected in Eq. (43). Here **D**, **T**, and **I** are the anomalous resistivity, anomalous viscosity, and anomalous inertial terms, which are given by

$$\mathbf{D} = -\frac{\langle \delta n \delta \mathbf{E} \rangle}{\langle n \rangle}, \quad (44)$$

$$\mathbf{T} = -\frac{\langle n \mathbf{u} \times \boldsymbol{B} \rangle}{\langle n \rangle} + \langle \mathbf{u} \rangle \times \langle \mathbf{B} \rangle, \quad (45)$$

$$\mathbf{I} = -\frac{m}{e \langle n \rangle} [\nabla \cdot (\langle n \mathbf{u} \mathbf{u} \rangle) - \nabla \cdot (\langle n \rangle \langle \mathbf{u} \rangle \langle \mathbf{u} \rangle)]. \quad (46)$$

This approach to calculate the anomalous terms corresponds to Reynolds averaging, which is often used to study fluid turbulence. The above terms were derived in Graham et al. (2022). Similar definitions are used in numerical simulations (Che et al. 2011 ; Price et al. 2016; Le et al. 2018; Price et al. 2020), although often the particle fluxes are treated as a single quantity, which can modify the contributions from the anomalous terms (Price et al. 2020).

The anomalous terms result from the correlations between fluctuating quantities associated with the waves. In numerical simulations the average is taken over the reconnection out-of-plane (*M*) direction (e.g., Price et al. 2016, 2020; Le et al. 2017), although the ensemble average can be performed over time (Le et al. 2018). With MMS an approximate average can also be obtained by averaging over the four spacecraft.

To calculate the anomalous contributions requires fields and particle measurements that can resolve the lower hybrid wave fluctuations. This is possible for the electron particle data using the highest time resolution electron distributions and moments, which can be sampled every 7.5 ms, rather than the nominal 30 ms sampling rate during burst mode (Pollock et al. 2016). Resolution of 7.5 ms is achieved by reducing the azimuthal resolution in the spacecraft spin plane of the electron measurements (Rager et al. 2018; Appendix A). Since we are interested in the bulk changes in the distributions rather than fine structures in computing the lower order moments, this reduced azimuthal resolution does not present a major problem.

The anomalous terms were calculated in Graham et al. (2022) and Figure 18 shows the steps to calculate the *M* component of **D**. We summarize the steps they used to calculate the anomalous terms:

(1) The vector quantities are converted to the LMN coordinate system.

(2) All electric and magnetic field data are resampled to the sampling frequency of the 7.5 ms electron moments.

(3) Four spacecraft timing analysis on $B_L$ at the current sheet to determine the velocity of the boundary in the normal direction and the time delays between the spacecraft.

(4) The time delays are used to offset the spacecraft times so all spacecraft cross the current sheet at the same time as MMS1.

(5) Quasi-stationary quantities $\langle Q \rangle$ are obtained by bandpass filtering the signals below 5 Hz, and averaging the data over the four spacecraft. For the magnetopause, as shown in Figure 18, the frequency of the lower hybrid waves is typically comparable to or above 10 Hz, so this removes most of the fluctuations due to these waves.

(6) The fluctuating quantities $\delta Q$ associated with lower hybrid waves are obtained by band pass filtering the data above 5 Hz (Figures 18b,c).

(7) The correlations between fluctuating quantities $\langle \delta Q \delta R \rangle$ are obtained by averaging $\delta Q \delta R$ over the four spacecraft; then low-pass filtering the product below 5 Hz. This removes most of the remaining higher-frequency components from these terms (Figure 18d). See Graham et al. (2022) for further details on the calculation of the anomalous terms (see also Table 5 in Appendix C).

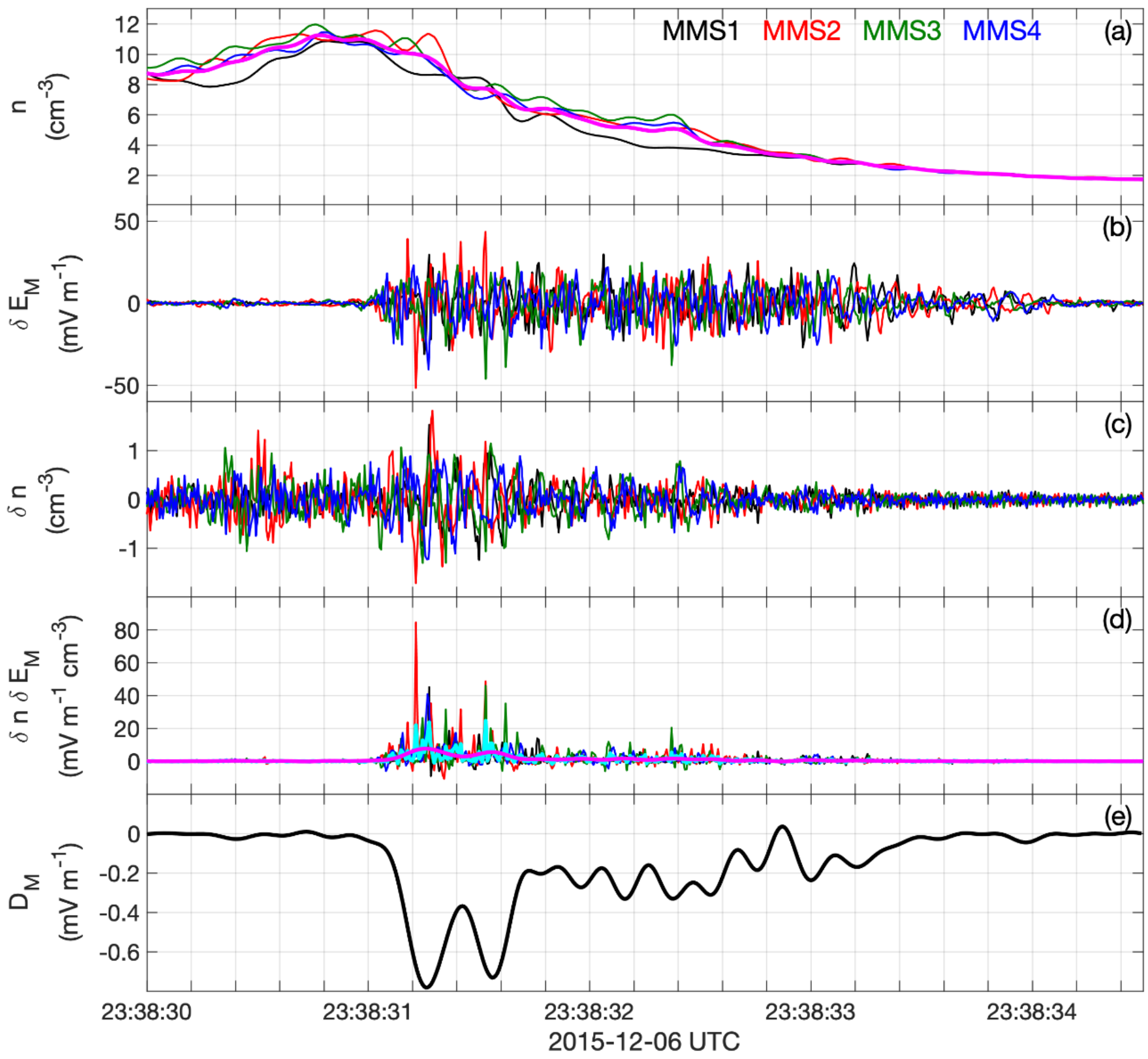

**Fig. 18** Example of the calculation of the $M$ component of the anomalous resistivity **D**. (**a**) Background (lowpass filtered) components of the electron number density $n$ (MMS1-MMS4 data are plotted in black, red, green, and blue, respectively) and $\langle n \rangle$ calculated by averaging over the four spacecraft (magenta line). (**b**) and (**c**) $\delta E_M$ and $\delta n$ for the four spacecraft. (**d**) $\delta n \delta E_M$ calculated for the four spacecraft, and the four-spacecraft averaged $\delta n \delta E_M$ (cyan), and $\langle \delta n \delta E_M \rangle$ (magenta) obtained by low-pass filtering the averaged $\delta n \delta E_M$. I $M$ component of **D** calculated from $\langle \delta n \delta E_M \rangle$ and $\langle n \rangle$. In all panels (**a**)-(**d**) the quantities from MMS2-MMS4 have been time shifted so they cross the current sheet at the same time.

### 3.2.2 Evaluation of Terms in the Electron Vlasov Equation

A wide variety of collisionless plasma phenomena have been studied and understood by utilizing the kinetic description provided by the Vlasov equation (e.g., Nicholson 1983; Califano et al. 2016; Gershman et al. 2017). Here, we summarize a methodology for utilizing the MMS Fast Plasma Investigation (FPI) Dual Electron Spectrometer (DES) data (Pollock et al. 2016) in order to compute each derivative of the electron PSD $f_e$ that appears in the electron Vlasov equation, given by

$$\frac{df_e}{dt} = \frac{\partial f_e}{\partial t} + \mathbf{v} \cdot \nabla f_e - \frac{e}{m_e}(\mathbf{E} + \mathbf{v} \times \mathbf{B}) \cdot \nabla_{\mathbf{v}} f_e = 0. \quad (47)$$

For example, since the generalized Ohm's law (electron momentum equation (41)) is derived from the moments of the Vlasov equation (47), the method, as explained below, allows us to discuss how and which part of the velocity-space distribution of each term of Eq. (47) contributes to each term of Eq. (41).

Figure 19 demonstrates the computation techniques needed for each Vlasov equation term in the context of an electron spatial-scale current sheet encountered by MMS on 23 December 2016 (Shuster et al. 2019, 2021a,b). This event is discussed in more detail in Chapter 1.3 of this issue (Norgren et al. 2023). The Vlasov equation $df_{\mathrm{e}}/dt = 0$ is a statement indicating that PSD is conserved along a particle's Lagrangian trajectory through phase space. In the Eulerian frame of the MMS spacecraft, it is necessary to consider and evaluate each term of the Vlasov equation (47) (see Shuster et al. (2019, 2023) for a more thorough discussion concerning the computation methods outlined here). Figures 19a-d, Figures 19e-h, and Figures 19i-m present a high-level, visual comparison of the three distinct methods for evaluating the terms $\partial f_{\mathrm{e}}/\partial t$, $\mathbf{v}\cdot\nabla f_{\mathrm{e}}$, and $-(e/m_{\mathrm{e}})(\mathbf{E} + \mathbf{v} \times \mathbf{B})\cdot\nabla_{\mathrm{v}} f_{\mathrm{e}}$, respectively.

For the event shown in Figure 19, the current layer's thickness was about 3 to 5 $d_{\mathrm{e}}$, where the local electron skin depth $d_{\mathrm{e}}$ was about 1.5 km. The normal velocity of the structure, $V_N$, was roughly 50 km/s, where $N$ indicates the direction normal to the current layer. The current layer passed by each MMS spacecraft in about a tenth of a second. Thus, the spatial thickness was roughly (50 km/s)·(0.1 s) = 5 km, comparable to the inter-spacecraft spacing of the four MMS spacecraft.

Shuster et al. (2023) explain how to use higher order finite difference approximations to obtain more accurate measures of the temporal and velocity-space derivative terms. Furthermore, Shuster et al. (2023) show how $\partial f_{\mathrm{e}}/\partial t$ may provide a useful estimate for $\nabla f_e$ in situations where the plasma is believed to be quasi-steady state:

$$\frac{Df_{\mathrm{e}}}{Dt} \equiv \frac{\partial f_{\mathrm{e}}}{\partial t} + \mathbf{V}_{\mathrm{str}} \cdot \nabla f_{\mathrm{e}} \approx 0 \quad \rightarrow \quad \frac{\partial f_{\mathrm{e}}}{\partial N} \approx \left(-\frac{1}{V_N}\right)\frac{\partial f_{\mathrm{e}}}{\partial t}, \quad (48)$$

where $\partial f_{\mathrm{e}}/\partial t$ is computed in the frame of the spacecraft. The temporal derivative notation $Df_{\mathrm{e}}/Dt$ is used to indicate a time derivative taken in a frame moving in position space with the velocity of the structure, $\mathbf{V}_{\mathrm{str}} = V_N\hat{\mathbf{e}}_N$. Here the structure is assumed to be planar, with spatial variations only in the $N$ direction (Figure 19). As noted by Shuster et al. (2019), we point out the connection between the spatial gradient term $\nabla f_{\mathrm{e}}$ and the bulk electron pressure divergence term $\nabla \cdot \mathbf{P}_{\mathrm{e}}$ via the integral identity utilized when deriving the electron momentum equation from the electron Vlasov equation (47):

$$m_{\mathrm{e}} \int \mathbf{v}(\mathbf{v} \cdot \nabla f_{\mathrm{e}}) d^3v = \nabla \cdot \mathbf{P}_{\mathrm{e}} + \nabla \cdot (m_{\mathrm{e}} n_{\mathrm{e}} \mathbf{u}_{\mathrm{e}} \mathbf{u}_{\mathrm{e}}). \quad (49)$$

For certain environments, such as magnetopause reconnection sites, the inertial term on the right-hand side of Eq. (49) is commonly negligible compared to $\nabla \cdot \mathbf{P}_e$. Thus, we can understand how velocity-space structures of each term of the Vlasov equation (47), as shown in Figure 19, contribute to collisionless plasma processes including the generation of a nonideal electric field (here, through the interrelationship among $\mathbf{v} \cdot \nabla f_e$, $\nabla \cdot \mathbf{P}_e$, and the electric field, based on Eq. (49) and the generalized Ohm's law) and field-to-electron energy conversion, which are key features of the EDR.

Regarding the velocity-space gradient term on the righthand side of Eq. (47), one may perform the derivative computations in the $\{E, \theta, \phi\}$ coordinates native to the FPI detectors, or one may choose to first interpolate the distribution to a Cartesian grid with $\{v_x, v_y, v_z\}$ coordinates before calculating the derivatives (see Shuster et al. (2023) for more details). As a consistency check, the results of each approach ought to be qualitatively consistent. We note also that the electric and magnetic fields may be averaged to the DES 30 ms cadence to obtain a measurement of the full velocity-space gradient term that appears in the Vlasov equation.

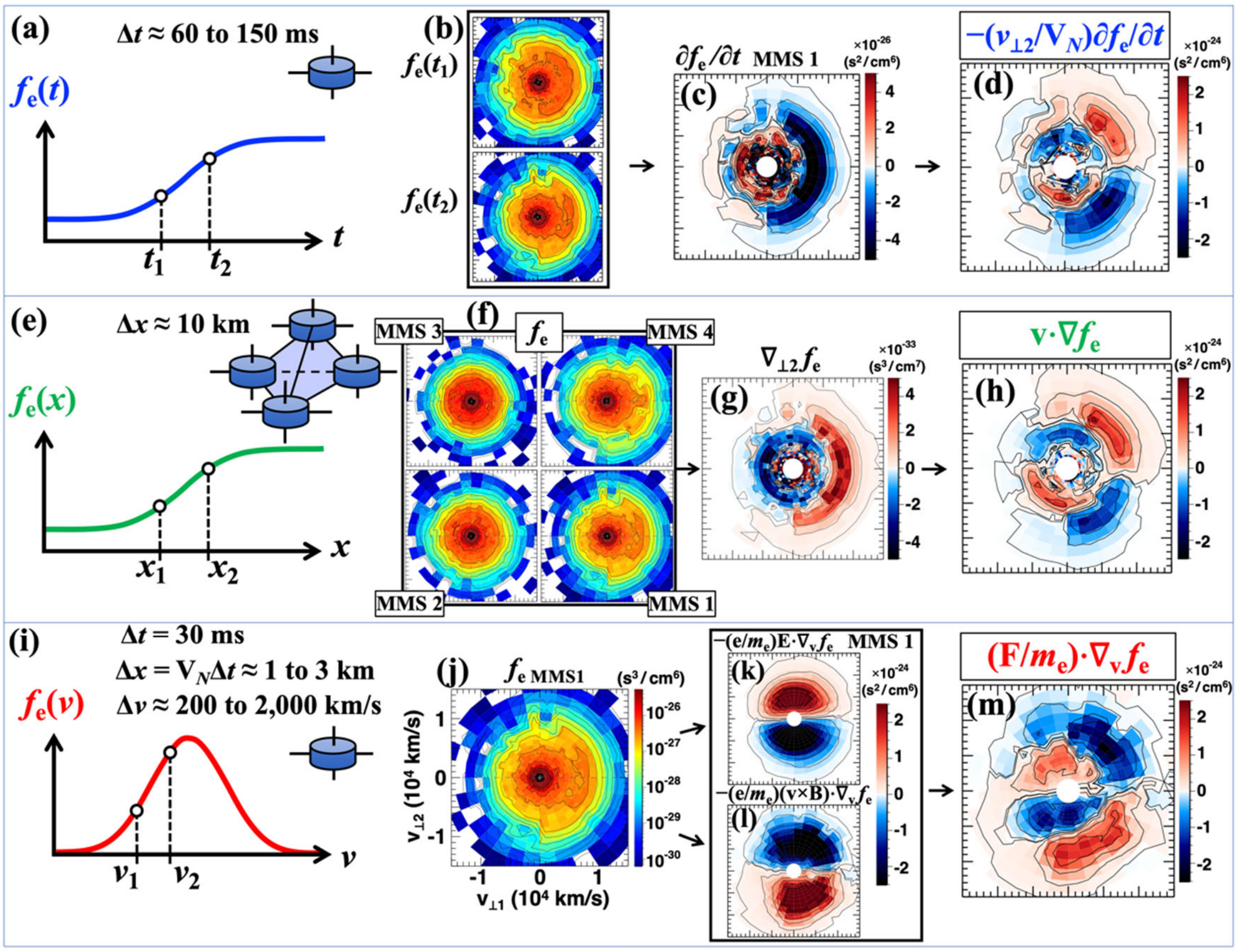

**Fig. 19** This figure presents a visual summary of the velocity-space structure and qualitative balance of the three Vlasov equation terms for an electron-scale current sheet. The computation

technique for (**a-d**) $\partial f_e/\partial t$, (**e-h**) $\mathbf{v}\cdot\nabla f_e$, and (**i-m**) $(\mathbf{F}/m_e)\cdot\nabla_v f_e$ is shown schematically. Each velocity-space panel represents a slice taken in the $v_{\perp 1}$-$v_{\perp 2}$ plane. In this example, $\hat{\mathbf{e}}_N$ points roughly along $\hat{\mathbf{e}}_{\perp 2}$, so that the quantities shown in panels (d) and (h) are roughly equivalent. Comparing panels (h) and (m), one can see a notable quadrupolar pattern but with a polarity difference, suggesting that Eq. (47) is roughly satisfied in velocity space under the assumption of negligible $\partial f_e/\partial t$ (panel (c)). Adapted from Shuster et al. (2023).

### 3.2.3 Non-Maxwellianity

In many plasmas, particle distributions can deviate significantly from thermal equilibrium, namely a Maxwellian distribution. Non-Maxwellian distributions develop during magnetic reconnection and can be unstable to a range of instabilities. Several scalar parameters have been defined to quantify the deviation of the observed distributions from a Maxwellian or bi-Maxwellian distribution function (Greco et al. 2012; Servidio et al. 2017; Liang et al. 2020; Graham et al. 2021; Argall et al. 2022; see also section 3.2.4 for other versions of non-Maxwellianity not mentioned in this section). Here we outline the one developed in Graham et al. (2021). In Greco et al. (2012), Servidio et al. (2017), and Graham et al. (2021) the definitions of non-Maxwelliantity are based on the magnitude of the differences between the observed distribution and a model Maxwellian distribution. In Liang et al. (2020) and Argall et al. (2022) the definition of non-Maxwellianity is based on the increase in kinetic entropy from a Maxwellian distribution. In Graham et al. (2021) the non-Maxwellianity parameter was defined as

$$\epsilon = \frac{1}{2n}\int_{v,\theta,\phi} |f(v,\theta,\phi) - f_{model}(v,\theta,\phi)| v^2 sin\theta dv d\theta d\phi, \qquad (50)$$

where $n$ is the number density, $f$ is the observed particle distribution function, $f_{model}$ is the model particle distribution function, $v$ is the speed, $\theta$ is the polar angle, and $\phi$ is the azimuthal angle in velocity space. The model distribution can be either a Maxwellian or bi-Maxwellian distribution with the same density, bulk velocity, and temperature as the observed distribution. The integral is performed in the same way as the particle moments calculations. The $1/(2n)$ factor normalizes ε to a dimensionless quantity with values between 0 and 1. A value of 0 indicates no deviation from the model distribution, while 1 corresponds to complete deviation from the model distribution. In Graham et al. (2021) a bi-Maxwellian distribution was used as the model distribution, given by:

$$f_{model}(\boldsymbol{v}) = \frac{n}{\pi^{3/2} v_{th,\parallel}^3} \frac{T_\parallel}{T_\perp} exp\left(-\frac{(v_\parallel - V_\parallel)^2}{v_{th,\parallel}^2} - \frac{(v_{\perp,1} - u_\perp)^2 + v_{\perp,2}^2}{v_{th,\parallel}^2 (T_\perp/T_\parallel)}\right), \qquad (51)$$

where $T_\parallel$ and $T_\perp$ are the parallel and perpendicular temperatures, $v_{th,\parallel} = \sqrt{2k_B T_\parallel/m}$ is the thermal speed, $k_B$ is Boltzman's constant, $m$ is the particle mass, and $u_\perp$ is the magnitude of the perpendicular bulk velocity. The velocity coordinates are defined such that $v_{\|}$ is aligned with the magnetic field, $v_{\perp,1}$ is aligned with the component of the bulk velocity perpendicular to the magnetic field, and $v_{\perp,2}$ is orthogonal to $v_{\|}$ and $v_{\perp,1}$. The calculation of ε corresponds to a zeroth order moment calculation, so the largest contributions to ε typically occur in the thermal energy range. The parameters used to calculate $f_{model}$ are obtained from the observed particle moments, so no fitting to the observed distribution is required. In Graham et al., the bi-Maxwellian distribution function was used rather than an isotropic Maxwellian so temperature anisotropies, which are simple to identify in the particle moments, are not the cause of non-Maxwellianity.

Figure 20 shows an example of ε calculated for an EDR observed on 22 October 2015. The figure shows a reconnection event, as indicated by the reversal in the direction of the magnetic field (panel a), an increase in density (panel b), and a northward ion outflow (panel c).

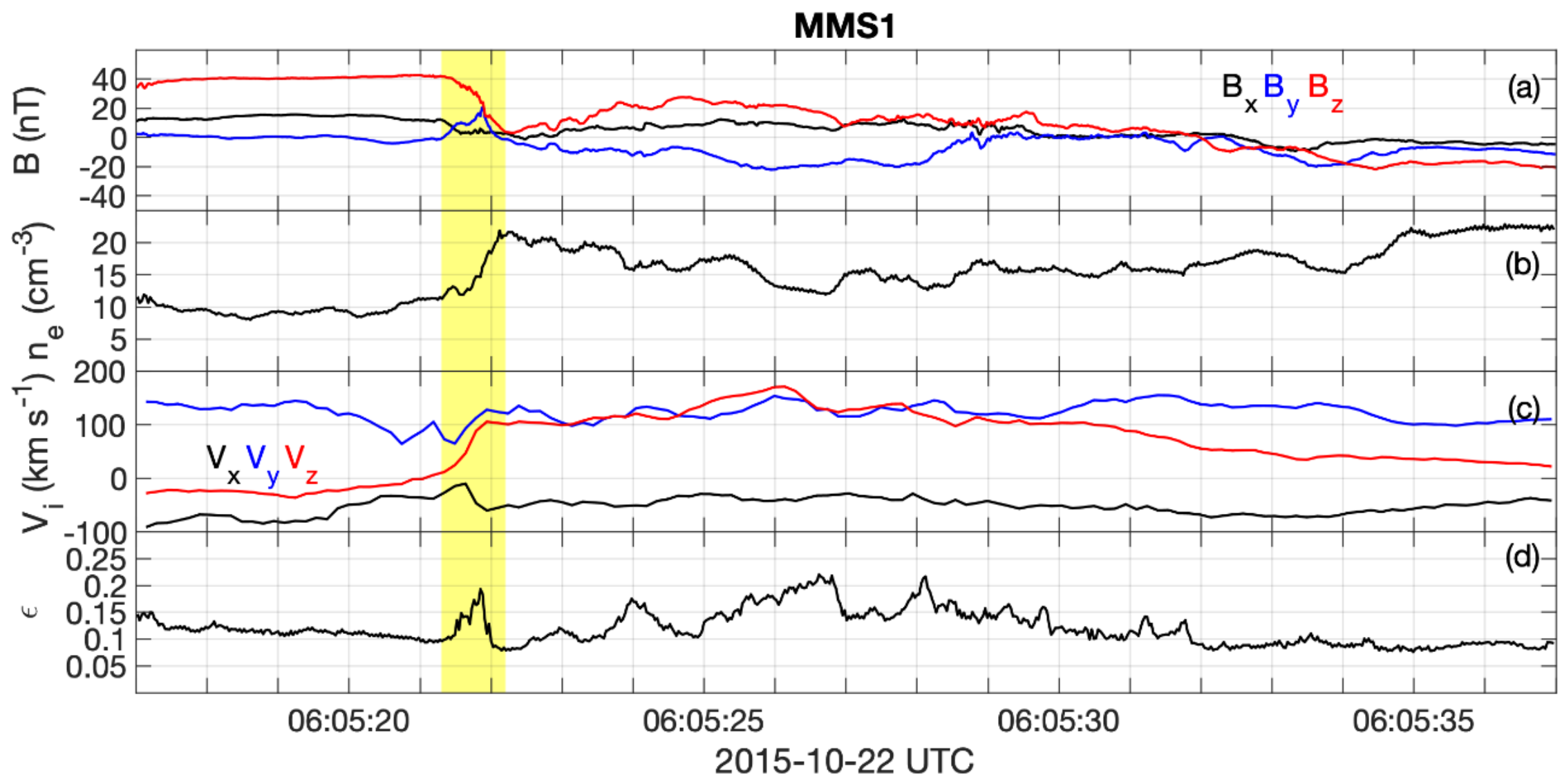

**Fig. 20** Example of electron non-Maxwellianity for the electron diffusion region crossing observed by MMS1 on 22 October 2015. (**a**) Magnetic field. (**b**) electron number density. (**c**) Ion bulk velocity. (**d**) Non-Maxwellianity ε of electrons. The EDR is indicated by the yellow-shaded region.

The EDR was observed at the time indicated by the yellow-shaded region (e.g., Phan et al. 2016; Toledo-Redondo et al. 2016). Figure 20d shows ε calculated from the above equations. There is a significant increase in ε, which peaks in the EDR, indicating that non-Maxwellian electron distributions develop there. Additionally, large ε occurs in the outflow region, indicating that non-Maxwellian electron distributions are not limited to the EDR. More generally, all the observed EDRs in the first phase of the MMS mission exhibited enhanced electron non-Maxwellianities (Graham et al. 2021). In Graham et al. (2021) a statistical analysis of electron distributions was performed on six months of data to determine which values of ε corresponded to enhanced non-Maxwellian electron distributions compared to typical values.

It should be noted that ε can be artificially large due to the low counting statistics from the particle detectors when the density is low, such as in Earth's magnetotail. This problem can be mitigated by averaging multiple particle distributions in such cases.

### 3.2.4 Kinetic Entropy

Due to the fact that collisions in space plasmas are weak, processes that involve kinetic physics often distort the distribution function so that it is no longer in equilibrium. Structures in the distribution function are linked to specific energy conversion processes (Egedal et al. 2010a, 2010b; Hoshino et al. 2001) and are used to create maps of the reconnection diffusion region (Chen et al. 2008, 2016) and study the spatio-temporal evolution of reconnection (Shuster et al. 2014, 2015; Egedal et al. 2016; Barbhuiya et al. 2022). Inhomogeneities in the distribution were quantified to provide indicators of when kinetic physics is important (Scudder and Daughton 2008; Aunai et al. 2013; Swisdak 2016; Section 3.1.2). The measured distribution was compared to an equivalent distribution in equilibrium to quantify the amount of free energy available to be dissipated (Greco et al. 2012; Servidio et al. 2017; Graham et al. 2021; Lindberg et al. 2022; section 3.2.3). Similar considerations were made with respect to kinetic entropy (Boltzmann 1877), noting that the measured distribution will be non-Maxwellian if its entropy is less than that of an equilibrium Maxwellian distribution with the same density and effective temperature (Kaufmann and Paterson 2009; Liang et al. 2019, 2020). The theoretical development of non-Maxwellianity (Liang et al. 2020) was adapted to the non-uniform velocity space grid of MMS plasma measurements, and then was applied to a magnetotail reconnection event to show that non-Maxwellianity is indeed linked to kinetic processes in the EDR (Argall et al. 2022).

To develop a theory of kinetic entropy for a distribution function in velocity space, we start with Boltzmann's equation $S = k_B \ln \Omega$, where $k_B$ is Boltzmann's constant and $\Omega = N_{tot}!/\prod_{j,k} N_{jk}!$ is the total number of microstates that correspond to a given macrostate, $N_{tot}$ is the total number of particles in the system and is assumed constant, $N_{jk}$ is the number of particles in the *j*th position-space and *k*th velocity-space cell of phase space, and the product over $j$ and $k$ is over all position- and velocity-space cells, respectively. We then break phase space up into discrete bins and separate the total entropy into position and velocity space entropy (Mouhot and Villani 2011) to obtain a value for entropy that is local in position space and hence is measurable by a single spacecraft (Liang et al. 2019),

$$S = S_r + S_V, \tag{52}$$

where

$$S_r = k_B \left\{ N_{tot} \ln\left(\frac{N_{tot}}{\Delta^3 r}\right) - \int d^3 r\, n(\mathbf{r})\, \ln[n(\mathbf{r})] \right\}, \tag{53}$$

$$S_V = \int d^3 r\, s_V(\mathbf{r}), \tag{54}$$

$$s_V(\mathbf{r}) = k_B \left\{ n(\mathbf{r}) \ln\left[\frac{n(\mathbf{r})}{\Delta^3 v}\right] - \int d^3 v\, f(\mathbf{r},\mathbf{v}) \ln[f(\mathbf{r},\mathbf{v})] \right\}. \tag{55}$$

Here, $\Delta^3 r$ and $\Delta^3 v$ are the position and velocity space volume elements of phase space, $f(\mathbf{r},\mathbf{v})$ is the distribution function, and $n(\mathbf{r})$ is the number density. Eq. (55) is the velocity-space entropy density and its last term is often referred to as the total kinetic entropy density, $s = -k_B \int d^3 v\, f(\mathbf{r},\mathbf{v}) \ln[f(\mathbf{r},\mathbf{v})]$ (Kaufmann and Paterson 2009).

The entropy of a measured distribution can be compared with that of an equivalent drifting Maxwellian distribution described by (omitting the dependence on $\mathbf{r}$)

$$f_M(\mathbf{v}) = n\left(\frac{m}{2\pi k_B T}\right)^{3/2} e^{[-m(\mathbf{v}-\mathbf{u})^2/(2k_B T)]}, \tag{56}$$

where $m$ is the particle mass, $\mathbf{u}$ is the bulk flow velocity, and $T$ is the effective temperature. Combining Eq. (56) with Eq. (55) to calculate the total and velocity-space entropy densities of a Maxwellian distribution ($s_M$ and $s_{M,V}$, respectively) yields

$$s_M = \frac{3}{2} k_B n \left[1 + \ln\left(\frac{2\pi k_B T}{m n^{3/2}}\right)\right], \tag{57}$$

$$s_{M,V} = \frac{3}{2} k_B n \left[1 + \ln\left(\frac{2\pi k_B T}{m(\Delta^3 v)^{3/2}}\right)\right]. \tag{58}$$

The difference between the equivalent Maxwellian and measured total and velocity-space entropy densities defines two non-Maxwellianity parameters

$$\bar{M}_{KP} = \frac{s_M - s}{(3/2)k_B T}, \tag{59}$$

$$\bar{M} = \frac{s_{M,V} - s_V}{s_{M,V}}, \tag{60}$$

where the Kaufmann and Paterson non-Maxwellianity, $\bar{M}_{KP}$, is normalized by the internal energy per particle of an ideal gas (Kaufmann and Paterson 2009), and $\bar{M}$ is normalized by the velocity space Maxwellian entropy density (Liang et al. 2020). While $\bar{M}$ is bounded between 0 and 1 and so provides a better measure for making comparisons of non-Maxwellianity between distributions, the normalization of $\bar{M}_{KP}$ better relates entropy to the energetics of the system (Cassak et al. 2023).

The development so far has considered uniform velocity space grids typical of theory and simulations. Particle detectors, on the other hand, have logarithmically spaced energy bins. Because of this, the velocity space volume element depends on velocity: $d^3v(\mathbf{v})$. Revisiting the above derivations in spherical velocity space coordinates leads to (Argall et al. 2022)

$$s_V = s + k_B n \ln n - k_B \int d^3v(\mathbf{v}) f(\mathbf{v}) \ln[d^3v(\mathbf{v})], \tag{61}$$

$$\bar{M} = \frac{s_M - s - k_B \int d^3v(\mathbf{v}) \ln[d^3v(\mathbf{v})][f_M(\mathbf{v}) - f(\mathbf{v})]}{1 + k_B n \ln n - k_B \int d^3v(\mathbf{v}) \ln[d^3v(\mathbf{v})] f_M(\mathbf{v})}. \tag{62}$$

A number of corrections are required when integrating the measured particle distribution functions. These include corrections for photoelectrons and spacecraft potential, and normalization of the energy and look-angle space of the instrument (FPI in the case of MMS). The subsequent spherical, normalized energy space is obtained by applying the transformations outlined in Moseev et al. (2019), and Argall et al. (2022) and its Supplemental Material. In addition, to calculate the equivalent Maxwellian distribution, it is essential to create a look-up table to minimize numerical errors (Argall et al. 2022).

Figure 21 shows calculations of non-Maxwellianity from MMS observations and PIC simulations within the EDR of a magnetotail reconnection event. The 2D profile of $\bar{M}$ (panel a) shows significant departures from Maxwellianity in the vicinity of the EDR, indicating that important kinetic effects are taking place. The trajectory (magenta dashed line) indicates the path that MMS took through the EDR and the corresponding data is shown in the following panels. Three different measures of non-Maxwellianity, including $\bar{M}$ and $\bar{M}_{KP}$ show qualitative agreement between observations and simulations.

Distribution functions observed by MMS (Figure 21f,g,h) and in PIC (Figure 21k) are taken at the black vertical dashed lines (and the "x" in panel a). PIC distributions (Figure 21i,j) outside the area shown in Figure 21a were taken from regions representative of the MMS locations within the reconnection domain. The electron distribution functions are from the upstream (Figure 21f,i), inflow (Figure 21g,j), and X-line (Figure 21h,k) regions. Upstream, the distributions are nearly Maxwellian and have the lowest non-Maxwellianity of the three regions sampled. The inflow distributions are elongated in the direction parallel to the magnetic field due to their bouncing within a parallel potential well (Egedal et al. 2010a), and X-line distributions are striated due to their meandering current sheet motion (Ng et al. 2011). This shows that entropy measures of non-Maxwellianity are able to identify regions where important kinetic effects are taking place.

Unfortunately, by breaking phase space up into discrete bins (i.e., by considering a distribution of particles instead of combinations of individual particles), we lose information about the system; $\overline{M}$ depends on the scale of the velocity space grids and their non-uniformity and thus is not equal to $\overline{M}_{KP}$. For details about how the difference between $\overline{M}$ and $\overline{M}_{KP}$ allows us to quantify the amount of information loss incurred by discretizing velocity space, see Argall et al. (2022). Implications include considering the thermal velocity of the target environment when designing plasma instruments and quantifying limits to observations of dissipation.

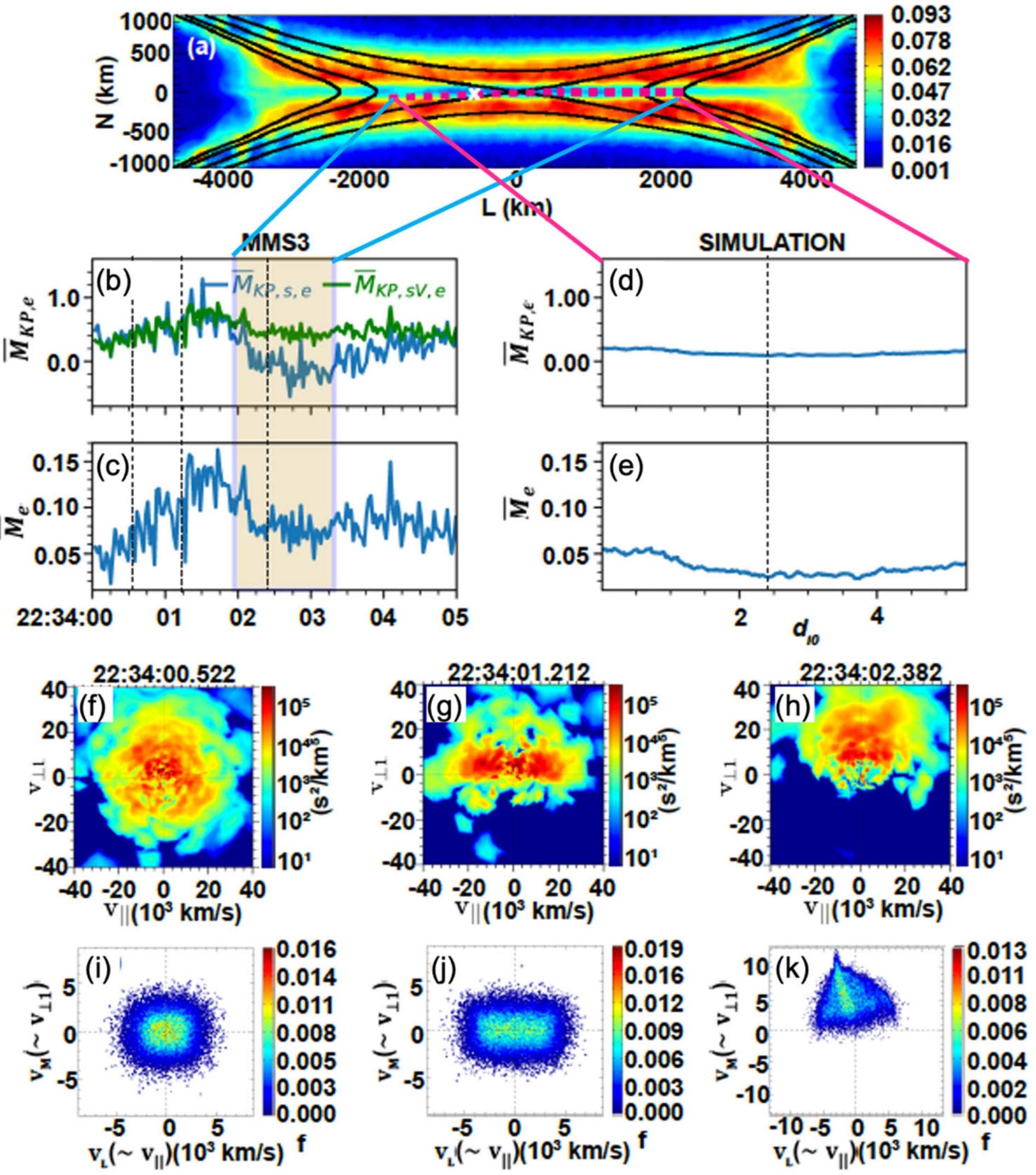

**Fig. 21** MMS and PIC simulation comparisons of non-Maxwellianity in a magnetotail electron diffusion region. (**a**) The 2D profile of non-Maxwellianity surrounding the EDR with the MMS trajectory shown (dotted magenta line). Different electron non-Maxwellianity quantities as observed (**b**,**c**) by MMS and (**d**,**e**) in PIC, where $\overline{M}_{KP}$ (Eq. (59)) is computed using $s$ (blue) and $s_V$ (Eq. 61) (green). The MMS panels show a larger view than is depicted in the simulation for comparison with other published studies.

Vertical dashed lines (and the "x" in panel **a**) indicate where electron distributions were obtained (**f**,**g**,**h**) by MMS and (**k**) in PIC simulation. PIC distributions in (**i**,**j**) were taken outside of the bounds of panel (**a**) from locations representative of the regions MMS sampled. Adapted from Argall et al. (Phys. Plasmas, 29, 022902, 2022; licensed under a Creative Commons Attribution (CC BY) license).

### 3.3 Reconnection Electric Field Estimation

#### 3.3.1 Extracting $E_M$ from Multiple Crescents in Electron Velocity Distributions

In the event on 11 July 2017 in the Earth's magnetotail (Torbert et al. 2018), MMS detected multi-crescent electron velocity distribution functions (VDFs) in the EDR. Multi-crescent VDFs are produced because the reconnection electric field $E_M$ accelerates electrons while they are meandering across the reconnecting current sheet. In the following, we explain how to extract the information about $E_M$ from a multi-crescent VDF.

Suppose electrons are meandering in a current sheet with a magnetic field $B_L=bN$ and a Hall electric field $E_N=-kN$, where $N$ is the position measured from the current sheet center ($B_L=0$) plane, and $b$ and $k$ are the slope of $B_L$ and $E_N$, respectively, in the $N$ direction. Consider an electron moving back and forth in the $N$ direction, starting from $N=0$ at $t=0$. The velocity $v_M$ is given as $v_M= v_{M0}-(eE_M/m_e)t-(eb/2m_e)N^2$, where $v_{M0}$ is the initial velocity. The $N$ motion is an oscillatory motion under $E_N=-kN$ and the magnetic force, and approximately described using Airy functions as $N=c_1\mathrm{Ai}(r)+c_2\mathrm{Bi}(r)$, where $c_1$ and $c_2$ are constants, and $r$ is defined as

$$r = -\left(\frac{e^2E_Mb}{m_\mathrm{e}^2}\right)^{1/3}\left[t-\frac{m_\mathrm{e}}{eE_M}\left(v_{M0}+\frac{k}{b}\right)\right]. \tag{63}$$

Bessho et al. (2018) derived the equation of $v_N$ as a function of $v_M$, as follows:

$$v_N = N\left(\frac{eb}{m_\mathrm{e}}\right)^{1/2}\left(-v_M-\frac{eb}{2m_\mathrm{e}}N^2-\frac{k}{b}\right)^{1/2}$$
$$\times\cot\left\{\frac{2}{3}\left(\frac{eb}{m_\mathrm{e}}\right)^{1/2}\frac{m_\mathrm{e}}{eE_M}\left[\left(-v_M-\frac{eb}{2m_\mathrm{e}}N^2-\frac{k}{b}\right)^{3/2}-\left(-v_{M0}-\frac{k}{b}\right)^{3/2}\right]\right\}, \tag{64}$$

which is for a case where $v_{M0}< -k/b$. In the other case, where $v_{M0}> -k/b$, we have

$$v_N = N\left(\frac{e^2E_yb}{m_\mathrm{e}^2}\right)^{1/3}\frac{\mathrm{Bi}(r_0)\frac{d\mathrm{Ai}(r)}{dr}-\mathrm{Ai}(r_0)\frac{d\mathrm{Bi}(r)}{dr}}{\mathrm{Ai}(r_0)\mathrm{Bi}(r)-\mathrm{Bi}(r_0)\mathrm{Ai}(r)}, \tag{65}$$

where $r_0$ is the initial value of $r$ at $N=0$. Both equations represent multiple curves in the $v_M$-$v_N$ plane, because the cot function and Airy functions Ai and Bi are oscillatory functions.

In the MMS observation of the Torbert event, the electron VDF at the neutral line shows $v_{M0}> -k/b$; therefore, in Bessho et al. (2018) (see Figure 22), to compare the observed multi-crescent VDF with the theory, Eq. (65) was used. In the following, we will explain how to extract the reconnection electric field $E_M$ by comparing the theory and the observed VDF.

**Procedure 1:** Deriving the field quantities from observed field data

From the magnetic field and electric field data, we obtain the slope $b$ for $B_L$ and the slope $k$ for $E_N$. Figure 22a shows magnetic fields (top), electric fields (middle), and the distance from the neutral line (bottom), obtained by MMS2 and MMS3. The $LMN$ coordinates were obtained by a hybrid method (Denton et al. 2016) of MDD (Shi et al. 2005; Section 2.2.1) and MVA (Sonnerup and Scheible, 1998). The $N$ distance was obtained by a method similar to Denton et al. (2016), from the time integral of $V_N = (\mathrm{d}B_L/\mathrm{d}t)/(\partial B_L/\partial N)$, which represents the MMS barycenter velocity relative to the current sheet. The values of $b$ and $k$ were obtained as $b$=9.0×10$^{-2}$ nT/km (from MMS3 data), and $k$=1.2×10$^3$ mV/km$^2$ (from MMS2 data).

**Procedure 2:** Obtaining $v_{M0}$ from the VDF at the neutral line $N$=0

There are two equations, Eq. (64) and Eq. (65), depending on the value of $v_{M0}$. Using the VDF on the neutral line, we identify the population of electrons that have just arrived at the neutral line and started the meandering motion. Figure 22b shows the VDF at the neutral line $N$=0 ($B_L$=0 line) by MMS3, and the population near the white vertical line shows that $v_{M0}$=-0.7× 10$^4$ km/s. In this event, the $E_N$×$B_L$ drift velocity, $-k/b$, is =-1.3× 10$^4$ km/s; therefore, the condition $v_{M0}>-k/b$ needs to be used, and we will use Eq. (65).

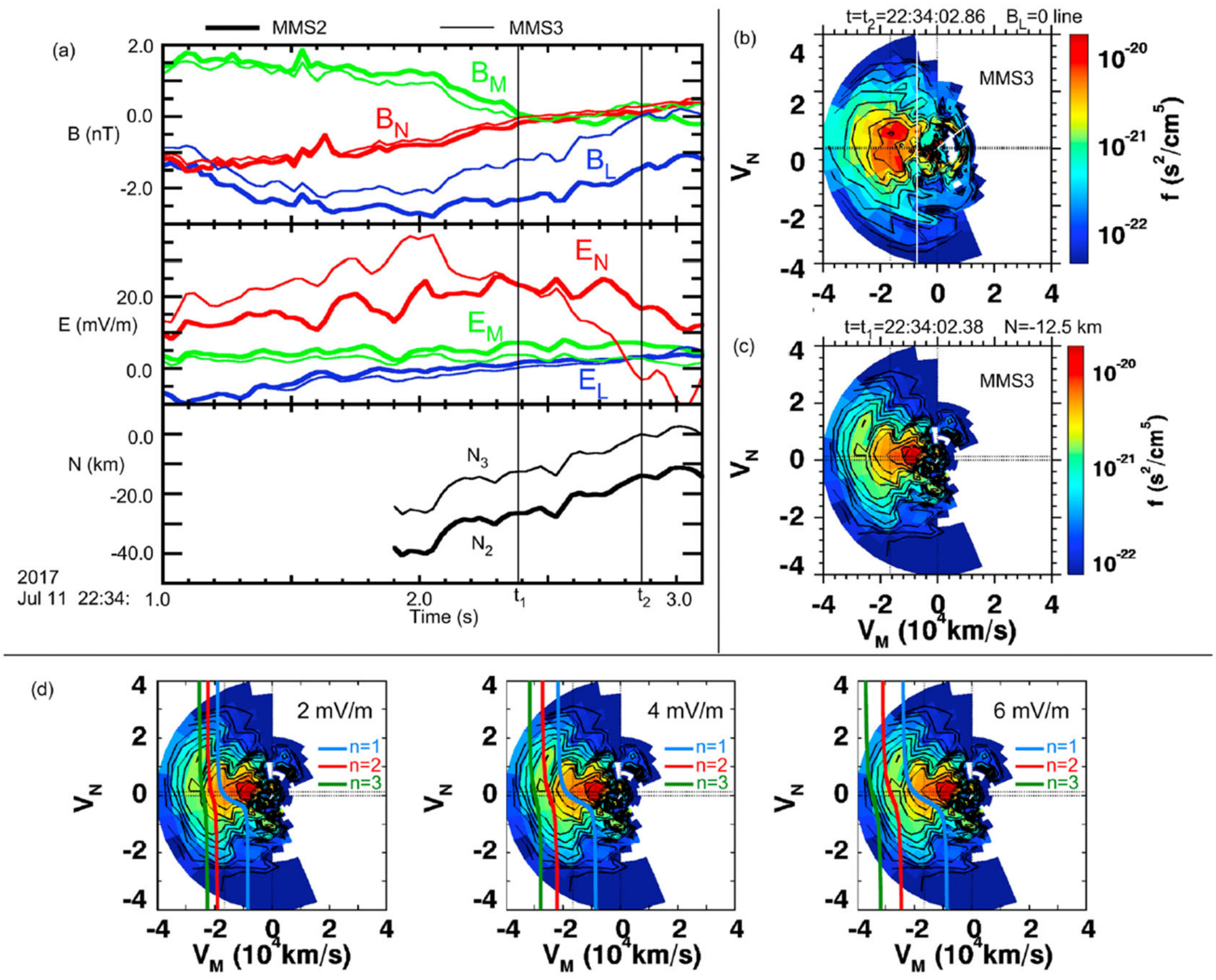

**Fig. 22** MMS data in the event on 11 July 2017. (**a**) MMS data for the field quantities and the distance $N$ from the neutral line. (**b**) Electron VDF at the neutral line. (**c**) VDF at $t$=$t_1$. (**d**) Comparison between the theory (blue, red, green curves by Eq. (65)) and the multi-crescent VDF. Adapted from Bessho et al. (2018).

**Procedure 3:** Compare the theory and the multi-crescent VDF, and determine $E_M$

Now, we have obtained all the required parameters, $b$, $k$, $N$, and $v_{M0}$ to draw the theoretical curves on the multi-crescent VDF, except for an undetermined parameter, $E_M$. In this event, Figure 22c is the VDF that was compared with the theory. In the VDF, there are three crescent-like stripes. Figure 22d shows three plots with different $E_M$ values, 2 mV/m, 4 mV/m, and 6 mV/m. Values of $E_M$ =2 mV/m and $E_M$ =6 mV/m are excluded because the separations between theoretical curves based on Eq. (65) do not match the separations in the observed VDF. In contrast, $E_M$ =4 mV/m is consistent with the observed VDF. In this way, we can determine the reconnection electric field from a multi-crescent VDF. Note that this value, $E_M$ =4 mV/m is close to the value of $E_M$ =3.2 mV/m determined by Genestreti et al. (2018), if we allow for an uncertainty of 1-2 mV/m in the MMS measurements.

### 3.3.2 Remote Sensing at the Reconnection Separatrix Boundary

**Background:** The change of magnetic field connectivity during magnetic reconnection occurs within the micro-scale diffusion region. Assuming that reconnection develops in the *x*-*z* plane (Figure 23), the *y*-component of the electric field $E_y$ around the diffusion region matches the convectional electric field at the inflow and outflow regions to conserve the magnetic flux, $E_y \sim -u_{in}B_{in} \sim -u_{out}B_{out}$ where $u_{in}$, $u_{out}$ and $B_{in}$, $B_{out}$ are the plasma flow speeds and the magnetic field strengths, respectively, at the boundaries of the diffusion region and subscripts "in" and "out" mean the inflow and outflow boundaries, respectively. The magnitude of $E_y$, which represents the flux transfer rate from the inflow region into the diffusion region, is called the reconnection electric field $E_r$ or the unnormalized reconnection rate. Near the center of the diffusion region called the EDR, $E_r$ is sustained by the non-ideal component of the generalized Ohm's law as $E_r \sim |E'_y| = |E_y + (\mathbf{u}_e \times \mathbf{B})_y|$. High-resolution in-situ observations by MMS successfully encountered the EDR and directly measured $E_y'$ near the EDRs in Earth's magnetopause (e.g., Burch et al. 2016b) and magnetotail (e.g., Torbert et al. 2018) current sheets. Statistically, it is relatively rare to encounter these small regions and measure $E_r$ directly. To meet this challenge, a new technique for estimating $E_r$ using in-situ measurements at the reconnection separatrix boundary was recently proposed (Nakamura et al., 2018a). Since the extent of the separatrix is longer than the micro-scale EDR, the probability for encountering the separatrix is much greater than that for the EDR.

**Methods:** Ignoring variations in the out-of-plane direction, then $E_r$, which corresponds to $E_y$ at the X-line, can be written as

$$E_r = -E_{y_x-line} = \frac{\partial A_{y,\mathrm{x-line}}}{\partial t} \sim \frac{\partial A_{ys}}{\partial t}, \qquad (66)$$

where $A_{y,\mathrm{x-line}}$ is the out-of-plane component of the vector potential at the X-line and $A_{ys}$ is the potential at the separatrix. Note that in the 2-D limit, the potential at the X-line is constant along the reconnection separatrix boundary (Vasylinuas 1975). Since the separatrix may be moving relative to observing spacecraft because of structure motion (see Figure 23), by sequentially obtaining the potential at the separatrix by two different probes that are separated in the boundary normal direction, $E_r$ can be estimated as $E_r \sim \Delta A_{ys} / \Delta t$. Here $\Delta t$ is the time difference between the separatrix detections by the two probes and $\Delta A_{ys}$ (corresponding to $A_3$-$A_2$ in Figure 23) is the difference of $A_{ys}$ between the probes during this separatrix crossing. Defining the separation of the two probes as $\Delta\mathbf{x}=(\Delta x, \Delta z)$, as illustrated in Figure 23, and assuming: (i) a constant reconnection rate during $\Delta$t, and (ii) the uniform electric and magnetic fields between the two probes, then $\Delta A_{ys}$ can be estimated as $\Delta A_{ys} \sim (\Delta\mathbf{X}\times\mathbf{B})_y$, where $\Delta\mathbf{X} = \Delta\mathbf{x} - \mathbf{V}_c\Delta t$ takes out the effect of background structural motion in the $\mathbf{E}\times\mathbf{B}$ drift velocity, $\mathbf{V}_c = (\mathbf{E}\times\mathbf{B})/B^2$. Then, from Eq. (66), $E_r$ can be described as,

$$E_r \sim \frac{(\Delta\mathbf{X}\times\mathbf{B})_y}{\Delta t} = [-(\mathbf{V}_{tim} - \mathbf{V}_c)\times\mathbf{B}]_y = -B_x\frac{\Delta z}{\Delta t} + B_z\frac{\Delta x}{\Delta t} + B_x\left(\frac{\mathbf{E}\times\mathbf{B}}{B^2}\right)_z - B_z\left(\frac{\mathbf{E}\times\mathbf{B}}{B^2}\right)_x, \quad (67)$$

where $\mathbf{V}_{tim} = \Delta\mathbf{x}/\Delta t$ is the separatrix velocity from the timing analysis. Eq. (67) suggests that if two probes that are separated in the normal direction sequentially detect the separatrix signatures, the reconnection electric field $E_r$ can be remotely estimated (Nakamura et al. 2018a).

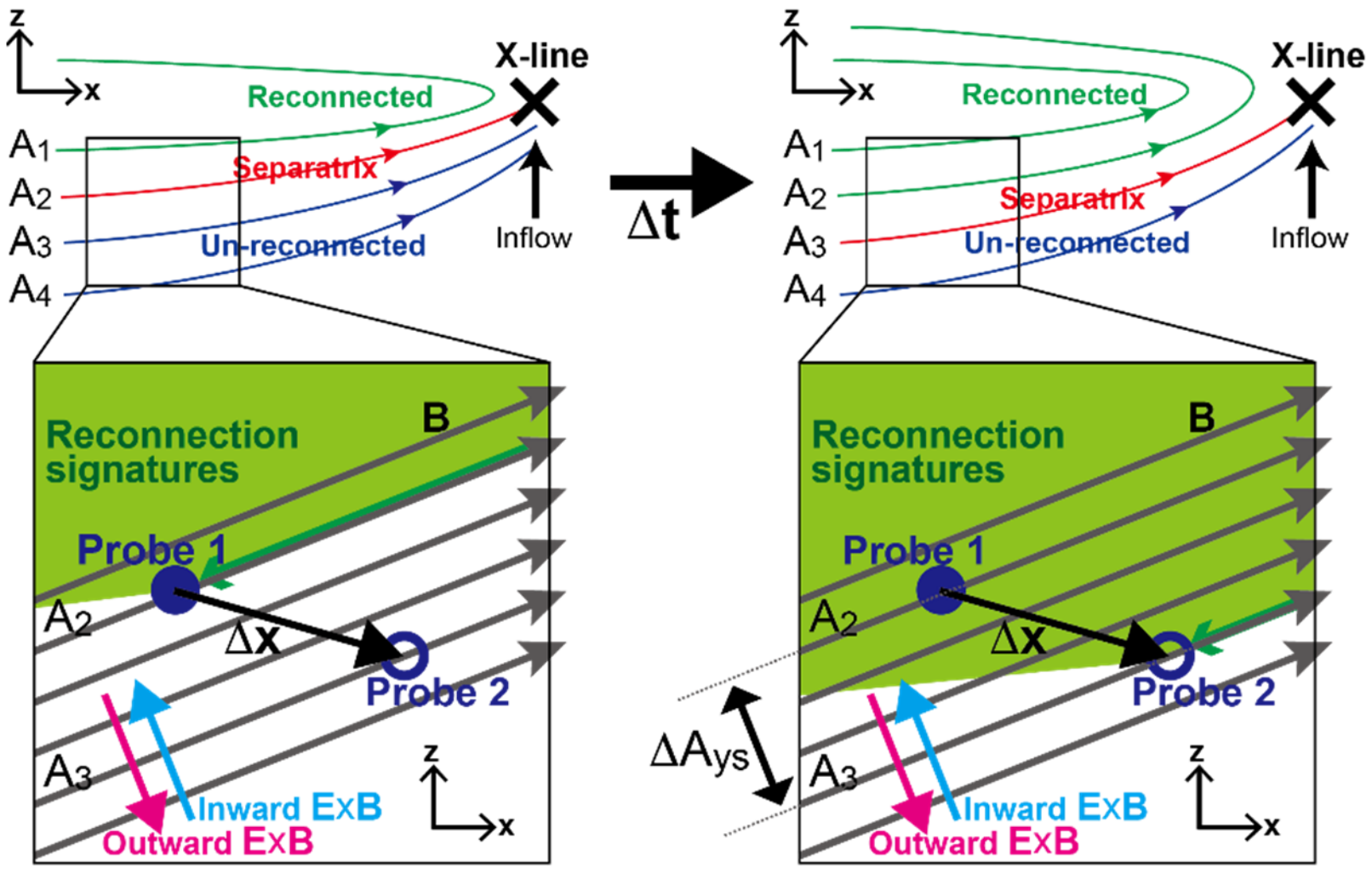

**Fig. 23** Schematic of a reconnecting current sheet in motion in the *x*-*z* plane during $\Delta t$, focusing on a region near the separatrix boundary where two probes separated by $\Delta\mathbf{x}$ sequentially detect the separatrix signatures (adapted from Nakamura et al. (2018a)). The field line motion (E×B drift velocity) can be inward (cyan arrow) or outward (magenta arrow), depending on the magnitude of the reconnection electric field and the magnitude and direction of the structure velocity.

**Applications:** In Nakamura et al. (2018a), the proposed technique was first tested using virtual satellites in a 2-D fully kinetic PIC simulation of reconnection without a guide field. To identify separatrix boundaries from observation data, Nakamura et al. (2018a) proposed to detect high-energy parallel electron beams that stream away from the X-line along the separatrix and resulting enhancements of the $B_y$ component (Hall fields). The remotely estimated $E_r$ from virtual observations for both electron beams and Hall fields indeed agree well with the directly obtained $E_r$ at the X-line, indicating the adequacy of this remote sensing technique.

In real in-situ observations, this technique requires sufficiently high-cadence magnetic and electric field data under an assumption that temporal and spatial variations of the fields while obtaining $\mathbf{V}_{tim}$ (i.e., while multi-probes sequentially detect the separatrix) are negligible. The MMS mission satisfies this requirement with its burst mode magnetic (Russell et al. 2016) and electric (Ergun et al. 2016; Lindqvist et al. 2016; Torbert et al. 2016a) field measurements and its small inter-spacecraft separation ($10^{1\text{-}2}$ km). The plasma particle measurements at burst mode cadences (Pollock et al., 2016) are also useful to detect signatures of the separatrix such as parallel electron beams (Varsani et al. 2017). Nakamura et al. (2018a) first applied this technique to an MMS observation event on 10 August 2016 of the plasma sheet crossing in the near-Earth region accompanied by a strong substorm, which was initially reported by Nakamura et al. (2017). By identifying the separatrix boundary from the Hall field enhancements, they estimated the reconnection electric field as $E_r \sim 15\pm5$ mV/m. Wellenzohn et al. (2021) applied this technique to another MMS plasma sheet crossing event accompanied by a small substorm on 12 July 2018. By identifying the separatrix boundary from high-energy parallel electron beams and resulting electric field disturbances, they estimated $E_r \sim 2\pm0.5$ mV/m. In another example, Nakamura et al. (2018b) focused on the magnetotail EDR crossing event by MMS on 11 July 2017 accompanied by a small substorm, which was initially reported by Torbert et al. (2018). In this event, $E_r$ can be obtained not only from the remote sensing technique at the separatrix near the EDR, but also from the direct measurement within the EDR. The obtained $E_r$ values from remote and direct methods both are in the range 2.5±0.5 mV/m. These initial results of the remote sensing technique performed using MMS observations suggest a positive correlation between $E_r$ and the intensity of substorms. A future statistical approach is required to comprehensively establish the relation between the local reconnection $E_r$ and the geomagnetic disturbances.

### 3.3.3 Separatrix Angle Related to Reconnection Rate

The reconnection electric field normalized by $V_{A0}B_0$, where $B_0$ is the background reconnecting magnetic field and $V_{A0}$ is the upstream Alfvén speed based on $B_0$ ($R \sim E_r/V_{A0}B_0$), measures how fast the connectivity changes during reconnection and it is generally called the normalized reconnection rate. Considering the force balance along the inflow and outflow directions at the meso-scale in the $\beta \ll 1$ limit, Liu et al. (2017) derived a general theory showing that for symmetric reconnection $R$ is related to the separatrix opening angle $\theta$ near the IDR as,

$$R \sim \frac{E_r}{V_{A0}B_0} \sim \tan\theta \left(\frac{1-\tan^2\theta}{1+\tan^2\theta}\right)^2 \sqrt{1-\tan^2\theta}. \qquad (68)$$

Here the separatrices are assumed to be straight from the center of the diffusion region through the observation points. Given that the adequacy of Eq. (68) was indeed confirmed by fully kinetic simulations (e.g., Liu et al. 2017, 2018a; Nakamura et al. 2018b), this theory implies that the reconnection rate $R$ can be obtained by measuring the opening angle θ at an ion-scale distance from the X-line.

As sketched in Figure 24a, the separatrix opening angle $\theta$ just outside the IDR needed in Eq. (68) matches the flaring angle $\tan^{-1}(|B_{Ns}|/|B_{Ls}|)$ made by the magnetic fields adjacent to the separatrix ($\mathbf{B}_\mathrm{s}$), which also matches the flaring angle $\tan^{-1}(|B_{N\mathrm{d}}|/|B_{L\mathrm{d}}|)$ made by the magnetic fields at the upstream/downstream edge of the diffusion region ($\mathbf{B}_\mathrm{d}$). In light of these relations, Nakamura et al. (2018b) introduced the following quantity $f_\mathrm{r}$, that is a function of $|B_N|/|B_L|$ measured along the spacecraft path,

$$f_r\left(\frac{|B_N|}{|B_L|}\right) \sim \frac{|B_N|}{|B_L|}\left(\frac{1-\left(\frac{|B_N|}{|B_L|}\right)^2}{1+\left(\frac{|B_N|}{|B_L|}\right)^2}\right)^2 \sqrt{1-\left(\frac{|B_N|}{|B_L|}\right)^2}\ . \qquad (69)$$

They then applied this function to estimate the reconnection rate $R$ in the magnetotail EDR crossing event by MMS on 11 July 2017 (Torbert et al. 2018). This function gives R whenever the spacecraft crosses the separatrix near the diffusion region; i.e., where $|B_N|/|B_L| = |B_{Ns}|/|B_{Ls}|$. Thus, $R$ can be estimated by detecting the separatrix signature and computing $f_\mathrm{r}$ at that time. In Nakamura et al. (2018b), the separatrix boundary was accurately deduced using close comparisons with a fully kinetic simulation of this MMS event. The reconnection rate $R$ was then successfully obtained as $R$~0.15-0.2 (Figure 24b), which indeed agrees well with the rate directly observed within the EDR (Genestreti et al. 2018).

Note that this technique of estimating the normalized reconnection rate $R$ would be applicable only in a limited region where the separatrix opening angle is sufficiently close to the exhaust opening angle critical to the rate. This condition would be satisfied just outside the edge of the IDR as sketched in Figure 24a. In addition, since the separatrix line in the $LN$ plane is nearly straight from outside the IDR toward the corner of the EDR (Nakamura et al. 2018b), this condition would be satisfied even within the IDR.

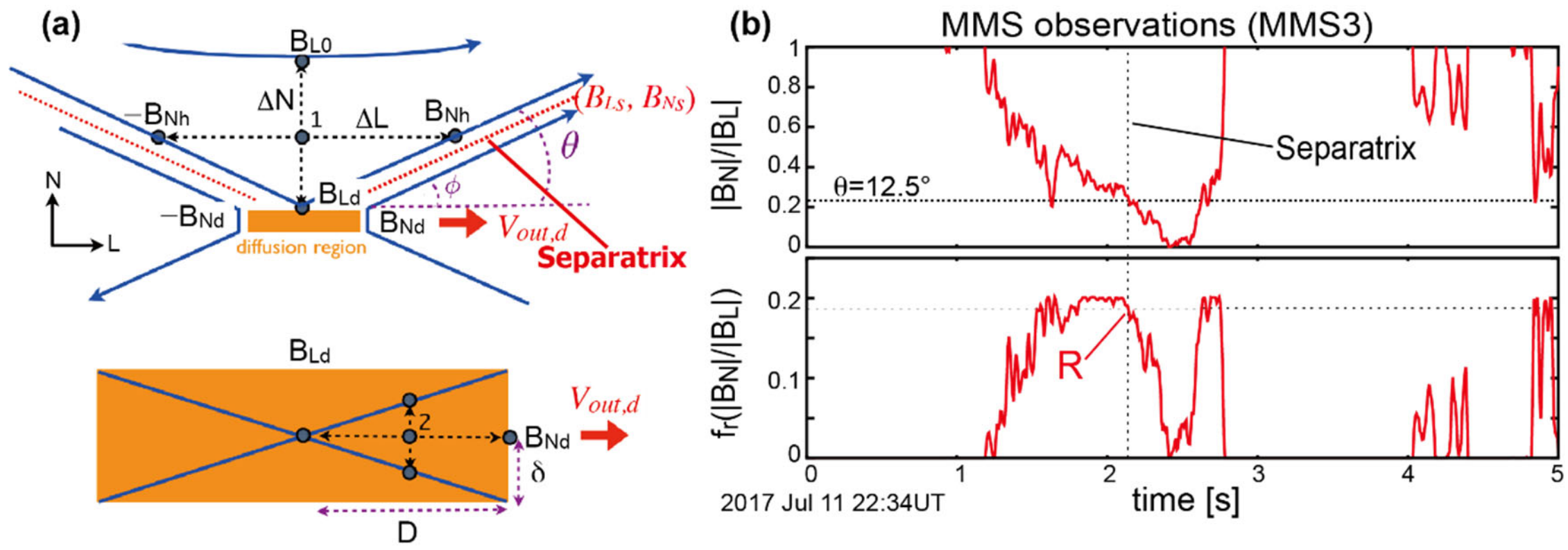

**Fig. 24** (**a**) Sketch of the reconnection geometry around the diffusion region (adapted from Liu et al. (2017)). (**b**) MMS observations of $|B_N|/|B_L|$ and $f_r$ (Eq. (69)) during the separatrix crossing on 11 July 2017 (adapted from Nakamura et al. (2018b)).

## 4 Summary and Outlook

Most of the methods discussed in the present review have not yet been extensively used in the analysis of in-situ data, including from the MMS mission. Thus, our hope is that the use of these methods will help advance our understanding of magnetic reconnection and associated processes in space, including wave excitation and particle acceleration, their feedback on the reconnection process, and coupling to macro-scale phenomena. Further improvement of the methodology would also be expected. In particular, since the MMS mission has focused essentially on electron-scale processes, cross-scale aspects of magnetic reconnection have not been well explored to date. Thus, there is vast room for the future development of data analysis methods to understand these multi-scale processes. See, for example, Broeren et al. (2021) for a potentially multi-scale method that can reconstruct a 3D magnetic field when more than four spacecraft are available, and Bard and Dorelli (2021) for a new type of reconstruction techniques based on Physics-Informed Neural Networks (Raissi et al. 2019).

## Appendix A Region and Current Sheet Identification

Magnetic reconnection is known to occur in thin current sheets often located at transitions between different plasma regions, as seen at the dayside magnetopause and in the nightside magnetotail. This has led to the assignment of regions of interest that define where the MMS spacecraft may encounter reconnection, parameters that identify where reconnection is likely to occur, and data management systems [the Automated Burst System (ABS), the Scientist-in-the-Loop (SITL), and the Ground Loop System (GLS)] to capture the right type of data. This appendix reviews the ways in which MMS locates, identifies, and captures reconnecting current sheets. A few methods based on machine learning for automated identification of regions in and around the magnetosphere are also summarized.

### A.1 Automated Burst System

Prior to MMS, observations of the EDR, where electrons are demagnetized and magnetic energy is converted to electron kinetic energy, had been enigmatic, with few direct observations (Nagai et al. 2011, 2013; Scudder et al. 2012; Tang et al. 2013; Oka et al. 2016). This was because spacecraft lacked the spatial and temporal resolution to resolve electron-scale dynamics. MMS has overcome these limitations by having four spacecraft in a tetrahedron configuration at unmatched spatial scales and sampling rates. Since launch, MMS has identified more than 50 EDRs (see Webster et al. (2018), Lenouvel et al. (2021), and Genestreti et al. (2022) for partial lists) and greatly expanded our knowledge of what catalyzes the global reconnection cycle (Chapter 3.1 of this issue; Fuselier et al. 2024). The amount of data required to obtain this success greatly exceeds the downlink allocations of the deep space network, which means that the satellites require an automated way of selecting time intervals for high time resolution burst data downlink, the ABS.

Figure 25a shows the amount of burst data held onboard MMS between 17 October and 15 November, 2022. Data is categorized into Category 0 (purple), 1 (red), 2 (yellow), 3 (green), and 4 (blue) based on the active science objectives (see Table 1 in Argall et al. (2020) for an example). Category 0 and Category 1 correspond to calibration and primary science data, respectively, while Category 2-4 correspond to lower-priority science goals. The orange dashed line indicates the threshold beyond which data will be overwritten, as the remaining buffers are reserved for new regions of interest. Figure 25b shows the amount of overwritten data in each category. Data classified into the lowest-priority category (Category 4) are overwritten first, unless some memory cleanup activity takes place, as on 15 November 2022. This figure demonstrates that MMS is able to store and downlink all of its highest priority data and effectively manage low-priority data to both achieve tertiary science objectives and make room for new data. Such categorization is important for the burst memory management system to be effective.

In an ABS, trigger data numbers (TDNs) are calculated by applying a look-up table of gains $G_i$ and offsets $O_i$ to each measured data quantity $x_i$ to create a scalar value that is a linear combination of the $x_i$ (Baker et al. 2016; Fuselier et al. 2016):

$$TDN = \sum_{i=0}^{N-1} G_i x_i + O_i. \quad (70)$$

When the TDNs exceed a threshold value, either the data are marked for downlink or the satellite is triggered into burst mode. Missions such as WIND, THEMIS, Cluster,

STEREO have burst mode schemes that operate only when triggered. Some WIND and THEMIS triggers used to detect plasma boundaries such as the magnetopause are documented by Phan et al. (2015). Triggers used on STEREO for shock detection, their evolution, and their efficacy are described by Jian et al. (2013). MMS has enough memory to capture burst mode data at all times within its science region of interest. This is so that the Scientist-in-the-Loop has a chance to review the low time resolution data and make their own selections before the unselected burst data is erased from memory. The TDNs are available publicly and the look-up tables of gains and offsets can be updated depending on the active science objectives.

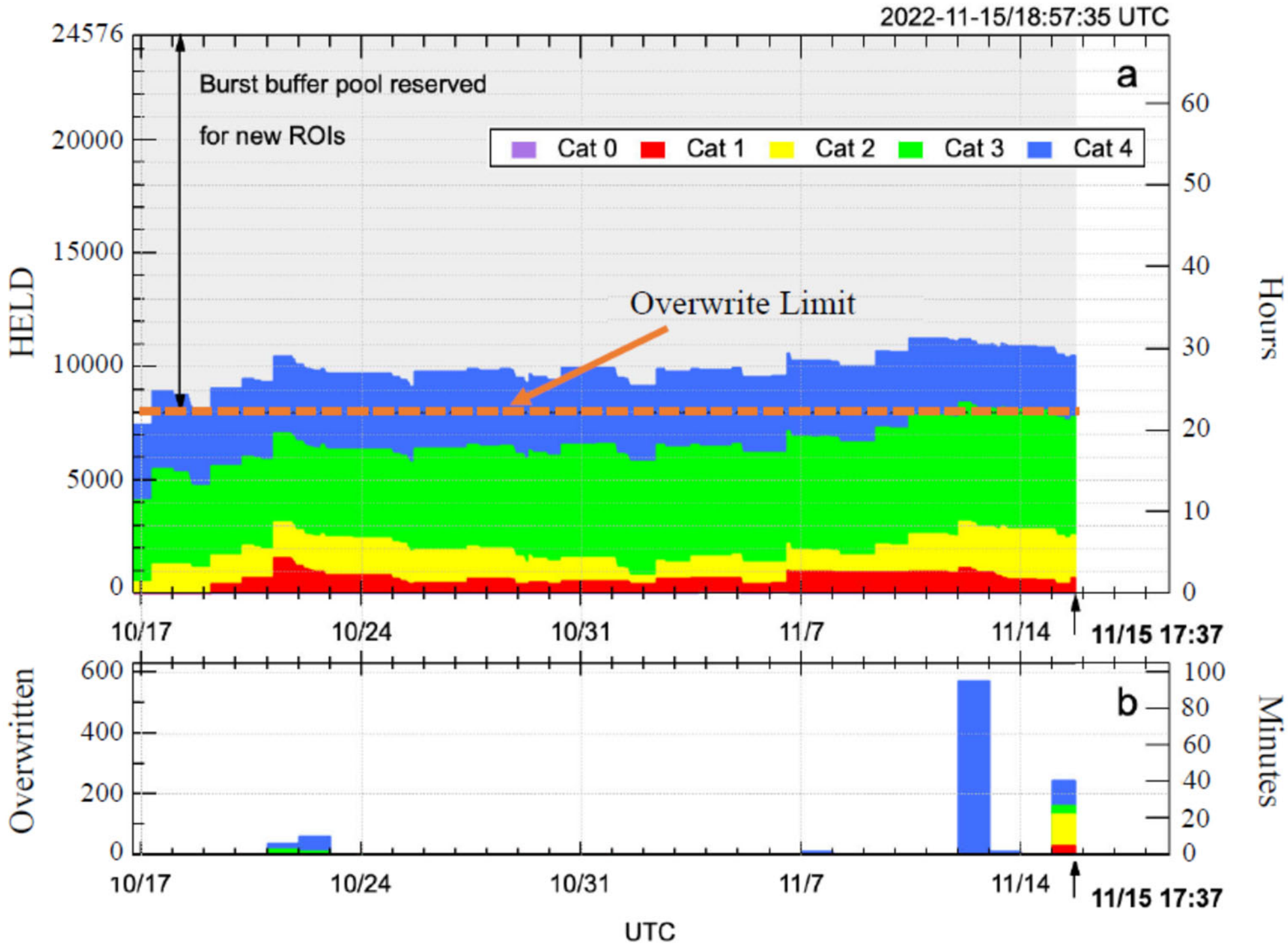

**Fig. 25** Burst memory buffers at risk of being overwritten. (**a**) Amount of Category 0 (purple), 1 (red), 2 (yellow), 3 (green), and 4 (blue) stored or HELD onboard MMS, shown in terms of the number of burst buffers (left axis) and the corresponding hours of data (right axis). (**b**) The amount of overwritten data in terms of Category (color), memory buffers (left axis) and total time (right axis). Category 0 data are for special operations (e.g., calibration) or critical science data; selections are rare and are transmitted to ground as soon as the downlink is available, so no Category 0 buffers are present in the figure.

## A.2 Scientist-in-the-Loop

The SITL is a role that is passed among the personnel on the MMS science team. The SITL scientist is responsible for manually selecting time intervals for burst data downlink by examining low time resolution data that is recorded simultaneously with the burst data and is downlinked at the end of each orbit. The SITL is guided by the mission-level science objectives, which are assigned a priority of 1-4 (Category 1-4) and a range of Figure-of-Merit values. The SITL assigns a Figure-of-Merit value to time intervals that fit the science objectives and the corresponding burst data is downlinked in priority-order. The SITL process has been successful and is being used as the primary method for burst data selections even during extended mission phases, despite the automated processes, as discussed in sections A.1 and A.3, having being operated.

Figure 26 shows a typical view of the data that the SITL scientist sees when using an interactive tool called "EVA" to make burst data selections (see Argall et al. (2020) for more details). Labels and scales have been removed on purpose except for the bottom two panels, which show the ABS and SITL selections, respectively. Note that while the SITL scientist selects most of the intervals selected by the ABS, the SITL makes more selections (vertical bars in the bottom panels) – a tedious process that could be alleviated, in part, by automated selection models. This is where the Ground Loop System (GLS) comes into play.

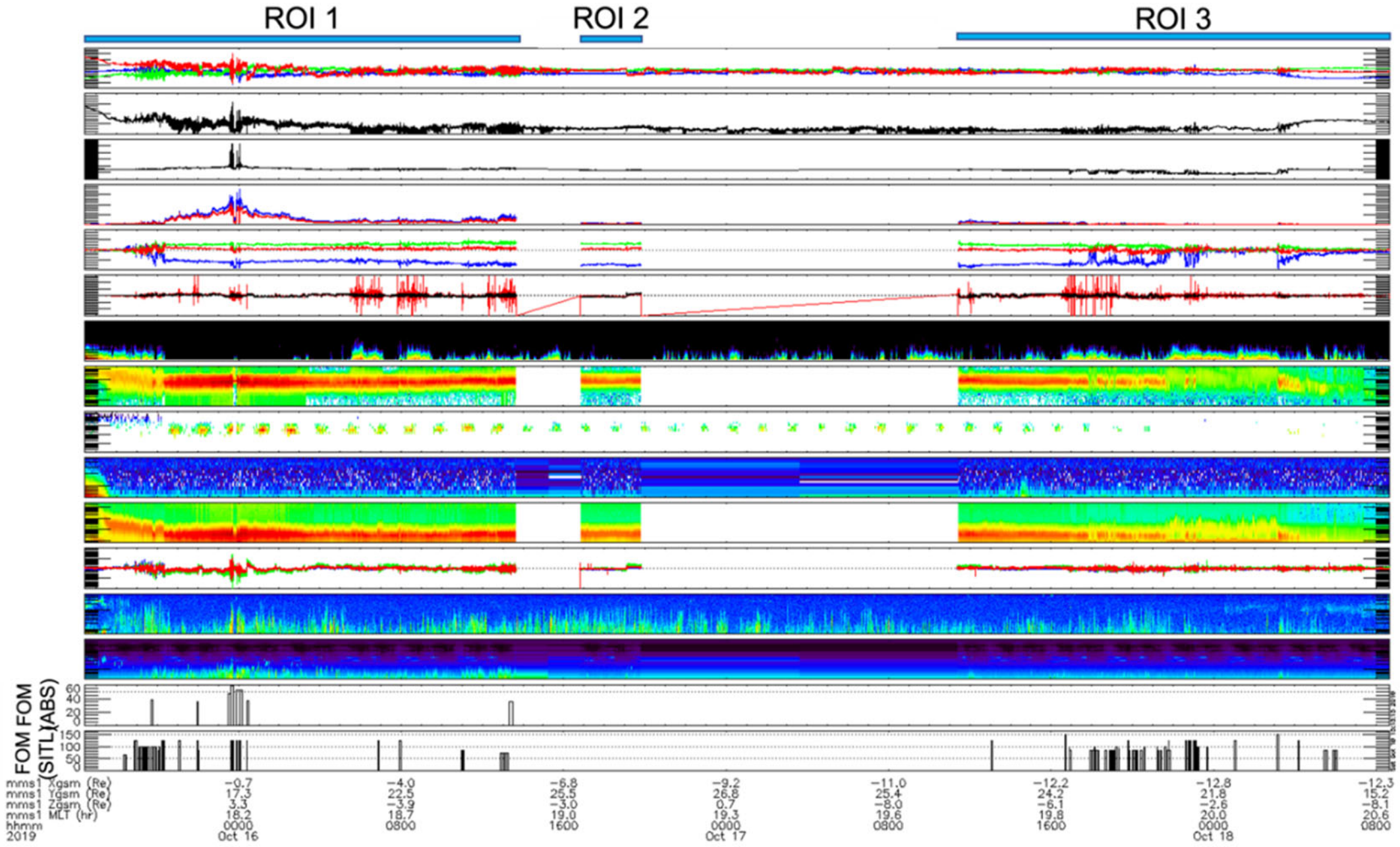

**Fig. 26** Data that the SITL uses to make selections. The SITL views data from the regions of interest, ROI 1, 2, and 3, as it becomes available, and uses a tool to interact with the plot, make selections, and submit them to the science data center. Quantities and scales are purposefully not shown except for the ABS (second to last panel) and SITL (last panel) selections. Note that while the SITL basically selected the same intervals (vertical bars)

as the ABS, the SITL selected many more.

### A.3 Ground Loop System

The GLS is designed to be a system of machine learning (ML) or empirical models that automate the event classification process using all of the data available to the SITL (much more than what is available to the ABS). Data available to the SITL is of restricted use because its quality is lower than of the science-quality (Level-2) data freely available to the public (Baker et al. 2016). Thus, ML models trained on SITL data may not perform as well when applied to Level-2 data, and vice versa. Argall et al. (2020) implemented the first GLS ML model for automated burst selections. Their model aimed to automate the SITL scientist's top priority – to select magnetopause crossings. Selecting magnetopause crossings was important because EDRs are not resolved in the low-resolution data available to the SITL; however, EDRs occur in reconnection events at the magnetopause and the magnetopause is easily identifiable in the SITL data.

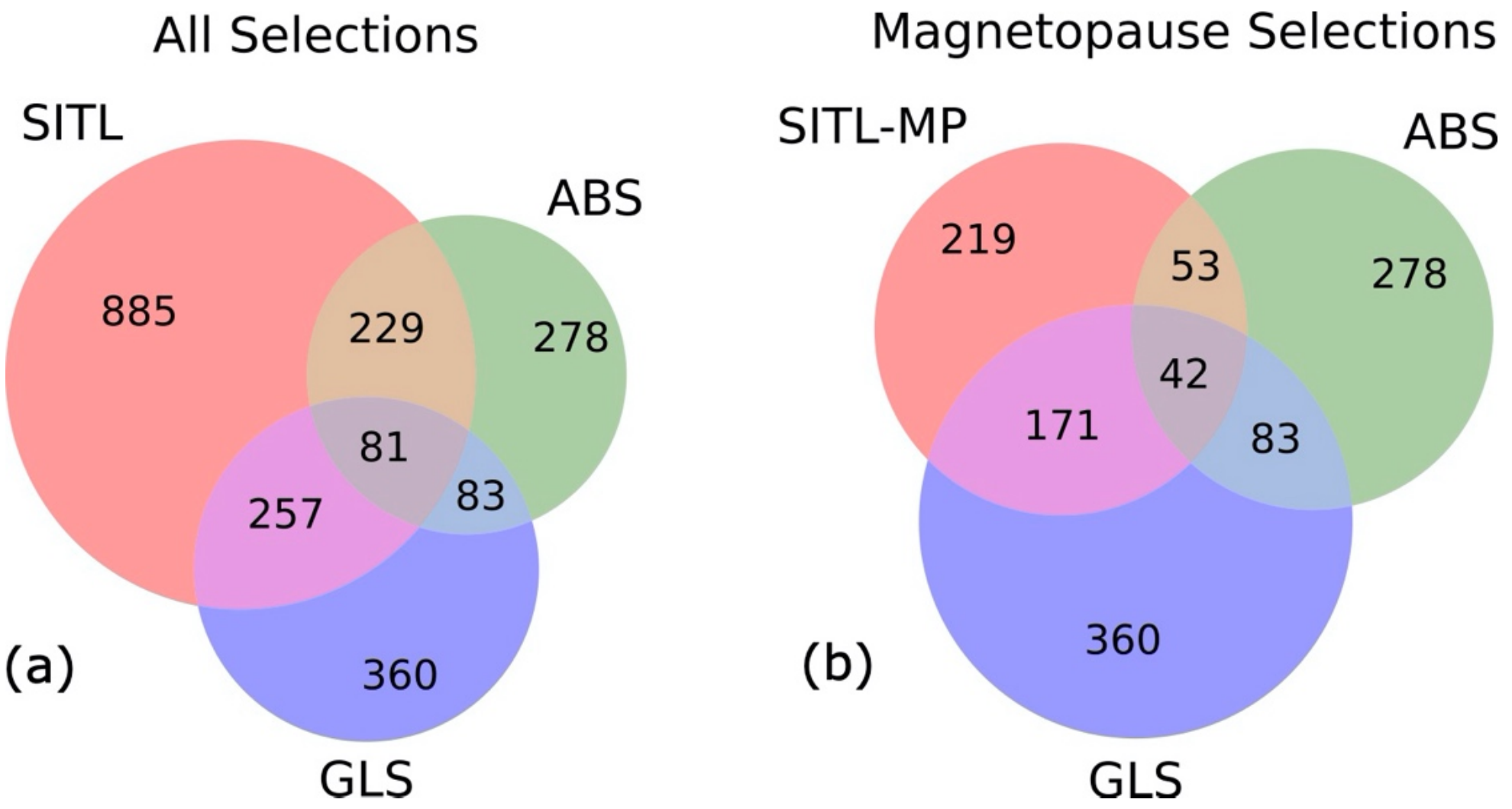

**Fig. 27** The GLS and ABS are complementary systems that make selections of high interest to the SITL. (a) Comparison of all selections; the SITL selects a significant number of selections made by both the GLS (71%) and ABS (64%). (b) Comparison of selections designated as magnetopause (MP) crossings by the SITL. In both cases, there is little overlap between the GLS and ABS. Adapted from Argall et al. (2020).

Figure 27a is a Venn diagram showing the overlap of SITL, ABS, and GLS selections using all selections made between 19 October 2019 and 25 March 2020. The SITL selects 71% of all GLS selections and 64% of all ABS selections, indicating that the ABS and GLS are making good selections. Figure 27b is the same format, but includes only those selections that were designated magnetopause crossings by the SITL. The GLS selects 78% (171/219) of all magnetopause crossings selected by the SITL. The other 22% of selections include intervals that were magnetopause-like (i.e., flux transfer events, reconnection jets), but were not specifically called magnetopause crossings by the SITL. This indicates that the GLS was effective at what it was designed to do. In both cases, the ABS and GLS had little overlap in selections – they are complementary systems that provide helpful information to the SITL.

The GLS was designed in such a way that new supervised ML models could be trained easily by searching and parsing the human-readable string SITL scientists assign to each of their selections. (A suggestion to future missions that wish to implement a similar system is to have a defined set of keywords to identify similar events.) These ML models could be implemented simultaneously in a hierarchical structure to automate science campaigns (Figure 8 in Argall et al. (2020)). The ground-level for the hierarchical campaigns is to implement region identifiers.

### A.4 Machine Learning-Based Region Identification

The selection of intervals of interest can be facilitated by the ability to identify key regions of the near-Earth environment. While historically identification of regions and boundaries was done through visual inspection of data, recent work has shown that modern methods relying on ML are an alternative method that can be more efficient. Novel methods are increasingly needed as the amount of data available nowadays precludes scientists from visually mining numerous and huge datasets accumulated over many different missions, sometimes over decades. Such a method was recently implemented by Nguyen et al. (2022a) and applied to statistical analysis and interpretation of the location and shape of the Earth's magnetopause (Nguyen et al. 2022b,c,d).

The identification of three key near-Earth regions, the magnetosphere, the magnetosheath, and the solar wind, was made based on an automatic classification method that uses in situ data (magnetic field and ion moments) from multiple spacecraft. The classification was performed with the so-called Gradient Boosting algorithm (Friedman 2001). While not offering as much flexibility as deep learning methods, this algorithm, based on the iterative fit of the residuals obtained by the successive training and predictions made by decision trees, has been recognized to perform well on complex, eventually imbalanced classification problems (Brown and Mues 2012). Furthermore, it typically needs less labeled data and is lighter to train than deep neutral networks. The prediction of the algorithm is shown to outperform routines based on manually set thresholds. Data from 11 different spacecraft (THEMIS, ARTEMIS, Cluster, one Double Star (TC-1), and one MMS spacecraft) were analyzed for a total of 83 cumulated years. A total of 15,062 magnetopause crossings and 17,227 bow shock crossings were identified. An example automated identification is illustrated in Figure 28 for an outbound orbit of the MMS mission. It highlights excellent agreement between label data and the method identification of the main regions and boundaries during this pass, except for magnetopause boundary layers (transition regions). The code is available online (Table 7 in Appendix C) and the datasets can easily be enhanced for future use by the community.

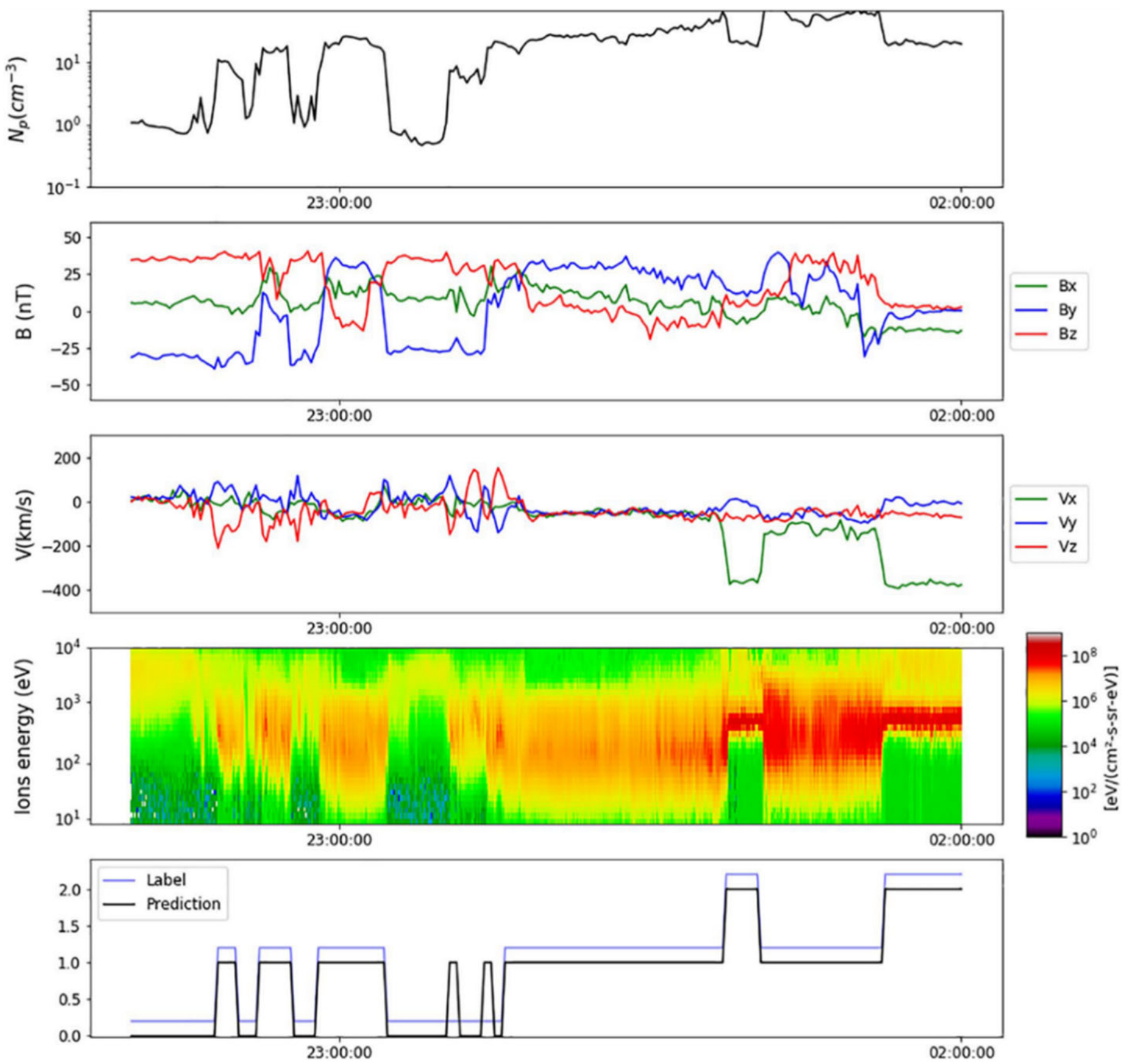

**Fig. 28** MMS observations on 31 December 2015 (adapted from Nguyen et al. (2022a)). The top to fourth panels show the ion density, magnetic field, ion velocity in GSM coordinates, and omnidirectional differential energy fluxes of ions. The bottom panel shows the evolution of the label (blue), intentionally shifted for visual inspection, and the prediction made by the ML algorithm (black), in which “0” means the magnetosphere, “1” the magnetosheath, and “2” the solar wind.

### A.5 Automated Region Classification Based on 3D Particle Velocity Distributions

Vast amounts of data produced by space missions make it difficult for scientists to identify interesting events manually, and automatic data classification methods can be convenient. Below we describe one approach that allows us to classify three-dimensional (3D) ion energy distribution samples according to the plasma regions (Olshevsky et al. 2021). The classified data can then be used for statistical studies and identifying boundaries, such as bow shocks and the magnetopause, where reconnection may occur. We will focus on the dayside magnetosphere and will define the following regions: solar wind (SW), ion foreshock (IF), magnetosheath (MSH), and magnetosphere (MSP), all of which have a distinct signature in ion VDFs. A training dataset representing "clean" samples (i.e., excluding boundaries) of the ion distributions was selected from the above four regions, and was used to train a 3D convolutional neural network classifier (Maturana and Scherer 2015) by a supervised ML approach. The classifier is then applied to ion distributions (other than the training dataset) and assigns a probability of each distribution belonging to one of the four classes.

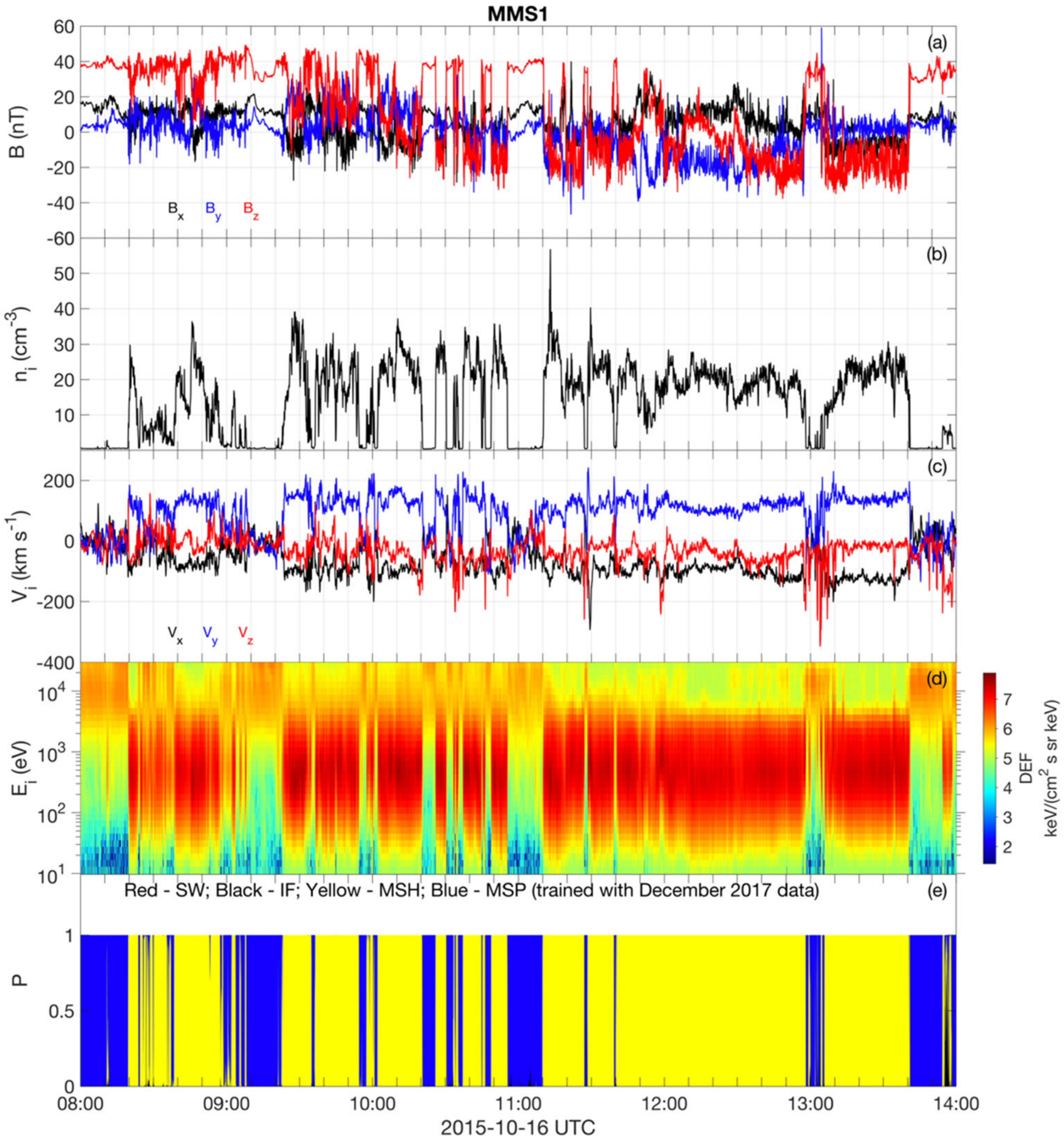

**Fig. 29** Example of plasma region classification of MMS data on 16 October 2015. The panels from top to bottom show (**a**) the magnetic field, (**b**) plasma density, (**c**) ion velocity, (**d**) ion energy spectrum, and (**e**) the probabilities provided by the classifier.

To illustrate the method, we show in Figure 29 MMS observations during a 6-hour interval on 16 October 2015 that contains multiple magnetopause crossings (transitions between MSH and MSP). The crossing at 13:06 UT contains an encounter with the EDR at the site of magnetic reconnection (Burch et al., 2016b). The MSP intervals are characterized by the northward magnetic field ($B_z$ dominant), low density and plasma flow, and increased ion flux at energies above 10 keV. The MSH intervals are, on the contrary, characterized by high density, fast plasma flow ($V_y$ ~ 150 km/s), and ion flux peaking at ~0.3 keV. The bottom panel shows the results of the neural network classifier; one can see that both the MSP (blue) and MSH (yellow) regions have been correctly classified except for boundary layers between the two regions.

Plasma regions can be identified using plasma moments and electromagnetic fields, as illustrated above and in several other ML approaches including the one introduced in section A.4 (Nguyen et al. 2022a; Cheng et al. 2022). However, ion VDFs contain more information than the moments of the distribution, having a unique footprint for each plasma region and thus possibly improving the reliability of classification. Trained scientists use plots of ion data for visual region classification. As it is difficult to visualize the time series of 3D VDFs, they would normally reduce the full 3D measurements in some way, for example, to omni-directional energy-time spectrograms (e.g., Figure 29d) where all look directions are summed up.

Another helpful representation is the angle-angle (azimuth-polar angle) plots of PSDs for a specific energy channel. The approach to the classification by Olshevsky et al. (2021), in its essence, is based on image recognition of such plots. One recognizes each of specific footprints in such images corresponding to different regions. During a human inspection, only a small number of energies can generally be analyzed simultaneously. ML has no such limitation, and can analyze all the data simultaneously. Thus, a 3D image recognition is performed, in which a data cube composed of a stack of 32 (number of energy channels for MMS FPI Dual Ion Spectrometers (DIS) (Pollock et al. 2016)) angle-angle plots is analyzed. This approach is enabled by the homogeneous dataset provided by DIS in fast mode, i.e., the numbers of angular bins and the energy ranges are fixed (except for the special solar wind mode), and thus there is no need to reduce or resample the data.

The results for classification can be used in many different ways. One use is to define times when MMS is in a particular plasma region, for example, in a pristine magnetosheath (e.g., Svenningsson et al. 2023). Another use is to analyze the probabilities' time series to identify the boundaries between the different plasma regions. For example, by analyzing the transition between the SW/IF and MSH, Lalti et al. (2022) identified ~3000 bow shock crossings by MMS. Similarly, magnetopause crossings can be identified from the transitions between the MSH and MSP classes. Another potential application is to use the probabilities (Figure 29e) to quantify plasma mixing in boundary layers, which, for example, can occur in Kelvin-Helmholtz vortices (Settino et al. 2022). As a final remark, we note that the method as described is based solely on the classification of individual ion VDFs, and can be further extended to ingest the information on time evolution and/or other data (e.g., magnetic field, spacecraft position, etc.), which can enable an even more robust classification.

## Appendix B The MMS "Quarter Moments" Data Product

The FPI on the MMS mission was designed to measure three-dimensional electron (ion) phase space densities every 30 (150) ms. This high data rate is achieved by arranging 8 electrostatic analyzers (ESA) for each species around the spacecraft, where the field-of-view of each ESA can be electrostatically deflected to four uniformly spaced look directions spanning 45 degrees. The high voltage power supply executes a full energy sweep (from ~10 eV to ~30 keV) for each of the four deflection states, so that the suite of analyzers simultaneously samples 8 uniformly spaced azimuthal angles for each energy sweep. This means that FPI samples a 32×16×8 regular (energy, zenith, azimuth) array every 7.5 (37.5) ms for electrons (ions), allowing for the possibility of recovering plasma moments a factor of 4 times faster than the nominal 30 (150) ms cadence. Figure 30 illustrates how the four deflection states are combined into the final FPI Level 2 phase space density "skymaps."

To recover 3D plasma moments at 7.5 (37.5) ms for electrons (ions), we use cubic spline interpolation to reconstruct 32 azimuthal samples – independently for each energy and zenith – from the 8 azimuthal samples corresponding to a single deflection state. To mitigate spline boundary condition issues, we interpolate an augmented set of data over the domain $[-2\pi, 3\pi]$ in which data from $[0, 2\pi]$ is copied to $[-2\pi, 0]$ and $[2\pi, 3\pi]$. The interpolated data is then passed to the production moments algorithm as a (32 energy) × (16 zenith) × (32 azimuth) array.

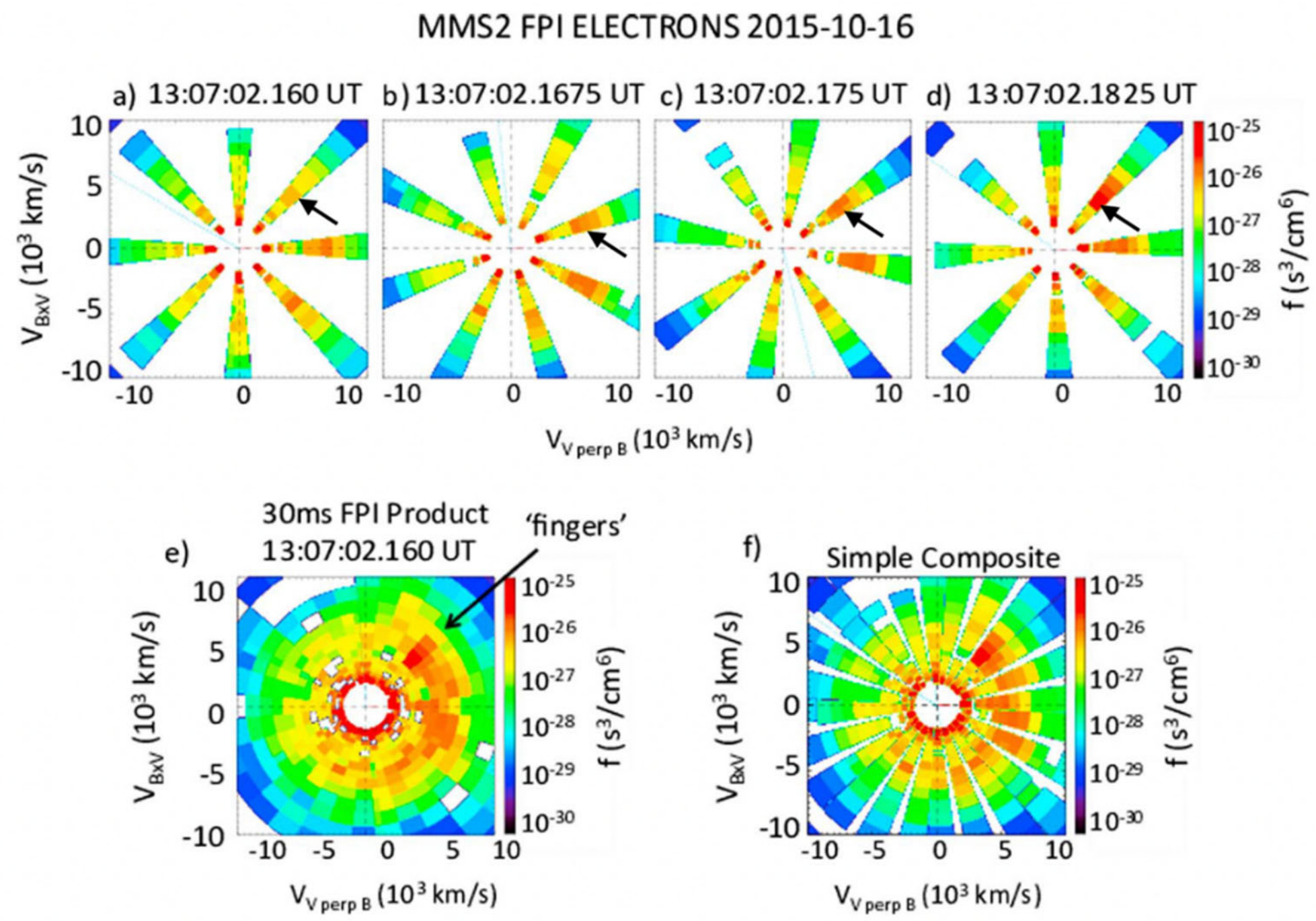

**Fig. 30** The four FPI electrostatic deflection states (top four images), each composed of a full 32-step energy sweep and obtained at 7.5 ms cadence for electrons, are combined to produce a full Level 2 "skymap" (bottom left images) every 30 ms (adapted from Rager et al. 2018).

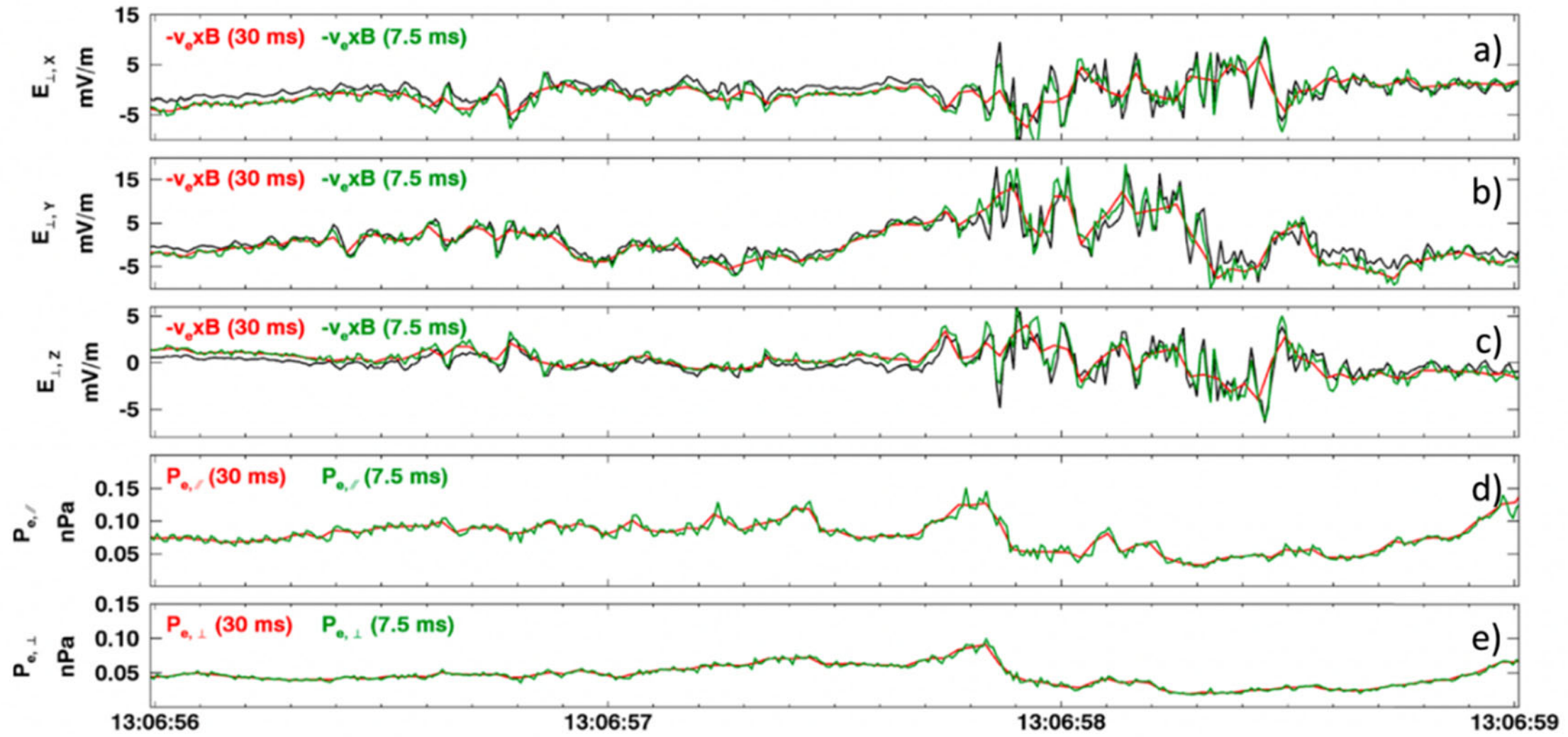

**Fig. 31** The spline interpolation technique does an excellent job of recovering electron bulk velocity at 7.5 ms resolution (green line) (adapted from Rager et al. 2018). The top three panels show three components of the electron convective electric field at 30 ms (red) and 7.5 ms (green) resolutions. The perpendicular electric field measured by the double probe instruments (Ergun et al. 2016; Lindqvist et al. 2016), averaged down to 7.5 ms, is shown in black. The bottom two panels show the electron pressures in the directions parallel and perpendicular to the magnetic field.

Figure 31

shows a validation test in which 7.5 ms perpendicular electron bulk velocity data is compared to the E×B drift velocity (from the electric field and magnetometer experiments averaged down to 7.5 ms). The agreement is excellent, demonstrating that this is a simple, robust method for extracting accurate plasma moments. These "quarter moment" products have been used in many MMS publications (e.g., Phan et al. 2018).

**Appendix C    Tables Summarizing Methods** (attached at the end)

**Acknowledgements**

We thank the reviewers whose comments helped improve this manuscript. The work by H.H. was supported by JSPS Grant-in-aid for Scientific Research KAKENHI 21K03504. Part of the work done while H.H. was at SwRI was supported by NASA Contract No. NNG04EB99C at SwRI. R.E.D. was supported by a NASA grant 80NSSC22K1109. The work by D.B.K. has been supported by the Austrian Science Fund (FWF): I 3506-N27 and Austrian FFG project ASAP15/873685. T.C.L. was supported by NSF award AGS-2000222. T.K.M.N. was supported by the Austrian Research Fund (FWF) P32175-N27. S.M.P was supported by Contract 499935Q. Y.Q. was supported by NASA MMS mission NNG04EB99C. The authors acknowledge the International Space Science Institute in Bern for supporting the international team of the workshop "Magnetic Reconnection: Explosive Energy Conversion in Space Plasmas".

**Compliance with Ethical Standards**

The authors declare they have no conflicts of interest.

**References**

V.A. Andreeva, N.A. Tsyganenko, Journal of Geophysical Research: Space Physics, **121**, 2249–2263 (2016). https://doi.org/10.1002/2015JA022242

M.R. Argall, C.R. Small, S. Piatt, et al., Frontiers in Astronomy and Space Sciences, **7**, 54 (2020). https://doi.org/10.3389/fspas.2020.00054

M.R. Argall, M.H. Barbhuiya, P.A. Cassak, et al., Phys. Plasmas, **29**, 022902 (2022). https://doi.org/10.1063/5.0073248

N. Aunai, M. Hesse, M. Kuznetsova, Phys. Plasmas, **20**, 092903 (2013). https://doi.org/10.1063/1.4820953
D.N. Baker, L. Riesberg, C.K. Pankratz, R.S. Panneton, B.L. Giles, F.D. Wilder, R.E. Ergun, Space Science Reviews, **199**, 545–575 (2016). https://doi.org/10.1007/s11214-014-0128-5
R. Bandyopadhyay, A. Chasapis, W. H. Matthaeus, et al., Phys. Plasmas, **28**, 112305 (2021). https://doi.org/10.1063/5.0071015
M.H. Barbhuiya, P.A. Cassak, Phys. Plasmas, **29**, 122308 (2022). https://doi.org/10.1063/5.0125256
M.H. Barbhuiya, P.A. Cassak, M.A. Shay, et al., Journal of Geophysical Research: Space Physics, **127**, e2022JA030610 (2022). https://doi.org/10.1029/2022JA030610
C. Bard, J. Dorelli, Front. Astron. Space Sci., **8**, 732275 (2021). doi:10.3389/fspas.2021.732275
N. Bessho, L.-J. Chen, M. Hesse, Geophysical Research Letters, **43**(5), 1828–1836 (2016). https://doi.org/10.1002/2016gl067886
N. B essho, L.-J. Chen, S. Wang, M. Hesse, Geophysical Research Letters, **45**, 12142–12152 (2018). https://doi.org/10.1029/2018GL081216
L. Boltzmann, Wiener Berichte, **76**, 373–435 (1877).
J.E. Borovsky, and K. Yakymenko, Journal of Geophysical Research: Space Physics, **122**(3), 2973-2998 (2017). https://doi.org/10.1002/2016JA023625
T. Broeren, K.G. Klein, J.M. TenBarge, I. Dors, O.W. Roberts, D. Verscharen, Front. Astron. Space Sci., **8**, 727076 (2021). https://doi.org/10.3389/fspas.2021.727076
I. Brown, C. Mues, Expert Systems with Applications, **39**, 3446-3453 (2012). https://doi.org/10.1016/j.eswa.2011.09.033
M. Buhmann, Radial Basis Functions: Theory and Implementations, Cambridge Univ. Press, Cambridge, U. K. (2003).
J.L. Burch, T.E. Moore, R.B. Torbert, B.L. Giles, Space Sci. Rev., **199**(1–4), 5-21 (2016a). https://doi.org/10.1007/s11214-015-0164-9
J.L. Burch, R.B. Torbert, T.D. Phan, et al., Science, **352**(6290), aaf2939 (2016b). https://doi.org/10.1126/science.aaf2939
J.L. Burch, J.M. Webster, M. Hesse, et al., Geophysical Research Letters, **47**, e2020GL089082 (2020). https://doi.org/10.1029/2020GL089082
F. Califano, G. Manfredi, F. Valentini, J. Plasma Phys., **82**, 701820603 (2016). https://doi.org/10.1017/S002237781600115X
P.A. Cassak, M.H. Barbhuiya, Phys. Plasmas, **29**, 122306 (2022). https://doi.org/10.1063/5.0125248

P.A. Cassak, M.H. Barbhuiya, H. Liang, M.R. Argall, Phys. Rev. Lett., **130**, 085201 (2023). https://doi.org/10.1103/PhysRevLett.130.085201

S.S. Cerri, Plasma turbulence in the dissipation range-theory and simulations (Universität Ulm) (2016). http://dx.doi.org/10.18725/OPARU-3355

H. Che, J.F. Drake, M. Swisdak, Nature, **474**, 184-187 (2011). https://doi.org/10.1038/nature10091G.-W. Chen, L.-N. Hau, Journal of Geophysical Research: Space Physics, **123**, 7358-7369 (2018). https://doi.org/10.1029/2018JA025842

L.-J. Chen, N. Bessho, B. Lefebvre, et al., Journal of Geophysical Research, **113**, A12213 (2008). https://doi.org/10.1029/2008JA013385

L.-J. Chen, M. Hesse, S. Wang, N. Bessho, W. Daughton, Geophysical Research Letters, **43**, 452–461 (2016). https://doi.org/10.1002/2016GL068243

W. Chen, X. Wang, N.A. Tsyganenko, V.A. Andreeva, V.S. Semenov, Journal of Geophysical Research: Space Physics, **124**, 10141-10152 (2019). https://doi.org/10.1029/2019JA027078

Y. Chen, G. Tóth, H. Hietala, et al., Earth and Space Science, **7**, e2020EA001331 (2020). https://doi.org/10.1029/2020EA001331

I.K. Cheng, N. Achilleos, A. Smith, Front. Astron. Space Sci., **9**, 1016453 (2022). https://doi.org/10.3389/fspas.2022.1016453

D.C. Chin, Opt. Eng., **38**, 606–611 (1999). https://doi.org/10.1117/1.602104

W. Daughton, J. Scudder, H. Karimabadi, Phys. Plasmas, **13**, 072101 (2006). https://doi.org/10.1063/1.2218817

F. De Hoffmann, and E. Teller, Phys. Rev., **80**(4), 692–703 (1950). https://doi.org/10.1103/physrev.80.692

R.E. Denton, B.U.Ö. Sonnerup, H. Hasegawa, et al., Geophys. Res. Lett., **43**, 5589-5596 (2016). https://doi.org/10.1002/2016GL069214

R.E. Denton, B.U.Ö. Sonnerup, C.T. Russell, et al., Journal of Geophysical Research: Space Physics, **123**, 2274-2295 (2018). https://doi.org/10.1002/2017JA024619

R.E. Denton, R.B. Torbert, H. Hasegawa, et al., Journal of Geophysical Research: Space Physics, **125**, e2019JA027481 (2020). https://doi.org/10.1029/2019JA027481

R.E. Denton, R.B. Torbert, H. Hasegawa, et al., J. Geophys. Res. Space Physics, **126**, e2020JA028705 (2021). https://doi.org/10.1029/2020JA028705

R.E. Denton, Y.-H. Liu, H. Hasegawa, R.B. Torbert, W. Li, S.A. Fuselier, J.L. Burch, J. Geophys. Res. Space Physics, **127**, e2022JA030512 (2022). https://doi.org/10.1029/2022JA030512

M.W. Dunlop, T.I. Woodward, Multi-spacecraft discontinuity analysis: Orientation and motion. in *Analysis methods for multi-spacecraft data*, edited by G. Paschmann & P. Daly, pp. 271–306, Switzerland: International Space Science Institute, SR-001 (1998).

M.W. Dunlop, D.J. Southwood, K.-H. Glassmeier, and F.M. Neubauer, Advances in Space Research, **8**, 273-277 (1988). https://doi.org/10.1016/0273-1177(88)90141-X

M.W. Dunlop, A. Balogh, K.-H. Glassmeier, P. Robert, J. Geophys. Res., **107**, 1384 (2002). https://doi.org/10.1029/2001JA005088

M.W. Dunlop, Q.-H. Zhang, Y.V. Bogdanova, et al., Ann. Geophys., **29**, 1683-1697 (2011). https://doi.org/10.5194/angeo-29-1683-2011

J. Egedal, A. Lê, Y. Zhu, et al., Geophysical Research Letters, **37**, L10102 (2010a). https://doi.org/10.1029/2010GL043487

J. Egedal, A. Lê, N. Katz, L.-J. Chen, B. Lefebvre, W. Daughton, A. Fazakerley, Journal of Geophysical Research: Space Physics, **115**, A03214 (2010b). https://doi.org/10.1029/2009JA014650

J. Egedal, A. Le, W. Daughton, et al., Phys. Rev. Lett., **117**, 185101 (2016). https://doi.org/10.1103/PhysRevLett.117.185101

J. Egedal, J. Ng, A Le, et al., Phys. Rev. Lett., **123**, 225101 (2019). https://doi.org/10.1103/PhysRevLett.123.225101

S. Ekawati, D. Cai, Journal of Geophysical Research: Space Physics, **128**, e2021JA029571 (2023). https://doi.org/10.1029/2021JA029571

R.E. Ergun, S. Tucker, J. Westfall, et al., Space Science Reviews, **199**(1-4), 167–188 (2016). https://doi.org/10.1007/s11214-014-0115-x

E. Eriksson, A. Vaivads, D.B. Graham, et al., Geophys. Res. Lett., **45**, 8081– 8090 (2018) https://doi.org/10.1029/2018GL078660

C.P. Escoubet, R. Schmidt, M.L. Goldstein, Space Science Reviews, **79**, 11-32 (1997). https://doi.org/10.1007/978-94-011-5666-0_1

S. Fadanelli, B. Lavraud, F. Califano, et al., Journal of Geophysical Research: Space Physics, **124**, 6850-6868 (2019). https://doi.org/10.1029/2019JA026747

S. Fadanelli, B. Lavraud, F. Califano, et al., Journal of Geophysical Research: Space Physics, **125**, e2020JA028333 (2020). https://doi.org/10.1029/2020JA028333

J.H. Friedman, Ann. Statist., **29**(5), 1189-1232 (2001). https://doi.org/10.1214/aos/1013203451

H.S. Fu, A. Vaivads, Y.V. Khotyaintsev, V. Olshevsky, M. André, J.B. Cao, S.Y. Huang, A. Retinò, G. Lapenta, J. Geophys. Res. Space Physics, **120**, 3758-3782 (2015), https://doi.org/10.1002/2015JA021082

S.A. Fuselier, S.M. Petrinec, K.J. Trattner, Geophysical Research Letters, **27**, 473-476 (2000). https://doi.org/10.1029/1999GL003706

S.A. Fuselier, H.U. Frey, K.J. Trattner, S.B. Mende, J.L. Burch, Journal of Geophysical Research, **107**(A7), 1111 (2002). https://doi.org/10.1029/2001JA900165

S.A. Fuselier, W.S. Lewis, C. Schiff, et al., Space Science Reviews, **199**(1–4), 77–103 (2016). https://doi.org/10.1007/s11214-014-0087-x

S.A. Fuselier, S.M. Petrinec, P.H. Reiff, et al., Space Sci Rev, **220**, 34 (2024). https://doi.org/10.1007/s11214-024-01067-0

K.J. Genestreti, T.K.M. Nakamura, R. Nakamura, et al., Journal of Geophysical Research: Space Physics, **123**, 9130–9149 (2018). https://doi.org/10.1029/2018JA025711

K.J. Genestreti, X. Li, Y.-H. Liu, et al., Phys. Plasmas, **29**, 082107 (2022). https://doi.org/10.1063/5.0090275

D.J. Gershman, A. F-Viñas, J.C. Dorelli, et al., Wave-particle energy exchange directly observed in a kinetic Alfvén-branch wave. Nature Communications, **8**, 14719 (2017). doi:10.1038/ncomms14719.

J.W. Gjerloev, Journal of Geophysical Research: Space Physics, **117**, A09213 (2012). https://doi.org/10.1029/2012JA017683

J.T. Gosling, M.F. Thomsen, S.J. Bame, R.C. Elphic, C.T. Russell, Journal of Geophysical Research, **95**, 8073 (1990). https://doi.org/10.1029/JA095iA06p08073

D.B. Graham, Y.V. Khotyaintsev, M. André, et al., Journal of Geophysical Research: Space Physics, **126**, e2021JA029260 (2021). https://doi.org/10.1029/2021JA029260

D.B. Graham, Y.V. Khotyaintsev, M. André, et al., Nature Communications, **13**(2954) (2022). https://doi.org/10.1038/s41467-022-30561-8

A. Greco, P. Chuychai, W.H. Matthaeus, S. Servidio, P. Dmitruk, Geophysical Research Letters, **35**, L19111 (2008). https://doi.org/10.1029/2008GL035454

A. Greco, F. Valentini, S. Servidio, W.H. Matthaeus, Physical Review E, **86**(6), 066405 (2012). https://doi.org/10.1103/PhysRevE.86.066405

R. Guo, Z. Pu, X. Wang, C. Xiao, J. He, Journal of Geophysical Research: Space Physics, **127**, e2021JA030248 (2022). https://doi.org/10.1029/2021JA030248

H. Hasegawa, Monogr. Environ. Earth Planets, **1**(2), 71-119 (2012). doi:10.5047/meep.2012.00102.0071

H. Hasegawa, B.U.Ö. Sonnerup, M.W. Dunlop, et al., Ann. Geophys., **22**, 1251-1266 (2004). https://doi.org/10.5194/angeo-22-1251-2004

H. Hasegawa, B.U.Ö. Sonnerup, R.E. Denton, et al., Geophys. Res. Lett., **44**, 4566-4574 (2017). https://doi.org/10.1002/2017GL073163

H. Hasegawa, R.E. Denton, R. Nakamura, et al., Journal of Geophysical Research: Space Physics, **124**, 122-138 (2019). https://doi.org/10.1029/2018JA026051

H. Hasegawa, T.K.M. Nakamura, R.E. Denton, Journal of Geophysical Research: Space Physics, **126**, e2021JA029841 (2021). https://doi.org/10.1029/2021JA029841

H. Hasegawa, R.E. Denton, K. Dokgo, et al., Journal of Geophysical Research: Space Physics, **128**, e2022JA031092 (2023). https://doi.org/10.1029/2022JA031092

L.-N. Hau, C.-K. Chang, G.-W. Chen, Astrophys. J., **900**, 97 (2020). https://doi.org/10.3847/1538-4357/aba2d0

L.-N. Hau, T.-D. Phan, B.U.Ö. Sonnerup, G. Paschmann, Geophys. Res. Lett., **20**, 2255-2258 (1993). https://doi.org/10.1029/93GL02491

M. Hesse, K. Schindler, J. Birn, M. Kuznetsova, Phys. Plasmas, **6**, 1781–1795 (1999). https://doi.org/10.1063/1.873436

M. Hesse, T. Neukirch, K. Schindler, M. Kuznetsova, S. Zenitani, Space Sci. Rev., **160**, 3-23 (2011). https://doi.org/10.1007/s11214-010-9740-1

M. Hesse, N. Aunai, D.G. Sibeck, J. Birn, Geophys. Res. Lett., **41**, 8673–8680 (2014). doi:10.1002/2014GL061586

S.V. Heuer, K.J. Genestreti, T.K.M. Nakamura, R.B. Torbert, J.L. Burch, R. Nakamura, Geophysical Research Letters, **49**, e2022GL100652. https://doi.org/10.1029/2022GL100652

M. Hoshino, T. Mukai, T. Terasawa, I. Shinohara, Journal of Geophysical Research: Space Physics, **106**, 25979–25997 (2001). https://doi.org/10.1029/2001JA900052

Q. Hu, B.U.Ö. Sonnerup, J. Geophys. Res., **107**(A7), (2002). https://doi.org/10.1029/2001JA000293

K.-J. Hwang, E. Choi, K. Dokgo, et al., Geophysical Research Letters, **46**, 6287-6296 (2019). https://doi.org/10.1029/2019GL082710

K.-J. Hwang, R. Nakamura, J.P. Eastwood, J.P. et al., Space Sci Rev, **219**, 71 (2023). https://doi.org/10.1007/s11214-023-01010-9

L.K. Jian, C.T. Russell, J.G. Luhmann, D. Curtis, P. Schroeder, AIP Conference Proceedings, **1539**, 195–198 (2013). https://doi.org/10.1063/1.4811021

H. Karimabadi, W. Daughton, J. Scudder, Geophys. Res. Lett., **34**, L13104 (2007). https://doi.org/10.1029/2007GL030306

R.L. Kaufmann, W.R. Paterson, Journal of Geophysical Research: Space Physics, **114**, A00D04 (2009). https://doi.org/10.1029/2008JA014030

A.V. Khrabrov, B.U.Ö. Sonnerup, DeHoffmann-Teller Analysis. in *Analysis methods for multi-spacecraft data*, edited by G. Paschmann & P. Daly, pp. 221–248, Bern, Switzerland: International Space Science Institute, SR-001 (1998).

A.S. Kingsep, K.V. Chukbar, V.V. Yan'kov, in *Reviews of Plasma Physics*, edited by B.B. Kadomtsev, **16**, 243–288, Consultants Bureau, New York (1990).

E. Kobel, E.O. Flückiger, Journal of Geophysical Research, **99**, 23617-23622 (1994). https://doi.org/10.1029/94JA01778

D.B. Korovinskiy, A.V. Divin, V.S. Semenov, N.V. Erkaev, S.A. Kiehas, I.V. Kubyshkin, Phys. Plasmas, **27**, 082905 (2020). https://doi.org/10.1063/5.0015240.

D.B. Korovinskiy, S.A. Kiehas, E.V. Panov, V.S. Semenov, N.V. Erkaev, A.V. Divin, I.V. Kubyshkin, Journal of Geophysical Research: Space Physics, **126**, e2020JA029045 (2021). https://doi.org/10.1029/2020JA029045

D. Korovinskiy, E. Panov, R. Nakamura, S. Kiehas, M. Hosner, D. Schmid, I. Ivanov, Front. Astron. Space Sci., **10**, 1069888 (2023). https://doi.org/10.3389/fspas.2023.1069888

A. Lalti, Y.V. Khotyaintsev, A.P. Dimmock, A. Johlander, D.B. Graham, V. Olshevsky, Journal of Geophysical Research: Space Physics, **127**, e2022JA030454 (2022). https://doi.org/10.1029/2022JA030454

A. Le, W. Daughton, L.-J. Chen, J. Egedal, Geophys. Res. Lett., **44**, 2096–2104 (2017). https://doi.org/10. 1002/2017GL072522

A. Le, W. Daughton, O. Ohia, L.-J. Chen, Y.-H. Liu, S. Wang, W.D. Nystrom, R. Bird, Physics of Plasmas, **25**, 062103 (2018). https://doi.org/10.1063/1.5027086

Y. Lecun, L. Bottou, Y. Bengio, P. Haffner, in Proceedings of the IEEE, 86(11), 2278-2324 (1998). https://doi.org/10.1109/5.726791

Q. Lenouvel, V. Génot, P. Garnier, et al., Earth and Space Science, **8**, e2020EA001530 (2021). https://doi.org/10.1029/2020EA001530

Q. Lenouvel, Identification by machine learning and analysis of electron diffusion regions at the Earth's magnetopause observed by MMS, PhD thesis (2022). https://theses.hal.science/tel-04075287v1

T.C. Li, Y.-H. Liu, M. Hesse, Y. Zou, J. Geophys. Res., **125**, e2019JA027094 (2020). https://doi.org/10.1029/2019JA027094

T.C. Li, Y.-H. Liu, Y. Qi, Astrophysical Journal Letters, **909**, L28 (2021). https://doi.org/10.3847/2041-8213/abea0b

T.C. Li, Y.-H. Liu, Y. Qi, M. Zhou, Phys. Rev. Lett., 131, 085201 (2023). https://doi.org/10.1103/PhysRevLett.131.085201

H. Liang, P.A. Cassak, S. Servidio, et al., Physics of Plasmas, **26**, 082903 (2019). https://doi.org/10.1063/1.5098888

H. Liang, M.H. Barbhuiya, P.A. Cassak, O. Pezzi, S. Servidio, F. Valentini, G.P. Zank, Journal of Plasma Physics, **86**, 825860502 (2020).

https://doi.org/10.1017/S0022377820001270

M. Lindberg, A. Vaivads, S. Raptis, P.-A. Lindqvist, B.L. Giles, D.J. Gershman, Entropy, **24**, 745 (2022). https://doi.org/10.3390/e24060745

P.-A. Lindqvist, G. Olsson, R.B. Torbert, et al., Space Science Reviews, **199**(1-4), 137–165 (2016). https://doi.org/10.1007/s11214-014-0116-9

Y.-H. Liu, M. Hesse, Phys. Plasmas, **23**, 060704 (2016). https://doi.org/10.1063/1.4954818

Y.-H. Liu, M. Hesse, F. Guo, W. Daughton, H. Li, P.A. Cassak, M.A. Shay, Physical Review Letters, **118**(8), 085101 (2017). https://doi.org/10.1103/PhysRevLett.118.085101

Y.-H. Liu, M. Hesse, F. Guo, H. Li, and T. K. M. Nakamura, Physics of Plasmas, **25**, 080701 (2018a). https://doi.org/10.1063/1.5042539

Y.-H. Liu, M. Hesse, T.C. Li, M. Kuznetsova, A. Le, J. Geophys. Res., **123**, 4908 (2018b). https://doi.org/10.1029/2018JA025410

Y.-H. Liu, T.C. Li, M. Hesse, W.J. Sun, J. Liu, J.L. Burch, J.A. Slavin, K. Huang, J. Geophys. Res., **124**, 2819 (2019). https://doi.org/10.1029/2019JA026539

Y.-H. Liu, H. Hesse, K. Genestreti, et al., submitted to Space Sci. Rev. (2024). arXiv:2406.00875

J.G. Lyon, J.A. Fedder, C.M. Mobarry, J. Atmos. Sol. Terr. Phys., **66**, 1333–1350 (2004). https://doi.org/10.1016/j.jastp.2004.03.020

A.T. Marshall, J.L. Burch, P.H. Reiff, J.M. Webster, R.B. Torbert, R.E. Ergun, et al., J. Geophys Res: Space Physics, **125**, e2019JA027296 (2020). https://doi.org/10.1029/2019JA027296

A.T. Marshall, J.L. Burch, P.H. Reiff, et al., Physics of Plasmas, **29**, 012905 (2022). https://doi.org/10.1063/5.0071159

D. Maturana, S. Scherer, in *IEEE/RSJ International Conference on Intelligent Robots and Systems (IROS)*, 922–928 (2015). https://doi.org/10.1109/IROS.2015.7353481

R.L. McPherron, C.T. Russell, M.P. Aubry, J. Geophys. Res., **78**(16), 3131-3149 (1973). https://doi.org/10.1029/JA078i016p03131

B. Michotte de Welle, N. Aunai, G. Nguyen, B. Lavraud, V. Génot, A. Jeandet, R. Smets, Journal of Geophysical Research: Space Physics, **127**, e2022JA030996 (2022). https://doi.org/10.1029/2022JA030996

D. Moseev, M. Salewski, Physics of Plasmas, **26**, 020901 (2019). https://doi.org/10.1063/1.5085429

C. Mouhot, C. Villani, Acta Math., **207**, 29–201 (2011). https://doi.org/10.1007/s11511-011-0068-9

T. Nagai, I. Shinohara, M. Fujimoto, A. Matsuoka, Y. Saito, T. Mukai, J. Geophys. Res., **116**, A04222 (2011), https://doi.org/10.1029/2010JA016283

T. Nagai, S. Zenitani, I. Shinohara, R. Nakamura, M. Fujimoto, Y. Saito, T. Mukai, Journal of Geophysical Research: Space Physics, **118**, 7703–7713 (2013). https://doi.org/10.1002/2013JA019135

R. Nakamura, T. Nagai, J. Birn, V.A. Sergeev, O. Le Contel, V. Varsani, A., et al., Earth Planets Space, **69**(1), 129 (2017). https://doi.org/10.1186/s40623-017-0707-2

T.K.M. Nakamura, R. Nakamura, A. Varsani, K.J. Genestreti, W. Baumjohann, and Y.-H. Liu, Geophysical Research Letters, **45**, 3829–3837 (2018a). https://doi.org/10.1029/2018GL078340

T.K.M. Nakamura, K.J. Genestreti, Y.-H. Liu, et al., Journal of Geophysical Research: Space Physics, **123**, 9150–9168 (2018b). https://doi.org/10.1029/2018JA025713

Y. Narita, Nonlin. Processes Geophys., **24**, 203-214 (2017). https://doi.org/10.5194/npg-24-203-2017

J. Ng, J. Egedal, A. Le, W. Daughton, L.-J. Chen, Phys. Rev. Lett., **106**, 065002 (2011). https://doi.org/10.1103/PhysRevLett.106.065002

J. Ng, L.-J. Chen, N. Bessho, J. Shuster, B. Burkholder, J. Yoo, Geophys. Res. Lett., **49**, e2022GL099544 (2022). https://doi.org/10.1029/2022GL099544

G. Nguyen, N. Aunai, B. Michotte de Welle, A. Jeandet, B. Lavraud, D. Fontaine, Journal of Geophysical Research: Space Physics, **127**, e2021JA029773 (2022a). https://doi.org/10.1029/2021JA029773

G. Nguyen, N. Aunai, B. Michotte de Welle, A. Jeandet, B. Lavraud, D. Fontaine, Journal of Geophysical Research: Space Physics, **127**, e2021JA029774 (2022b). https://doi.org/10.1029/2021JA029774

G. Nguyen, N. Aunai, B. Michotte de Welle, A. Jeandet, B. Lavraud, D. Fontaine, Journal of Geophysical Research: Space Physics, **127**, e2021JA030112 (2022c). https://doi.org/10.1029/2021JA030112

G. Nguyen, N. Aunai, B. Michotte de Welle, A. Jeandet, B. Lavraud, D. Fontaine, Journal of Geophysical Research: Space Physics, **127**, e2021JA029776 (2022d). https://doi.org/10.1029/2021JA029776

D.R. Nicholson, Introduction to Plasma Theory. New York: Wiley (1983).

C. Norgren, L.-J. Chen, N. Bessho, et al., submitted to Space Sci. Rev. (2024).

M. Oka, T.-D. Phan, M. Øieroset, V. Angelopoulos, Journal of Geophysical Research: Space Physics, **121**, 1955–1968 (2016). https://doi.org/10.1002/2015JA022040

M. Oka, J. Birn, J. Egedal, et al., Space Sci. Rev., **219**, 75 (2023). https://doi.org/10.1007/s11214-023-01011-8

V. Olshevsky, D.I. Pontin, B. Williams, et al., Astron. Astrophys., **644**, A150 (2020). https://doi.org/10.1051/0004-6361/202039182

V. Olshevsky, Y.V. Khotyaintsev, A. Lalti, et al., Journal of Geophysical Research: Space Physics, **126**, e2021JA029620 (2021). https://doi.org/10.1029/2021JA029620

T.G. Onsager, M.F. Thomsen, J.T. Gosling, S.J. Bame, Geophysical Research Letters, **17**(11), 1837–1840 (1990). https://doi.org/10.1029/GL017i011p01837

T.G. Onsager, M.F. Thomsen, R.C. Elphic, J.T. Gosling, Journal of Geophysical Research, **96**, 20999-21011 (1991). https://doi.org/10.1029/91JA01983

G. Paschmann, P.W. Daly, Analysis methods for multi-spacecraft data. ISSI Scientific Report SR-001, ESA Publ. Div., Noordwijk, Netherlands (1998)

G. Paschmann, P.W. Daly, Multi-Spacecraft Analysis Methods Revisited. ISSI Scientific Report SR-008, ESA Publ. Div., Noordwijk, Netherlands (2008)

G. Paschmann, S. Schwartz, C.P. Escoubet, S. Haaland, Outer magnetospheric boundaries: Cluster results, Space Sciences Series of ISSI, **20**, Springer (2005).

G. Paschmann, M. Øieroset, T.D. Phan, Space Sci. Rev., **47**, 309-341 (2013). https://doi.org/10.1007/978-1-4899-7413-6_12

D.S. Payne, C.J. Farrugia, R.B. Torbert, K. Germaschewski, A.R. Rogers, M.R. Argall, Phys. Plasmas, **28**, 112901 (2021). https://doi.org/10.1063/5.0068317

T.D. Phan, C.P. Escoubet, L. Rezeau, et al., Space Sci. Rev., **118**, 367-424 (2005). https://doi.org/10.1007/s11214-005-3836-z

T.D. Phan, H. Hasegawa, M. Fujimoto, M. Øieroset, T. Mukai, R.P. Lin, W. Paterson, Geophysical Research Letters, **33**, L09104 (2006). https://doi.org/10.1029/2006GL025756

T.D. Phan, M.A. Shay, J.P. Eastwood, V. Angelopoulos, M. Øieroset, M. Oka, M. Fujimoto, Space Science Reviews, **199**, 631–650 (2015). https://doi.org/10.1007/s11214-015-0150-2

T.D. Phan, J.P. Eastwood, P.A. Cassak, et al., Geophysical Research Letters, **43**(12), 6060–6069 (2016). https://doi.org/10.1002/2016GL069212

T.D. Phan, J.P. Eastwood, M.A. Shay, et al., Nature, **557**, 202-206 (2018). https://doi.org/10.1038/s41586-018-0091-5

C. Pollock, T.E. Moore, A. Jacques, et al., Space Science Reviews, **199**(1-4), 331–406 (2016). https://doi.org/10.1007/s11214-016-0245-4

D.I. Pontin, E.R. Priest, Living Rev. Sol. Phys., **19**, 1 (2022). https://doi.org/10.1007/s41116-022-00032-9

L. Price, M. Swisdak, J.F. Drake, P.A. Cassak, J.T. Dahlin, and R.E. Ergun, Geophys. Res. Lett. **43**, 6020–6027 (2016). https://doi.org/10.1002/2016GL069578

L. Price, M. Swisdak, J.F. Drake, and D.B. Graham, Journal of Geophysical Research: Space Physics, **125**(4) (2020). https://doi.org/10. 1029/2019JA027498

P.L. Pritchett, J. Geophys. Res., **106**, 3783–3798 (2001). https://doi.org/10.1029/1999JA001006

Y. Qi, T.C. Li, C.T. Russell, R.E. Ergun, Y.-D. Jia, M. Hubbert, Astrophysical Journal Letters, **926**, L34 (2022). https://doi.org/10.3847/2041-8213/ac5181

J. Qiu, Q. Hu, T.A. Howard, et al., Astrophys. J., **659**(1), 758–772 (2007). doi:10.1086/512060

J. Raeder, W.D. Cramer, K. Germaschewski, J. Jensen, Space Science Reviews, **206**, 601-620 (2017). https://doi.org/10.1007/s11214-016-0304-x

A.C. Rager, J.C. Dorelli, D.J. Gershman, et al., Geophysical Research Letters, **45**, 578–584 (2018). https://doi.org/10.1002/2017GL076260

M. Raissi, P. Perdikaris, G.E. Karniadakis, J. Comput. Phys., **378**, 686–707 (2019). https://doi.org/10.1016/j.jcp.2018.10.045

P.H. Reiff, T.W. Hill, J.L. Burch, Journal of Geophysical Research, **82**, 479–491 (1977). doi:10.1029/JA082i004p00479

P.H. Reiff, A.G. Daou, S.Y. Sazykin et al., Geophys. Res. Lett., **43,** 7311–7318 (2016). https://doi.org/10.1002/2016GL069154

P.H. Reiff, J.M. Webster, A.G. Daou, et al., CCMC Modeling of Magnetic Reconnection in Electron Diffusion Regions. in *Space Weather of the Heliosphere: Processes and Forecasts*, Proceedings IAU Symposium No. 335, p. 142-146, Eds. C. Foullon and O. Malandraki (2017). https://doi.org/10.1017/S1743921317010845

P.H. Reiff, A. Marshall, J. Webster, S. Sazykin, C.T. Russell, L. Rastaetter, MMS observations and CCMC modeling of field line stretching at separator lines. Fall AGU e-Lightning poster (2018). https://agu2018fallmeeting-agu.ipostersessions.com/default.aspx?s=B2-10-20-70-BD-2D-A2-4E-35-27-A4-FE-DC-C0-6D-DA, doi: 10.1002/essoar.10502075.1

P.I. Reyes, V.A. Pinto, P. S. Moya, Space Weather, **19**(9), e2021SW002766 (2021). https://doi.org/10.1029/2021SW002766

L. Rezeau, G. Belmont, R. Manuzzo, N. Aunai, J. Dargent, J. Geophys. Res. Space Phys. **123**(1), 227–241 (2018). https://doi.org/10.1002/2017ja024526

C.T. Russell, B.J. Anderson, W. Baumjohann, et al., Space Science Reviews, **199**(1-4), 189–256 (2016). https://doi.org/10.1007/s11214-014-0057-3

H. Rosenbauer, H. Grünwaldt, M.D. Montgomery, G. Paschmann, N. Sckopke, Journal of Geophysical Research, 80, 2723-2737 (1975). doi:10.1029/JA080i019p02723

D. Rumelhart, G. Hinton, R. Williams, Nature, 323, 533-536 (1986). https://doi.org/10.1038/323533a0

J. Safrankova, Z. Nemecek, S. Dusik, L. Prech, D.G. Sibeck, N.N. Borodkova, Annales Geophysicae, 20, 301-309 (2002). https://doi.org/10.5194/angeo-20-301-2002

H. Schlichting, Boundary Layer Theory, McGraw-Hill, p.817 (1979).

J.M. Schroeder, J. Egedal, G. Cozzani, Y.V. Khotyaintsev, W. Daughton, R.E. Denton, J.L. Burch, Geophysical Research Letters, **49**, e2022GL100384 (2022). https://doi.org/10.1029/2022GL100384

J. Scudder, W. Daughton, J. Geophys. Res., **113**, A06222 (2008). https://doi.org/10.1029/2008JA013035

J. Scudder, W. Daughton, Journal of Geophysical Research: Space Physics, **113**, 1–16 (2008). https://doi.org/10.1029/2008JA013035

J.D. Scudder, R.D. Holdaway, W.S. Daughton, H. Karimabadi, V. Roytershteyn, C.T. Russell, J.Y. Lopez, Physical Review Letters, **108**, 225005 (2012). https://doi.org/10.1103/PhysRevLett.108.225005

V. Sergeev, V. Angelopoulos, M. Kubyshkina, et al., Journal of Geophysical Research: Space Physics, **116**, A00I26 (2011). https://doi.org/10.1029/2010JA015689

S. Servidio, A. Chasapis, W.H. Matthaeus, et al., Phys. Rev. Lett., **119**, 205101 (2017). https://doi.org/10.1103/PhysRevLett.119.205101

A. Settino, Y.V. Khotyaintsev, D.B. Graham, D. Perrone, F. Valentini, Journal of Geophysical Research: Space Physics, **127**, e2021JA029758 (2022). https://doi.org/10.1029/2021JA029758

M.A. Shay, J.F. Drake, M. Swisdak, Phys. Rev. Lett., **99**, 155002 (2007). https://doi.org/10.1103/PhysRevLett.99.155002

E.G. Shelley, A.G. Ghielmetti, H. Balsiger, et al., Space Sci. Rev., **71**, 497–530 (1995). https://doi.org/10.1007/BF00751339

C. Shen, X. Li, M. Dunlop, Q.Q. Shi, Z.X. Liu, E. Lucek, Z.Q. Chen, J. Geophys. Res., **112**, A06211 (2007). https://doi.org/10.1029/2005JA011584

Q.Q. Shi, C. Shen, Z.Y. Pu, et al., Geophys. Res. Lett., **32**, L12105 (2005). https://doi.org/10.1029/2005GL022454

Q.Q. Shi, C. Shen, M.W. Dunlop, et al., Geophys. Res. Lett. **33**, L08109 (2006). https://doi.org/10.1029/2005GL025073

Q.Q. Shi, A.M. Tian, S.C. Bai, et al., Space Science Reviews, **215**(4), 35 (2019). https://doi.org/10.1007/s11214-019-0601-2

J.R. Shuster, L.-J. Chen, W. Daughton, et al., Geophysical Research Letters, **41**, 5389–5395 (2014). https://doi.org/10.1002/2014GL060608

J.R. Shuster, L.-J. Chen, M. Hesse, M.R. Argall, W. Daughton, R.B. Torbert, N. Bessho, Geophysical Research Letters, **42**, 2586–2593 (2015). https://doi.org/10.1002/2015GL063601

J.R. Shuster, M.R. Argall, R.B. Torbert, et al., Geophysical Research Letters, **44**, 1625–1633 (2017). https://doi.org/10.1002/2017GL072570

J.R. Shuster, D.J. Gershman, L.-J. Chen, et al., Geophysical Research Letters, **46**, 7862–7872 (2019). https://doi.org/10.1029/2019GL083549

J.R. Shuster, D.J. Gershman, J.C. Dorelli, et al., Nat. Phys. **17**, 1056–1065 (2021a). https://doi.org/10.1038/s41567-021-01280-6

J.R. Shuster, N. Bessho, S. Wang, J. Ng, Phys. Plasmas, **28**, 122902 (2021b). https://doi.org/10.1063/5.0069559

J.R. Shuster, D.J. Gershman, B.L. Giles, et al., Journal of Geophysical Research: Space Physics, **128**, e2022JA030949 (2023). https://doi.org/10.1029/2022JA030949

D.G. Sibeck, R.E. Lopez, E.C. Roelof, Journal of Geophysical Research, **96**, 5489-5495 (1991). doi:10.1029/90JA02464

M.I. Sitnov, A.S. Sharma, K. Papadopoulos, D. Vassiliadis, Phys. Rev. E, **65**, 016116 (2001). https://doi.org/10.1103/PhysRevE.65.016116

M.I. Sitnov, N.A. Tsyganenko, A.Y. Ukhorskiy, P.C. Brandt, Journal of Geophysical Research: Space Physics, **113**(A7), A07218 (2008). https://doi.org/10.1029/2007JA013003

M.I. Sitnov, G.K. Stephens, N.A. Tsyganenko, A.Y. Ukhorskiy, S. Wing, H. Korth, B.J. Anderson, Spatial structure and asymmetries of magnetospheric currents inferred from high-resolution empirical geomagnetic field models. In *Dawn-dusk asymmetries in planetary plasma environments*, edited by S. Haaland, A. Runov and C. Forsyth, p. 199-212, American Geophysical Union (AGU) (2017). https://doi.org/10.1002/9781119216346.ch15

M.I. Sitnov, G.K. Stephens, N.A. Tsyganenko, et al., Journal of Geophysical Research: Space Physics, **124**(11), 8427-8456 (2019). https://doi.org/10.1029/2019JA027037

M.I. Sitnov, G.K. Stephens, N.A. Tsyganenko, et al., Space Weather, **18**, e2020SW002561 (2020). https://doi.org/10.1029/2020SW002561

M. Sitnov, G. Stephens, T. Motoba, M. Swisdak, Frontiers in Physics, **9** (2021). https://doi.org/10.3389/fphy.2021.644884

B.U.Ö. Sonnerup, L.J. Cahill Jr., J. Geophys. Res., **72**(1), 171–183 (1967). doi:10.1029/JZ072i001p00171

B.U.Ö. Sonnerup, M. Guo, Geophys. Res. Lett., **23**(25) 3679-3682 (1996)

B.U.Ö. Sonnerup, M. Scheible, Minimum and maximum variance analysis. in *Analysis methods for multi-spacecraft data*, edited by G. Paschmann and P.W. Daly, ISSI. Sci. Rep. SR-001 (pp. 185–220). Noordwijk, Netherlands: ESA Publ. (1998)

B.U.Ö. Sonnerup, W.-L. Teh, J. Geophys. Res., **113**(A5), A05202 (2008). https://doi.org/10.1029/2007JA012718

B.U.Ö. Sonnerup, W.-L. Teh, J. Geophys. Res., **114**, A04206 (2009). https://doi.org/10.1029/2008JA013897

B.U.Ö. Sonnerup, S. Haaland, G. Paschmann, M.W. Dunlop, H. Rème, A. Balogh, J. Geophys. Res., **111**, A05203 (2006a). https://doi.org/10.1029/2005JA011538

B.U.Ö. Sonnerup, H. Hasegawa, W.-L. Teh, L.-N. Hau, J. Geophys. Res. Space Physics, **111**, A09204 (2006b). https://doi.org/10.1029/2006JA011717

B.U.Ö. Sonnerup, W.-L. Teh, H. Hasegawa, Grad-Shafranov and MHD reconstructions. in *Multi-spacecraft analysis methods revisited*, edited by G. Paschmann and P.W. Daly, ISSI SR-008, pp. 81-90, ESA Publications Division (2008)

B.U.Ö. Sonnerup, H. Hasegawa, R.E. Denton, T.K.M. Nakamura, J. Geophys. Res. Space Physics, **121**(5), 4279–4290 (2016). https://doi.org/10.1002/2016ja022430

J.C. Spall, Johns Hopkins APL Tech. Dig., **19**, 482–492 (1998).

J.C. Spall, Introduction to Stochastic Search and Optimization: Estimation, Simulation, and Control. New York: John Wiley & Sons, 595 (2003).

G.K. Stephens, M.I. Sitnov, Frontiers in Physics, **9**, 653111 (2021). https://doi.org/10.3389/fphy.2021.653111

G.K. Stephens, M.I. Sitnov, H. Korth, N.A. Tsyganenko, S. Ohtani, M. Gkioulidou, A.Y. Ukhorskiy, Journal of Geophysical Research: Space Physics, **124**(2), 1085-1110 (2019). https://doi.org/10.1029/2018JA025843

G.K. Stephens, S.T. Bingham, M.I. Sitnov, et al., Space Weather, **18**(12), e2020SW002583 (2020). https://doi.org/10.1029/2020SW002583

G.K. Stephens, M.I. Sitnov, R.S. Weigel, et al., Journal of Geophysical Research: Space Physics, **128**, e2022JA031066 (2023). https://doi.org/10.1029/2022JA031066

D.P. Stern, Reviews of Geophysics and Space Physics, **14**(2), 200 (1976). https://doi.org/10.1029/RG014i002p00199

I. Svenningsson, E. Yodradnova, Y.V. Khotyaintsev, M. André, G. Cozzani, EGU General Assembly 2023, EGU23-8664 (2023). https://doi.org/10.5194/egusphere-egu23-8664

M. Swisdak, Geophys. Res. Lett., **43**, 43-49 (2016). https://doi.org/10.1002/2015GL066980

B.-B. Tang, W.Y. Li, D.B. Graham, et al., Geophysical Research Letters, **46**, 3024-3032 (2019). https://doi.org/10.1029/2019GL082231

X. Tang, C. Cattell, J. Dombeck, et al., Geophysical Research Letters, **40**, 2884–2890 (2013). https://doi.org/10.1002/grl.50565

W.-L. Teh, Journal of Geophysical Research: Space Physics, **124**, 1644–1650 (2019). https://doi.org/10.1029/2018JA026416

W.-L. Teh, S. Zenitani, Earth and Space Science, **7**, e2020EA001449 (2020). https://doi.org/10.1029/2020EA001449

T. Terasawa, H. Kawano, I. Shinohara, et al., J. Geomagn. Geoelectr., **48**(5–6), 603–614 (1996)

A.M. Tian, K. Xiao, A.W. Degeling, Q.Q. Shi, J.-S. Park, M. Nowada, T. Pitkanen, Astrophys. J., **889**, 35 (2020). https://doi.org/10.3847/1538-4357/ab6296

S. Toledo-Redondo, M. André, Y.V. Khotyaintsev, et al., Geophysical Research Letters, **43**(13), 6759–6767 (2016). https://doi.org/10.1002/2016gl069877

R.B. Torbert, C.T. Russell, W. Magnes, et al., Space Science Reviews, **199**(1-4), 105–135 (2016a). https://doi.org/10.1007/s11214-014-0109-8

R.B. Torbert, H. Vaith, M. Granoff, et al., Space Sci. Rev., **199**, 285-305 (2016b). https://doi.org/10.1007/s11214-015-0182-7

R.B. Torbert, J.L.Burch, M.R. Argall, et al., J. Geophys. Res.: Space Physics, **122**, 11901–11916 (2017). https://doi.org/10.1002/2017JA024579

R.B. Torbert, J.L. Burch, T.D. Phan, et al., Science, **362**(6421), 1391–1395 (2018). https://doi.org/10.1126/science.aat2998

R.B. Torbert, I. Dors, M.R. Argall, et al., Geophysical Research Letters, **47**, e2019GL085542 (2020). https://doi.org/10.1029/2019GL085542

G. Tóth, I.V. Sokolov, T. Gombosi, et al., J. Geophys. Res., **110**, A12226 (2005). doi:10.1029/2005JA011126

K.J. Trattner, S.A. Fuselier, S.M. Petrinec, T.K. Yeoman, C. Mouikis, H. Kucharek, H. Rème, Journal of Geophysical Research, **110**, A04207 (2005). https://doi.org/10.1029/2004JA010722

K.J. Trattner J.S. Mulcock, S.M. Petrinec, S.A. Fuselier, Journal of Geophysical Research, **112**, A08210 (2007). https://doi.org/10.1029/2007JA012270

K.J. Trattner, S.M. Petrinec, S.A. Fuselier, N. Omidi, D.G. Sibeck, Journal of Geophysical Research, **117**, A01213 (2012). https://doi.org/10.1029/2011JA017080

K.J. Trattner, J.L. Burch, R.E. Ergun, et al., Journal of Geophysical Research, **122**, 11991– 12005 (2017). https://doi.org/10.1002/2017JA024488

K.J. Trattner, J.L. Burch, R.E. Ergun, et al., Journal of Geophysical Research, Space

Physics, **123**, 10177– 10188 (2018). https://doi.org/10.1029/2018JA026081

K.J. Trattner, S.M. Petrinec, S.A. Fuselier, Space Science Review, **217**, 41 (2021). https://doi.org/10.1007/s11214-021-00817-8

L. Trenchi, M.F. Marcucci, G. Pallocchia, et al., Journal of Geophysical Research, **113**, A07S10 (2008). https://doi.org/10.1029/2007JA012774

L. Trenchi, M.F. Marcucci, G. Pallocchia, et al., Mem. Soc. Astron. Ital., **80**, 287 (2009)

N. A. Tsyganenko, Journal of Geophysical Research, **100**, 5599-5612 (1995)

N. Tsyganenko, Planetary and Space Science, **39**(4), 641–654 (1991). http://dx.doi.org/10.1016/0032-0633(91)90058-I

N.A. Tsyganenko, M.I. Sitnov, Journal of Geophysical Research: Space Physics, **110**, A03208 (2005). https://doi.org/10.1029/2004JA010798

N.A. Tsyganenko, M.I. Sitnov, Journal of Geophysical Research: Space Physics, **112**, A06225 (2007). https://doi.org/10.1029/2007JA012260

N.A. Tsyganenko, V.A. Andreeva, Journal of Geophysical Research: Space Physics, **121**, 10,786–10,802 (2016). https://doi.org/10.1002/2016JA023217

A. Varsani, R. Nakamura, V.A. Sergeev, et al., Journal of Geophysical Research: Space Physics, **122**, 10891–10909 (2017). https://doi.org/10.1002/2017JA024547

V.M. Vasyliunas, Rev. Geophys., **13**, 303-336 (1975). https://doi.org/10.1029/RG013i001p00303

J.M. Webster, J.L. Burch, P.H. Reiff, et al., Journal of Geophysical Research: Space Physics, **123**(6), 4858–4878 (2018). https://doi.org/10.1029/2018JA025245

S. Wellenzohn, R Nakamura, T.K.M. Nakamura, et al., Journal of Geophysical Research: Space Physics, **126**, e2020JA028917 (2021). https://doi.org/10.1029/2020JA028917

D. Wettschereck, D.W. Aha, T. Mohri, Artificial Intelligence Review, **11**(1), 273-314 (1997). https://doi.org/10.1023/A:1006593614256

F.D. Wilder, R.E. Ergun, J.L. Burch, et al., Journal of Geophysical Research: Space Physics, **123**, 6533-6547 (2018). https://doi.org/10.1029/2018JA025529

Y. Yang, W.H. Matthaeus, T.N. Parashar, et al., Phys. Rev. E, **95**, 061201(R) (2017). https://doi.org/10.1103/PhysRevE.95.061201

S. Zenitani, M. Hesse, A. Klimas, M. Kuznetsova, Phys. Rev. Lett., **106**, 195003 (2011). https://doi.org/10.1103/PhysRevLett.106.195003

S. Zenitani, M. Hesse, A. Klimas, C. Black, M. Kuznetsova, Phys. Plasmas, **18**, 122108 (2011). https://doi.org/10.1063/1.3662430

S. Zenitani, I. Shinohara, T. Nagai, Geophys. Res. Lett., **39**, L11102 (2012). https://doi.org/10.1029/2012GL051938

X. Zhu, J.C. Spall, Int. J. Adapt. Control. Signal. Process., **16**, 397–409 (2002). doi:10.1002/acs.715

X. Zhu, I.J. Cohen, B.H. Mauk, R. Nikoukar, D.L. Turner, R.B. Torbert, Front. Astron. Space Sci., **9**, 878403 (2022). doi:10.3389/fspas.2022.878403

**Appendix C Tables Summarizing Methods Reviewed in the Paper**

**Table 1** Methods for large-scale context

| Method | # of spacecraft (SC) needed | Underlying theory, concept, or name of data mining method | Input | Output | Assumptions | Spatial or temporal scale of interest | Other requirements if any | References |
|---|---|---|---|---|---|---|---|---|
| Maximum magnetic shear model | 1-SC in solar wind | Time-of-flight analysis | ion moments ($n$, $\mathbf{u}_\mathrm{i}$), **B** in solar wind | Magnetopause magnetic shear plot; dayside reconnection location | | Entire magnetopause | T96 & IMF draping models | Trattner et al. (2007); Kobel and Flückiger (1994); Tsyganenko (1995) |
| Global MHD model[1] | 1-SC in solar wind | MHD equations | MHD parameters in solar wind | MHD quantities everywhere in space and time | MHD | MHD | | Tóth et al. (2005); see the CCMC website[1] for other codes |
| Data mining (DM) reconstruction [2] | Multi-mission | DM using distance weighted kNN method and basis function | Magnetometer archives; Solar wind $u_{sw}B_z$; geomagnetic | 3D magnetic field parametrized by ~$10^3$ | Magnetostatic | >1 $R_E$ and >5 min | Global activity indices availability | Stephens et al. (2019); Sitnov et al. (2008); |
| | | magnetic field architectures | indices SML, SMR & their time derivatives | parameters derived from data | | | | Tsyganenko &Sitnov (2007) |
| Global in situ data reconstruction | Multi-mission | kNN statistical method | Solar wind parameters | 3D global reconstruction of outer magnetosphere and magnetosheath properties | SW propagation from L1, magnetopause and bow shock models for normalizing data location | Dayside magnetosphere, magnetopause, and magnetosheath | | Michotte et Welle et al. (2022) |

[1]The code can be run at CCMC (https://ccmc.gsfc.nasa.gov/).

[2]$u_{sw}$: solar wind speed; $B_z$: north-south component of the interplanetary magnetic field in geocentric solar magnetospheric (GSM) coordinates

**Table 2** Methods for coordinate systems, frame velocity, and spacecraft trajectory estimation

| Method | # of spacecraft (SC) needed | Underlying theory, concept | Input | Output | Assumptions | Spatial or temporal scale of interest | References |
|---|---|---|---|---|---|---|---|
| MVAB | 1 | $\nabla \cdot \mathbf{B} = 0$ | $\mathbf{B}$ | Variance directions; normal direction; LMN coordinates | 1D structure to get a good estimate for the normal ( $\mathbf{e}_N$ ) direction | N/A | Sonnerup and Scheible (1998) |
| MVAVe | 1 | None | $\mathbf{u}_e$ | Variance directions | Presence of electron outflow | Sub-ion scale | Genestreti et al. (2018) |
| | | | | | jets | | |
| MDDB | 4 | None | **B**, SC position | Gradient directions; LMN coordinates | Steady structure | >SC separation | Shi et al. (2019) |
| Dimensionality | 4 | MDDB | MDDB eigenvalues | Dimensionality indices | Steady structure | >SC separation | Rezeau et al. (2018) |
| Hybrid method | 4 | MDDB + MVAB or MVAVe | **B**, SC position (and $\mathbf{u}_e$) | LMN coordinates | Large maximum eigenvalues | >SC separation | Denton et al. (2018); Heuer et al. (2022) |
| HT analysis | 1 | $\partial\mathbf{B}/\partial t = 0$ | **B**, **u** (or **E**) | Structure velocity | Not in diffusion regions | Ion or MHD scale | deHoffmann and Teller (1950); Khrabrov and Sonnerup (1998) |
| MFR | 1 | Faraday's law, $\partial\mathbf{B}/\partial t = 0$ | **E** (or **u**), **B** | Normal direction and velocity | 1D structure, $d/dt = 0$, constant velocity | >SC separation | Dunlop and Woodward (1998) |
| 4SC timing | 4 | None | Various | Normal direction and velocity | 1D structure, $d/dt = 0$, constant velocity | >SC separation | Dunlop and Woodward (1998) |
| STD | 4 | B changes from convection | **B**(t), SC position | Structure velocity | Partial time derivative=0 | >SC separation | Shi et al. (2006, 2019) |
| Reconstructed X-line motion | 4 | $\nabla\cdot\mathbf{B} = 0$; Ampère's law | **B**, **j**, SC position | Velocity in reconnection | Various depending on model | Spacecraft spacing | Denton et al. (2021) |
| | | | | plane | | | |
| Comparison with simulation | 1 or more | PIC code applicable | Various | Path through simulation domain | Simulation is realistic | Simulation scale | Shuster et al. (2017); Schroeder et al. (2022) |
| MCA | 4 | None | **B**, SC position | B-variance eigenvectors; planarity; elongation; dimensionality | Steady structure | >SC separation | Fadanelli et al. (2019) |

**Table 3** Methods for reconstructing 2D/3D structures

| Method | # of spacecraft (SC) needed | Underlying theory, concept | Input | Output | Assumptions | Spatial or temporal scale of interest | Other requirements if any | References |
|---|---|---|---|---|---|---|---|---|
| 3D B-field quadratic reconstruction | 4 | Maxwell equations | **B**, **j** at 4 SC locations | Full 3D vector B-field near tetrahedron | No quadratic terms in minimum variance direction | Spacecraft spacing | | Torbert et al. (2020) |
| 3D polynomial reconstruction[1] | 4 | $\nabla \cdot \mathbf{B} = 0$ ; Ampère's law | **B**, **j** at 4 SC locations | Full 3D vector B-field near tetrahedron | Various depending on model | Spacecraft spacing | | Denton et al. (2020) |
| 3D Polynomial Reconstruction with | 4 | $\nabla \cdot \mathbf{B} = 0$ ; Ampère's law | **B**, **j** at 4 SC locations | Full 3D vector B-field near | Various depending on model | Spacecraft spacing | | Denton et al. (2022) |
| multiple input times[2] | | | | tetrahedron | | | | |
| 3D empirical reconstruction using stochastic optimization method | 4 | Simultaneous perturbation stochastic approximation | **B**, **j** at 4 SC locations | Full 3D vector B-field near tetrahedron | $\nabla \cdot \mathbf{B} = 0$ ; Ampère's law | Spacecraft spacing | | Zhu et al. (2022) |
| Simulation-assisted 2D reconstruction of an EDR | Multi-SC | Comparison with kinetic simulation | Various | 2D fields of all fluid quantities | Simulation is realistic | Ion skin depth | | Schroeder et al. (2022) |
| Grad-Shafranov reconstruction | 1 | Static magnetic field, $\nabla \cdot \mathbf{B} = 0$ , Ampère's law | **B**, pressure | 2D maps of **B** & pressure | 2D, MHD equilibrium | MHD scale | Predefined structure velocity | Sonnerup et al. (2006) |
| MHD reconstruction | 1 | MHD equations | MHD parameters (occasionally incl. *P*-anisotropy) | 2D maps of MHD quantities | 2D, steady structure; various depending on model | MHD scale | Predefined structure velocity | Sonnerup et al. (2006); Chen and Hau (2018); Teh (2019); Tian et al. (2020) |
| EMHD reconstruction[3] | 1 | EMHD equations | **B**, **E**, electron moments ($n$, $\mathbf{u}_e$, $T_e$) | 2D maps of B, E, and electron moments | 2D, steady structure; various depending on model | Electron scale | Predefined structure velocity | Sonnerup et al. (2016); Hasegawa et al. (2021); Korovinskiy et al. (2021, 2023) |
| 3D RBF-based reconstruction | Multi-SC | Euler potentials expressed by modified radial basis functions | **B** | Full 3D vector B-field near tetrahedron | Steady structure; $\nabla \cdot \mathbf{B} = 0$ | Spacecraft spacing | | Chen et al. (2019) |

[1]The code is available at Zenodo (https://doi.org/10.5281/zenodo.3906853).

[2]The code is available at Zenodo (https://doi.org/10.5281/zenodo.6395044).

[3]The code is available at Zenodo (https://doi.org/10.5281/zenodo.5144478).

**Table 4** Methods for diffusion region identification

| Method | # of spacecraft (SC) needed | Underlying theory, concept, or name of machine learning method | Input | Output | Assumptions | Spatial or temporal scale of interest | Other requirements if any | References |
|---|---|---|---|---|---|---|---|---|
| Electron-frame dissipation measure[1] | 1 | $\mathbf{j}\cdot(\mathbf{E}+\mathbf{u}_e\times\mathbf{B}) - \rho_c(\mathbf{u}_e\cdot\mathbf{E})$ | $\mathbf{j}$, $\mathbf{E}$, $\mathbf{B}$, $\mathbf{u}_e$ [$\rho_c$] | Scalar in units of [W/m$^3$] | Maxwell's equations | | | Zenitani et al. (2011) |
| Agyrotropy | 1 | Particle distribution function | $\mathbf{P}$, $\hat{\mathbf{b}}$ | Scalar | None | Diffusion region | | Swisdak (2016) |
| Pressure strain | 4 | $-(\mathbf{P}\cdot\nabla)\cdot\mathbf{u}$ | $\mathbf{P}$, $\mathbf{u}$ | Ion and electron dissipation rate in scalar | Differentiable velocity; curlometer | | | Yang et al. (2017); Bandyopadhyay et al. (2021) |
| Electron vorticity | 4 | $\Omega_e \sim \omega_{ce}$ (electron gyrofrequency) defines the characteristic frequency of the EDR | $\mathbf{v}_e$ | Electron vorticity ($\mathbf{\Omega}_e = \nabla\times\mathbf{u}_e$) | None | EDR | | Hwang et al. (2019) |
| Magnetic flux transport (MFT)[2] | 1 or 4 | $\partial_t\psi + \mathbf{U}_\psi\cdot\nabla_\perp\psi = 0$ (Advection equation of magnetic flux) | $\mathbf{B}$, $\mathbf{E}$, SC position | MFT velocity ( $\mathbf{U}_\psi$ ) or its divergence ($\nabla\cdot$ | Quasi-2D reconnection ( $k_M \ll$ | Diffusion region | Predefined LMN coordina | Li et al. (2021); Qi et al. (2022) |
| | | | | $\mathbf{U}_{\psi}$) | $k_{\perp}$ ); no magnetic field diffusion or generation | | tes and frame velocity | |
| Machine learning 1 to classify regions | 1 | feed-forward multilayer perceptron | Plasma and field quantities, reduced electron distribution functions | 4 classes (IDR, EDR, boundary layers, magnetosphere) | None | IDR, EDR, boundary layers, magnetosphere | | Lenouvel et al. (2021) |
| Machine learning 2 to identify EDRs | 1 | Convolutional Neural Network | Electron distribution functions | EDR / no EDR | None | Diffusion regions | | Lenouvel (2022) |

[1]$\rho_c$: electric change density

[2]$k_M$: wave number corresponding to the scale of the spatial variation in the direction of the reconnection X-line; $k_{\perp}$: wave number corresponding to the scale of the spatial variation in the plane transverse to the X-line

**Table 5** Methods for analyzing the electron diffusion region

| Method | # of spacecraft (SC) | Underlying theory, concept | Input | Output | Assumptions | Spatial or temporal scale of | Other requirements if any | References |
|---|---|---|---|---|---|---|---|---|
| | needed | | | | | interest | | |
| Anomalous transport terms estimation from lower hybrid waves[1] | 4 | Fluid equations, wave-particle interactions | **B**, **E**, particle moments ($n$, $\mathbf{u}$, $\mathbf{P}$) | Anomalous transport terms | None | Sub-ion scales | Particle moments that resolve the waves | Graham et al. (2022) |
| Computing $\partial f_\mathrm{e}/\partial t$ in the spacecraft frame | 1 | Electron Vlasov equation | Electron distribution function $f_\mathrm{e}$ at multiple times | $\partial f_\mathrm{e}/\partial t$ (sometimes an approximation for $\partial f_\mathrm{e}/\partial N$) | Cadence of electron measurements is sufficient for $\partial f_\mathrm{e}/\partial t$ estimation | Electron scale for EDR | Consult with the FPI team for proper implementation | Shuster et al. (2019, 2021a,b) |
| Computing the spatial gradient $\mathbf{v}\cdot\nabla f_\mathrm{e}$ term | 4 | Electron Vlasov equation | $f_\mathrm{e}$, velocity-space coordinates for $f_\mathrm{e}$, SC position | $\nabla f_\mathrm{e}$ and $\mathbf{v}\cdot\nabla f_\mathrm{e}$ | SC separation is sufficient for estimating $\nabla f_\mathrm{e}$ | Electron scale for EDR | Consult with the FPI team for proper implementation | Shuster et al. (2019, 2021a,b) |
| Computing the $\mathbf{F}\cdot\nabla_\mathbf{v} f_\mathrm{e}$ term in the spacecraft frame | 1 | Electron Vlasov equation | $f_\mathrm{e}$, velocity-space coordinates for $f_\mathrm{e}$, **E**, **B** | $\nabla_\mathbf{v} f_\mathrm{e}$ and each of the 3rd term of the Vlasov equation | Velocity-space resolution is sufficient for estimating $\nabla_\mathbf{v} f_\mathrm{e}$ | Electron scale for EDR | Consult with the FPI team for proper implementation | Shuster et al. (2019, 2021a,b) |
| Non- | 1 | Particle | **B**, particle | Scalar non- | None | Kinetic scale | None | Greco et al. |
| Maxwellianity[2] | | distribution function | distribution, particle moments | Maxwellianity value | | of particles of interest | | (2012); Servidio et al. (2017); Graham et al. (2021) |
| Kinetic entropy | 1 | Boltzmann's entropy theory | Particle distribution, particle moments | Entropy densities, scalar non-Maxwellianity value | None | Kinetic scale of particles of interest | Velocity space is properly resolved by particle measurements | Argall et al. (2022) |

[1]The scripts and data required to reproduce the figures in Graham et al. (2022) can be found at https://zenodo.org/records/6370048, and the routines require the irf-matlab software package: https://github.com/irfu/irfu-matlab.

[2]The routines to compute non-Maxwellianity can be found at https://github.com/irfu/irfu-matlab.

**Table 6** Methods for reconnection electric field ($E_M$) estimation

| Method | # of spacecraft (SC) needed | Underlying theory or concept | Input | Output | Assumptions | Spatial or temporal scale of interest | Other requirements if any | References |
|---|---|---|---|---|---|---|---|---|
| Direct measurement | 1 | Faraday's law | **E** | $E_M$ | 2D, steady | N/A | Accurate LMN | Genestreti et al. (2018); Burch et al. (2020) |
| Inflow velocity | 1 | | $\mathbf{u}_e$, $n$, **B** | Inflow velocity; normalized reconnection rate | 2D, steady | EDR | Accurate LMN | Burch et al. (2020) |
| Multi-crescent VDF[1] | 1 | Equations of motion | $B_L$, $E_N$, $N$, $u_{eM}$ | $E_M$ | $B_L$=b$N$, $E_N$=-k$N$, No guide field | Electron scale | $N$=0 crossing | Bessho et al. (2018) |
| Remote sensing at separatrix | More than 2 | Magnetic flux conservation | Timing velocity at separatrix, **E**, **B** | $E_M$ | 2D, constant reconnection rate, uniform field between multi-probes | Ion to MHD scale | Separatrix identification | Nakamura et al. (2018a) |
| Separatrix angle | 1 or 4 (depending on how separatrix is detected) | Flux and force balance along inflow and outflow direction | $B_N$, $B_M$ (or $B_z$, $B_x$ for tail) | Normalized reconnection rate | 2-D, low beta, separatrix angle=exhaust opening angle | Electron to ion scale | Separatrix identification | Liu et al. (2017); Nakamura et al. (2018b) |

[1]$N$: distance in the normal ($N$) direction from the current sheet center ($B_L$=0) plane; see section 3.3.1 for the definitions of b and k.

**Table 7** Methods for region and current sheet identification

| Method | # of spacecraft (SC) | Underlying theory, concept, or name of machine | Input | Output | Assumptions | Spatial or temporal scale of | Other requirements if any | References |
|---|---|---|---|---|---|---|---|---|
| | needed | learning method | | | | interest | | |
| ML-based region identification[1] | 1 | Gradient boosting/machine learning | **B**, ion moments ($n$, $\mathbf{u}_i$, $T_i$) | Near-Earth region labelling, magnetopause and bow shock crossings | None | MHD[2] and larger | N/A | Nguyen et al. (2022a) |
| Region classification using 3D particle distributions | 1 | Convolutional Neural Network/supervised machine learning | 3D ion velocity distributions | Near-Earth region labelling, magnetopause and bow shock crossings | None | MHD and larger | N/A | Olshevsky et al. (2021) |

[1]The code is available at github (https://github.com/gautiernguyen/in-situ_Events_lists).

[2]Magnetohydrodynamics